\documentclass[journal]{IEEEtran}

\usepackage{amsmath}
\usepackage{amssymb}
\usepackage{cite}
\usepackage{hyperref}
\usepackage{graphicx}
\usepackage{comment}
\usepackage{color}
\usepackage{mathtools}
\usepackage{enumitem}
\usepackage{subfig}
\usepackage[dvipsnames]{xcolor}
\usepackage{booktabs}
\usepackage{multirow}
\usepackage{rotating}
\usepackage{floatpag}
\usepackage{times}
\usepackage{epsfig}
\usepackage{wrapfig}
\usepackage{soul}

\definecolor{electricultramarine}{rgb}{0.25, 0.0, 1.0}
\newcommand{\ch}[1]{\textcolor{black}{#1}}
\newcommand{\chh}[1]{\textcolor{black}{#1}}
\newcommand{\red}[1]{\textcolor{red}{#1}}
\newcommand{\magenta}[1]{\textcolor{magenta}{#1}}

\newcommand{\green}[1]{\textcolor{PineGreen}{#1}}
\newcommand{\blue}[1]{\textcolor{blue}{#1}}

\begin{document}
\title{JPD-SE: High-Level Semantics for Joint Perception-Distortion Enhancement in Image Compression}

\author{Shiyu~Duan*, Huaijin~Chen, and Jinwei~Gu
\thanks{SD (\href{mailto:michaelshiyu3@gmail.com}{michaelshiyu3@gmail.com}) is with the Department of Electrical and Computer Engineering, University of Florida, Gainesville, FL, 32611. HC (\href{mailto:chenhuaijin@sensebrain.site}{chenhuaijin@sensebrain.site}) and JG (\href{mailto:gujinwei@sensebrain.site}{gujinwei@sensebrain.site}) are with SenseBrain Technology, San Jose, CA, 95131. \newline *Work done during internship at SenseBrain.
}
}

\markboth{}
{Duan, Chen, and Gu: Semantics for Perception-Accuracy-Distortion Enhancement in Compression}

\maketitle

\begin{abstract}
    While humans can effortlessly transform complex visual scenes into simple words and the other way around by leveraging their high-level understanding of the content, conventional or the more recent learned image compression codecs do not seem to utilize the semantic meanings of visual content to their full potential.
    Moreover, they focus mostly on rate-distortion and tend to underperform in perception quality especially in low bitrate regime, and often disregard the performance of downstream computer vision algorithms, which is a fast-growing consumer group of compressed images in addition to human viewers.
    In this paper, we (1) present a generic framework that can enable any image codec to leverage high-level semantics and (2) \chh{study the joint optimization of perception quality and distortion}.
    Our idea is that given \textit{any} codec, we utilize high-level semantics to augment the low-level visual features extracted by it and produce essentially a new, semantic-aware codec.
    \chh{We propose a three-phase training scheme that teaches semantic-aware codecs to leverage the power of semantic to jointly optimize rate-perception-distortion (R-PD) performance.
    As an additional benefit, semantic-aware codecs also boost the performance of downstream computer vision algorithms.} 
    To validate our claim, we perform extensive empirical evaluations and provide both quantitative and qualitative results.
\end{abstract}


\section{Introduction}
\IEEEPARstart{T}{he} pervasive use of mobile devices equipped with powerful cameras has pushed the generation of high-definition images and videos to an unprecedented rate.
Such constant and ubiquitous data collection poses an urgent motivation for better visual data compression techniques.

\begin{figure}
    \includegraphics[width=1\columnwidth]{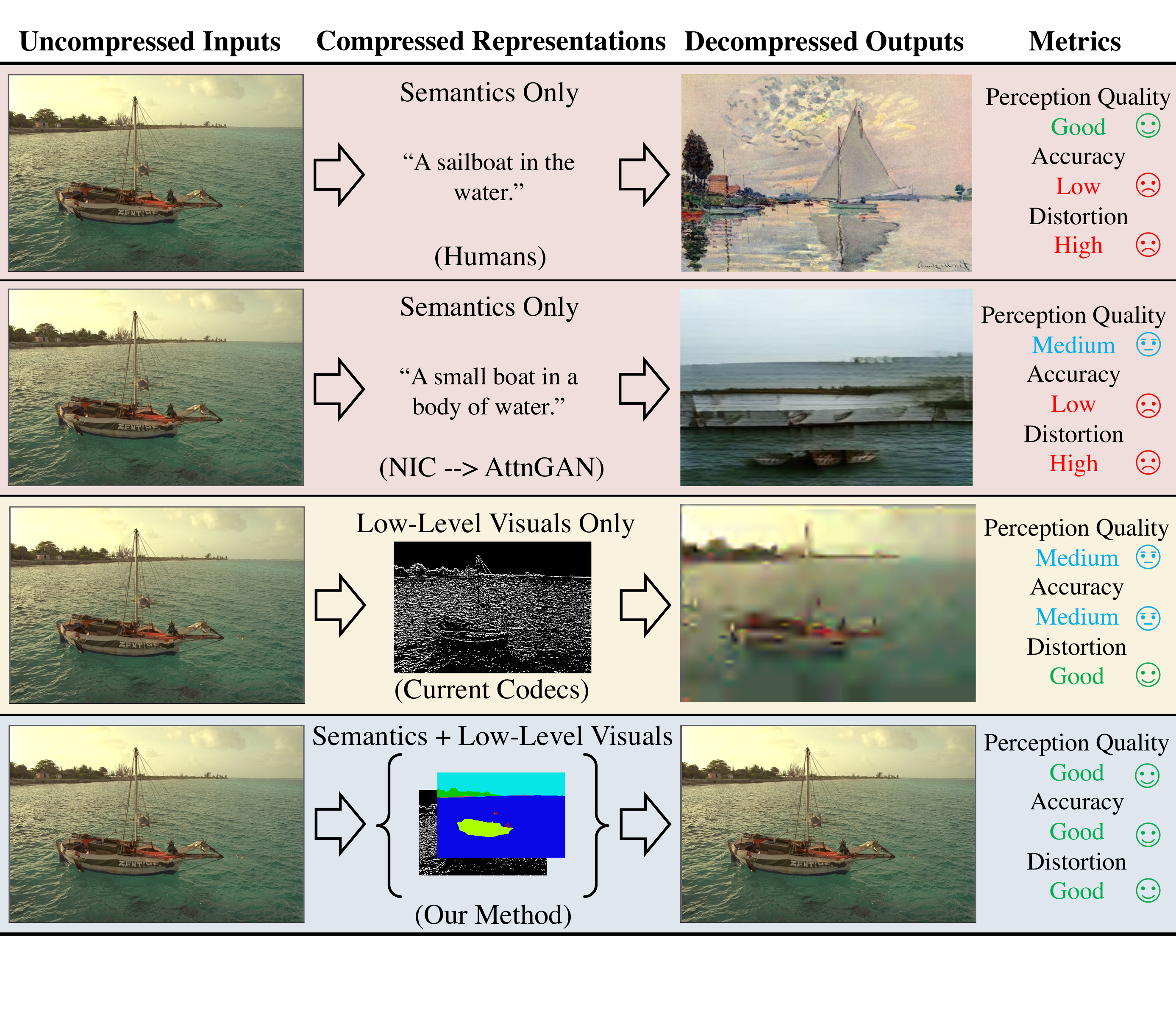}
    \caption{
        Human can leverage high-level semantics to compress efficiently (top row).
        While this behavior can be replicated by deep learning models~\cite{vinyals2015show,xu2018attngan} (the second row from the top), existing compression codecs underutilize semantics (the third row).
        \ch{We enhance existing codecs by augmenting the low-level visuals with semantics and consequently let the codecs leverage a high-level ``understanding'' to compress better along three performance axes: perception quality~\cite{blau2019rethinking}, performance of downstream vision algorithms (accuracy), and distortion (bottom row).}
    }
    \label{fig0}
\end{figure}

Consider how humans compress visual input (Fig.~\ref{fig0}): The sender first comprehends the visual content and then extracts high-level semantic descriptions.
He or she then describes them in a few words and the receiver, upon obtaining this information, can reconstruct visually complex scenes.
The fact that we can all leverage our understanding and associate concise high-level concepts such as ``boat'' with pixel intensities and arrangements enables us to condense even tens of millions of pixels into several words and the other way around.
Being able to effortlessly leverage high-level meanings of visual content to increase the efficiency of its processing has been a long-sought goal since at least the advent of MPEG-7~\cite{manjunath2002introduction}.
However, efforts in this direction have seen limited success due to the sheer difficulties in computationally implementing how humans extract and leverage the high-level meanings of low-level visual features.
As of today, existing image compression algorithms still underutilize high-level semantics~\cite{webp,bpg,taubman2012jpeg2000,agustsson2017soft,agustsson2019generative,toderici2016variable,toderici2017full,balle2016end,balle2017end,balle2018variational,rippel2017real,minnen2018joint,li2018learning,mentzer2018conditional,johnston2018improved,tschannen2018deep,theis2017lossy,li2020learning}.

Another limitation in existing codec design paradigms is that the models are typically optimized only for rate-distortion (R-D), yet the traditional R-D metrics are insufficient especially in low bitrate~\cite{blau2018the,blau2019rethinking}, causing the codecs to produce visually unappealing results in low-rate compression.\footnote{As a sign of the image compression community realizing the severity of this issue, the low-rate track of the 2020 Workshop and Challenge on Learned Image Compression (CLIC) is for the first time using human ratings over PSNR/MS-SSIM~\cite{wang2003multiscale} as the primary metric~\cite{clic2020}.}
Apart from human viewers, existing codecs also ignore ``machine viewers'', i.e., the downstream computer vision algorithms performing tasks such as face recognition, object tracking, etc. on the decoded content.
With the fast-expanding use of machine learning tools in image understanding tasks, rate-accuracy (R-A), where ``accuracy'' refers to the performance of downstream machine viewers, must become another important axis when evaluating image codec.

\begin{figure*}[t] 
    \centering
    \subfloat{
        \rotatebox{90}{\scriptsize{(a) JPEG (0.24 bpp)}}
        \includegraphics[trim={25cm 0 0 0},clip,width=.28\columnwidth]{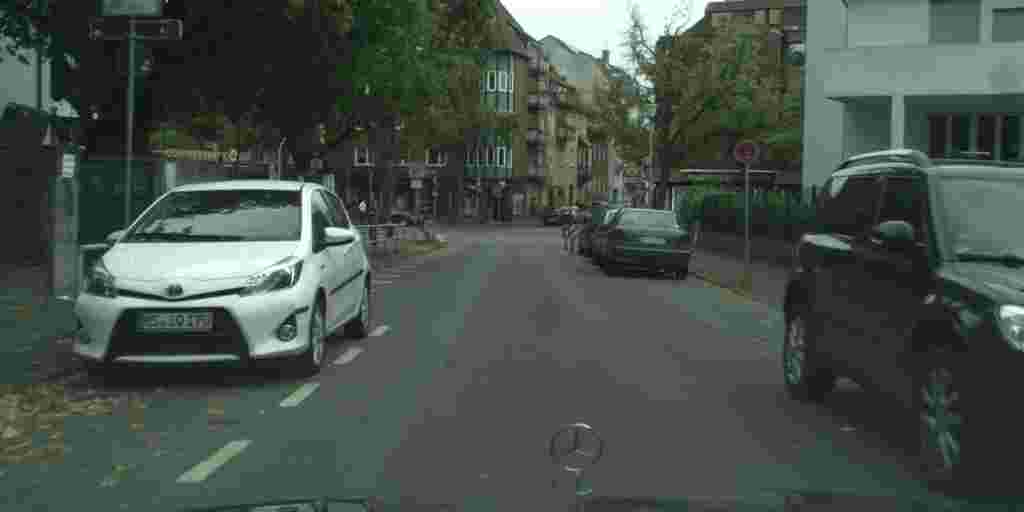}
    }
    \subfloat{
        \hspace{-.095in}
        \includegraphics[trim={25cm 0 0 0},clip,width=.28\columnwidth]{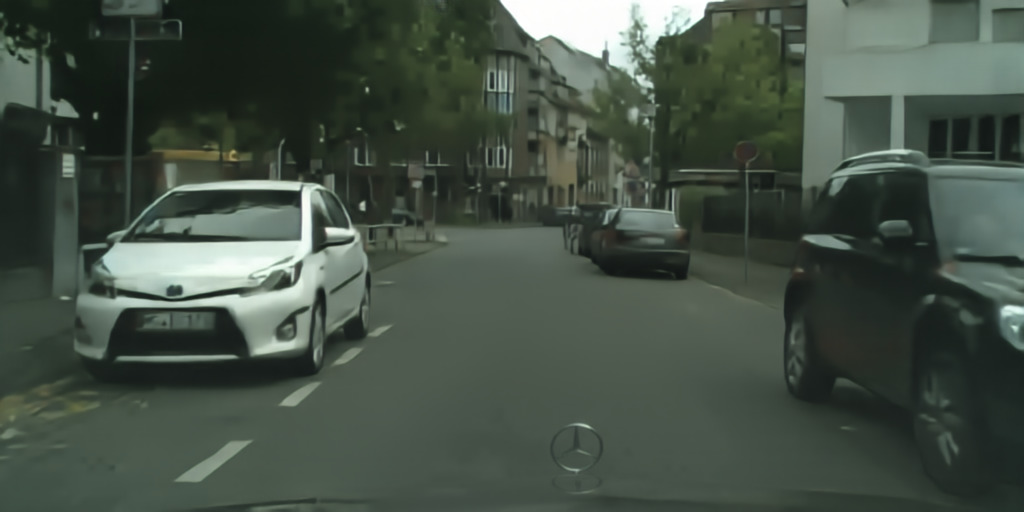}
        \rotatebox{90}{\scriptsize{(b) JPEG-SE (0.21 bpp)}}
    }
    \subfloat{
        \rotatebox{90}{\scriptsize{(c) BPG (0.05 bpp)}}
        \includegraphics[trim={25cm 0 0 0},clip,width=.28\columnwidth]{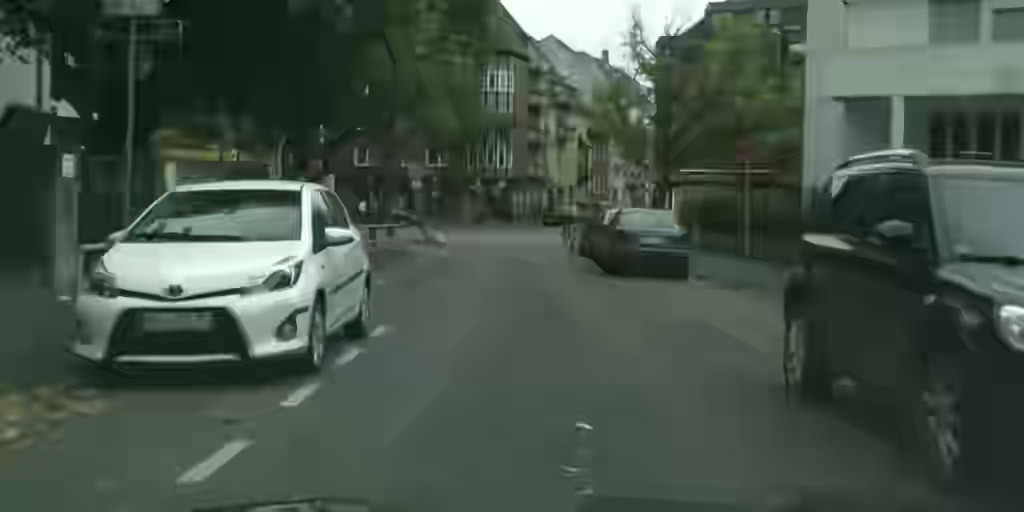}
    }
    \subfloat{
        \hspace{-.095in}
        \includegraphics[trim={25cm 0 0 0},clip,width=.28\columnwidth]{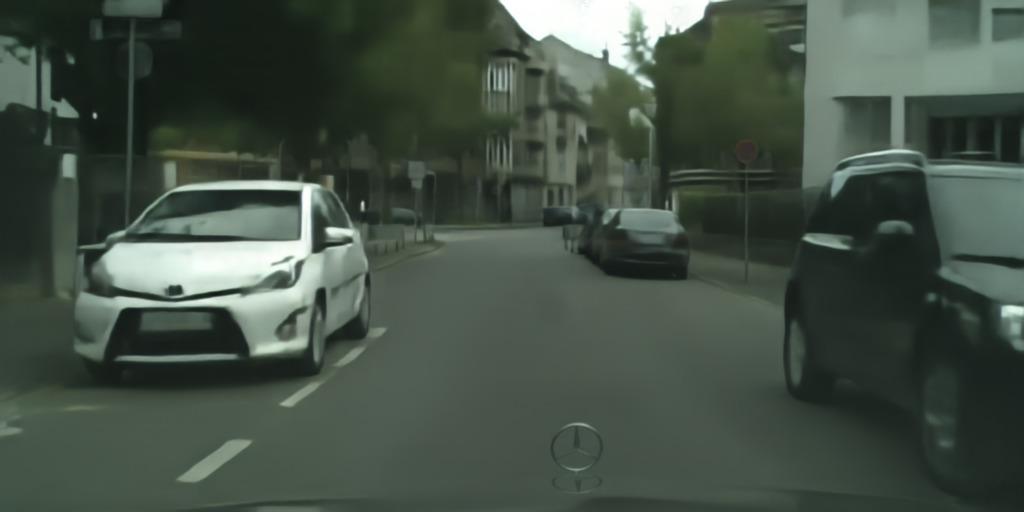}
        \rotatebox{90}{\scriptsize{(d) BPG-SE (0.05 bpp)}}
    }
    \subfloat{
        \rotatebox{90}{\scriptsize{(e) WebP (0.11 bpp)}}
        \includegraphics[trim={25cm 0 0 0},clip,width=.28\columnwidth]{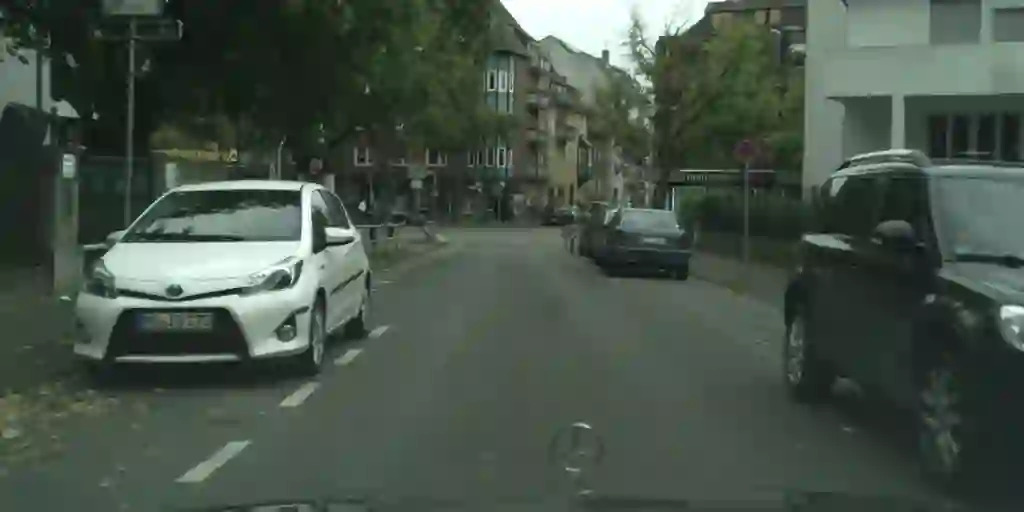}
    }
    \subfloat{
        \hspace{-.095in}
        \includegraphics[trim={25cm 0 0 0},clip,width=.28\columnwidth]{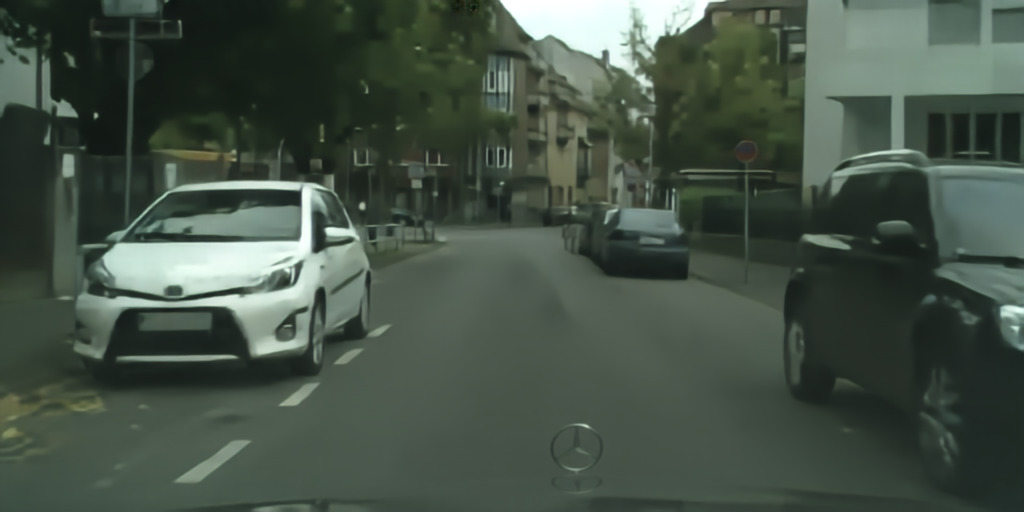}
        \rotatebox{90}{\scriptsize{(f) WebP-SE (0.10 bpp)}}
        \label{fig3f1}
    }
    \vspace{-10pt}
    \subfloat{
        \rotatebox{90}{\scriptsize{(g) learned (0.21 bpp)}}
        \label{fig3g1}
        \includegraphics[trim={25cm 0 0 0},clip,width=.28\columnwidth]{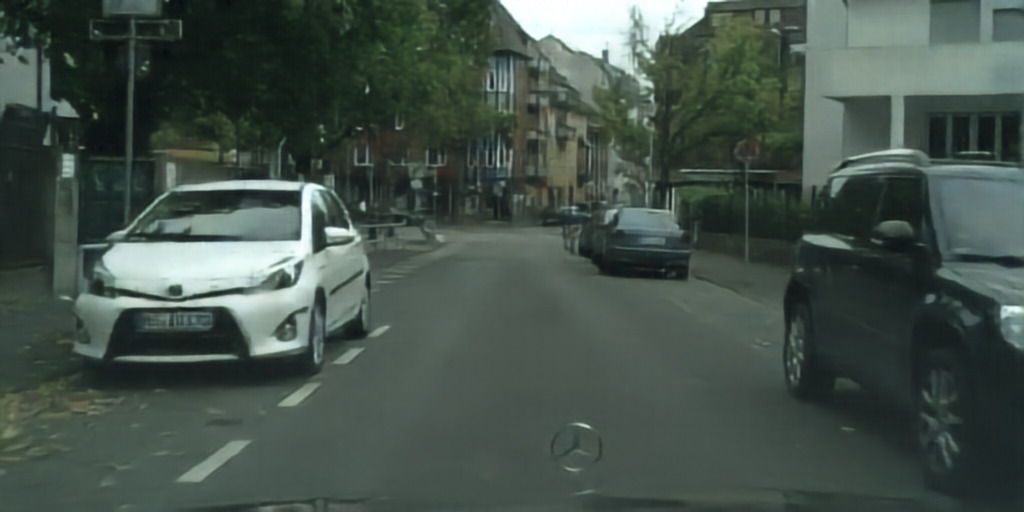}
    }
    \subfloat{
        \hspace{-.095in}
        \includegraphics[trim={25cm 0 0 0},clip,width=.28\columnwidth]{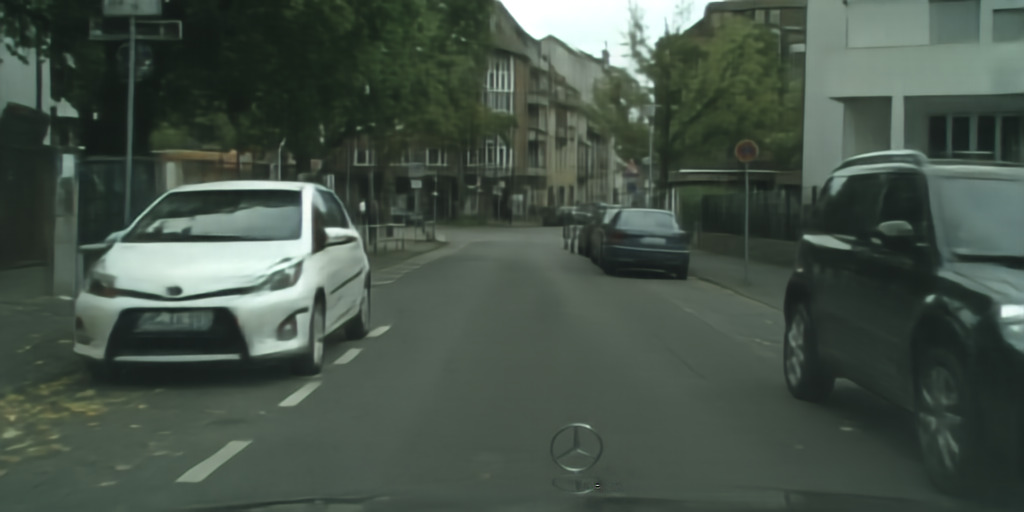}
        \rotatebox{90}{\scriptsize{(h) learned-SE (0.14 bpp)}}
        \label{fig3h1}
    }
    \subfloat{
        \rotatebox{90}{\scriptsize{(i) JP2 (0.12 bpp)}}
        \label{fig3g}
        \includegraphics[trim={25cm 0 0 0},clip,width=.28\columnwidth]{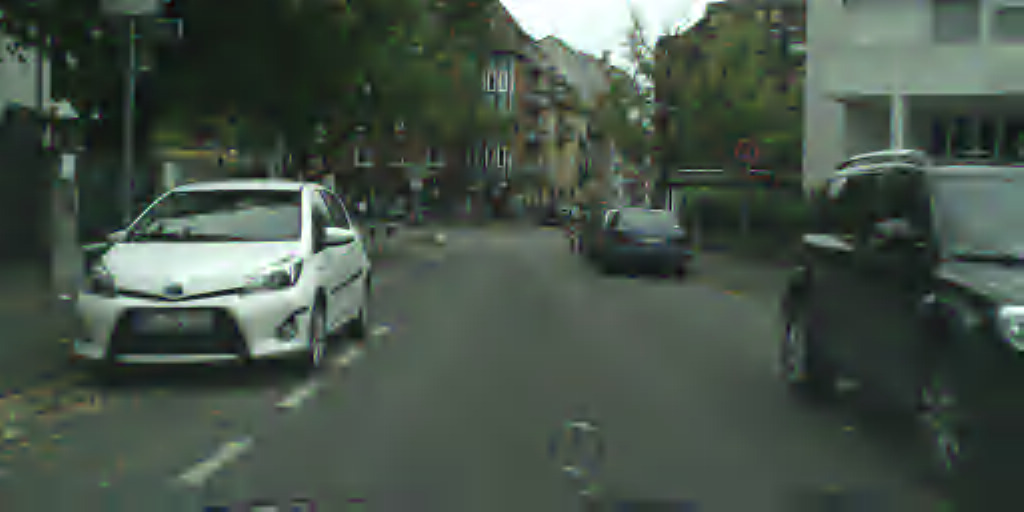}
    }
    \subfloat{
        \hspace{-.095in}
        \includegraphics[trim={25cm 0 0 0},clip,width=.28\columnwidth]{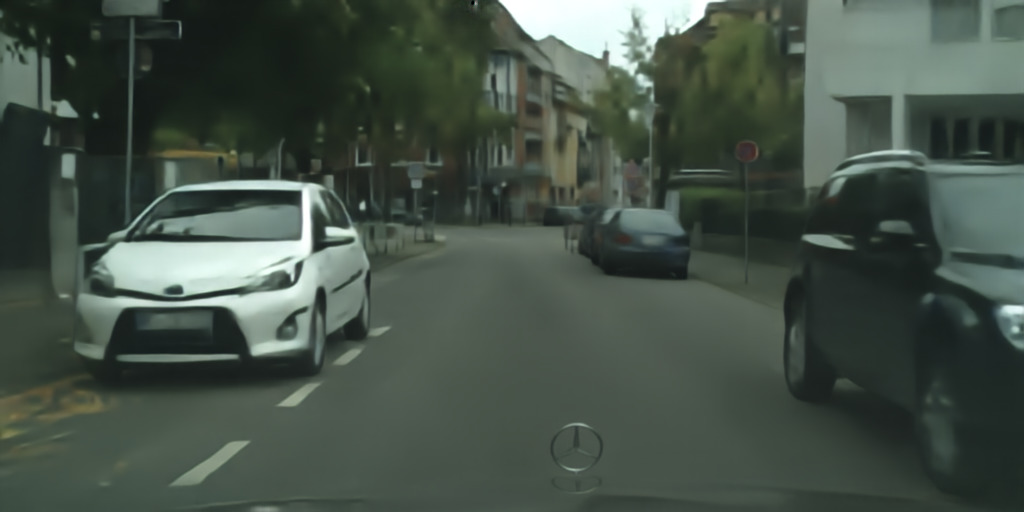}
        \rotatebox{90}{\scriptsize{(j) JP2-SE (0.11 bpp)}}
        \label{fig3h}
    }
    \subfloat{
        \rotatebox{90}{\scriptsize{(k) JP2-SE (-s; test)}}
        \label{fig3f}
        \includegraphics[trim={25cm 0 0 0},clip,width=.28\columnwidth]{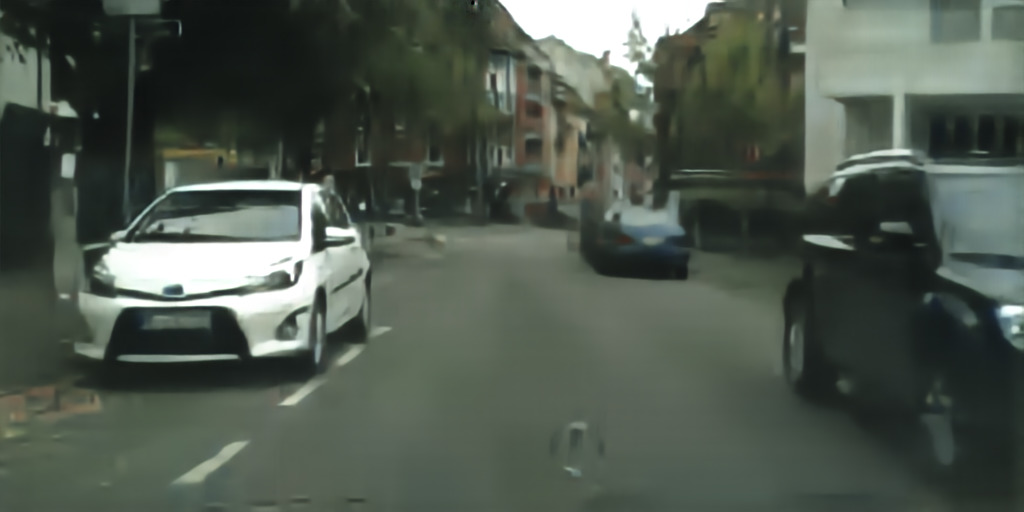}
    }
    \subfloat{
        \hspace{-.095in}
        \includegraphics[trim={25cm 0 0 0},clip,width=.28\columnwidth]{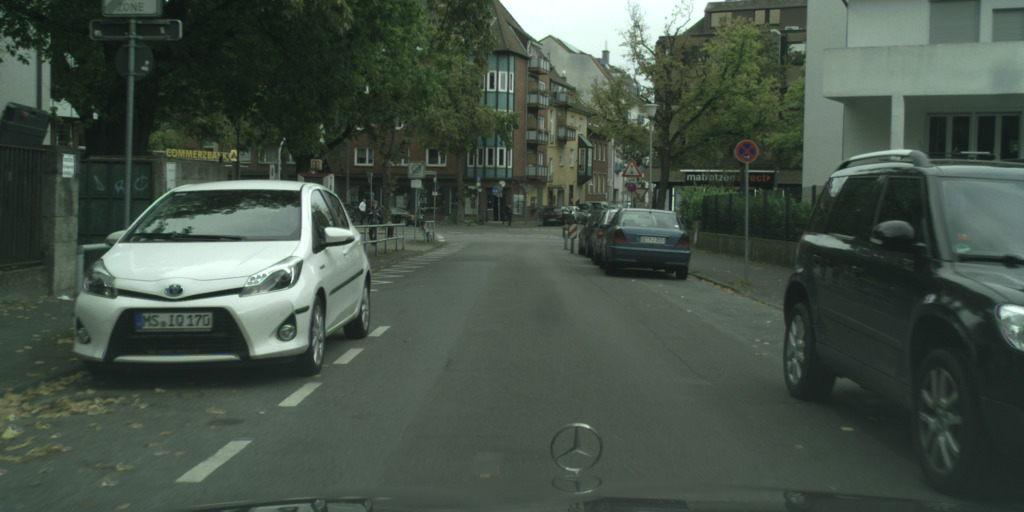}
        \rotatebox{90}{\scriptsize{(l) Original}}
        \label{fig3l}
    }
    \caption{
        Semantically-enhanced (SE) codecs vs. originals.
        Note how the SE codecs are able to ``understand'' an input image and produce a more natural and visually appealing reconstruction in places where the originals failed (e.g., the boundary between the tires and the road).
        Fig.~\ref{fig3f} was produced by the same SE codec that produced~\ref{fig3h} using dummy semantics during test (-s; test), illustrating the importance of semantics.
        Shown are cropped images.
        Full images are in the supplementary materials.
    }
    \label{fig3}
\end{figure*}

\begin{figure*}[t]
    \centering
    \includegraphics[width=1\textwidth]{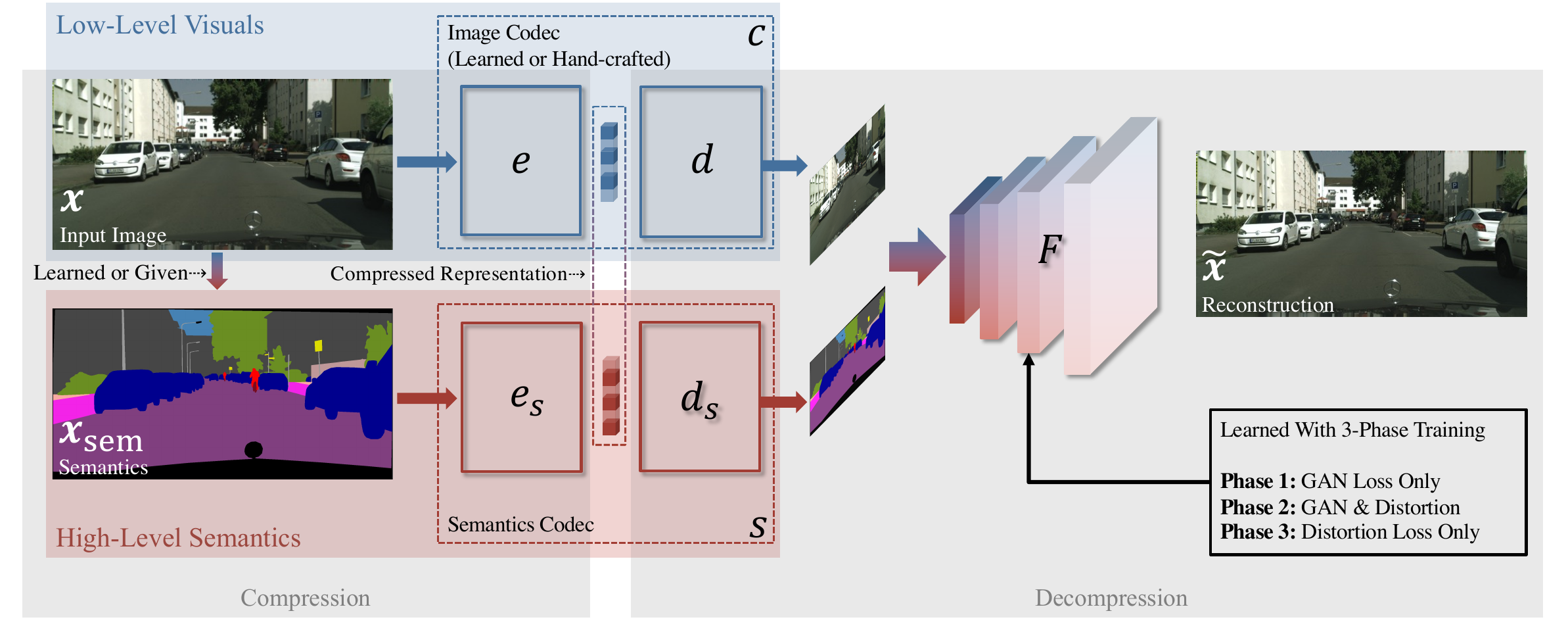}
    \caption{
        Our proposed framework consists of a given backbone image codec \(c\), a semantics codec \(s\), and a fusion network \(F\).
        To compress, we use \(e\) and \(e_s\) together as the encoder that constructs a hidden representation of the image with both high-level semantics and low-level visuals.
        To decompress, \(d,\, d_s\), and \(F\) work jointly as the decoder.
        \ch{\(d\) and \(d_s\) reconstruct visuals and semantics from the compressed representation, respectively.}
        \(F\) fuses their outputs to produce the final result.
        \ch{\(F\) is learned, whereas the other modules are not necessarily so.}
        In particular, we tested with both learned and engineered codecs as the backbone \(c\).
        For the learned components, a three-phase training scheme is proposed to effectively optimize for both distortion and perception quality. 
        We focus on segmentation maps as semantics for this work although the framework is generic.
    }
    \label{fig1}
\end{figure*}

We propose a universal framework to achieve \chh{\textbf{j}oint \textbf{p}erception-\textbf{d}istortion optimization with \textit{any} given image codec through \textbf{s}emantic \textbf{e}nhancement (JPD-SE)}.
\ch{Given any image codec, the idea is to augment its hidden representations constructed from low-level visuals with high-level semantics. 
Then a synthesis network leverages this semantics to enhance the codec output (Fig.~\ref{fig1}).}
\chh{A three-stage training scheme is used to teach the model to best leverage semantics to achieve joint perception-distortion optimization.}

\chh{By simply feeding semantics to the codec, we give the codec more leverage to improve perception-distortion-accuracy performance.}
First, just like humans know that the sky is usually blue and grass green, a semantic-aware codec can learn what low-level features are more likely to appear in different semantic regions.
This ``prior knowledge'' enables the decoder to create more natural and perceptually pleasing reconstructions.
This also helps remove noise on the low-level visual features without training on noisy examples since noise can be considered as unnatural visuals.
Second, since complex low-level features can often be packed into concise high-level semantic concepts, more information can be conveyed using a blend of semantics and low-level visuals than using only the latter.
Therefore, a semantic-aware codec can leverage this extra degree of freedom and save bits.
Finally, this high-level comprehension of the image helps the codec retain important semantic information that is critical for downstream computer vision tasks such as object detection or tracking.

\chh{To teach the model to utilize semantics, we propose a three-stage training scheme aiming at jointly optimize perception-distortion performance.
The first stage trains the model with a GAN loss, guiding it to hallucinate photorealistic pixels from semantics.
The second stage combines the GAN loss with a distortion loss to jointly improve perceptual quality and distortion.
Finally, training with both GAN and distortion empirically leads to models that overly emphasize on perceptual quality, generating hallucination artifacts.
Therefore, we train for a third stage with distortion loss alone, producing better joint perception-distortion performance.
Although our training does not directly optimize rate-accuracy performance, we demonstrate through a posteriori evaluation that semantically-enhanced codecs can boost the performance of downstream vision algorithms.}

We test on Cityscapes~\cite{cordts2016the} and ADE20k~\cite{zhou2016semantic,zhou2017scene} at 1024\(\times\)512 resolution with learned and conventional codecs (JPEG, JPEG 2000~\cite{taubman2012jpeg2000}, WebP~\cite{webp}, and BPG~\cite{bpg}) as the backbone image codec and semantic and instance segmentation maps as the semantics.
We both quantitatively and qualitatively show that our semantically-enhanced (SE) codecs achieve favorable rate-perception-accuracy-distortion performance as quantified via a large-scale user study, accuracy of a downstream bounding-box object detector, and three distortion measures (LPIPS~\cite{zhang2018the}, MS-SSIM~\cite{wang2003multiscale}, and PSNR).
Extensive ablation study was performed to validate the usefulness of the proposed components.
In particular, we demonstrate that the improvements are beyond what could be achieved with a post-processing network using purely low-level visuals, proving the value of semantics.

To sum up, the main contributions of this work include:
\begin{itemize}[label=$\bullet$]
    \setlength\itemsep{0em}
    \item We conduct a systematic study on the efficacy of semantics as a fundamental supplement to visuals in the general image compression setting.
    \item We propose a generic framework that enables any given codec, learned or hand-crafted, to leverage high-level semantics. \chh{As part of this framework, a three-stage training scheme guides the codec to leverage semantics to jointly optimize perception-distortion.} 
    \item We \chh{evaluate} perception-accuracy-distortion performance on full-resolution images quantified by a large-scale user study, accuracy of downstream bounding-box object detection, and three distortion metrics.
    \item We hypothesize and verify that enhancing codecs with high-level semantics improves perception-accuracy-distortion.
\end{itemize}

Code is available at \href{https://github.com/SenseBrain/JPD-SE}{https://github.com/SenseBrain/JPD-SE}.

\section{Related Work}
\paragraph{Image Compression}
The key components of an image codec include an encoder, which transforms the original image into a more compressible representation, and a decoder, which reconstructs the image from a possibly quantized version of this new representation.
Traditionally, these components were hand-crafted by experts.
Some commonly used engineered image codecs include JPEG, JPEG 2000~\cite{taubman2012jpeg2000}, WebP~\cite{webp}, BPG~\cite{bpg}, PNG~\cite{png}, and FLIF~\cite{flif}.
PNG and FLIF are designed only for lossless compression.
Many works on learned compression have emerged over recent years and the proposed codecs with learnable components produced favorable results compared to the traditional ones but are conceptually much simpler due to their end-to-end nature~\cite{agustsson2017soft,toderici2016variable,toderici2017full,balle2016end,balle2017end,balle2018variational,rippel2017real,minnen2018joint,li2018learning,mentzer2018conditional,johnston2018improved,tschannen2018deep,theis2017lossy}.
Our proposed method can work with any learned or engineered image codec, leveraging high-level semantics to improve compression quality as measured by joint PAD performance.

\paragraph{Semantics in Image Compression}
Most existing codecs do not explicitly utilize high-level semantics during encoding or decoding.
\cite{prakash2017semantic,li2018learning} proposed methods for content-weighted bitrate control without explicitly utilizing high-level semantics.

\cite{agustsson2019generative,akbari2019dsslic} explored explicitly utilizing semantics in image compression but both to a limited extent.
\ch{Specifically, \cite{agustsson2019generative} only used semantics for bitrate allocation in a somewhat constrained setting, requiring that users choose to preserve some semantic regions while ignoring others.
\cite{akbari2019dsslic} used semantics only for up/downsampling. 
Also, BPG was used to compress the residuals of its semantics-aware codec, making it difficult to assess the contributions from semantics.
Apart from the different compression objectives (to be discussed in the next section), our work differs from the previous two in the following aspects.
First, we study how semantics can serve as a fundamental supplement to visuals that boosts PAD performance instead of only as some auxiliary side information.
Second, our work considers the general compression setting without any limiting assumption such as that distinct semantic regions are of different levels of importance to the user, which is essential to~\cite{agustsson2019generative}.}

\paragraph{Compression Objective}
\ch{Most existing works focus only on R-D and ignore perception quality~\cite{taubman2012jpeg2000,webp,bpg,agustsson2017soft,toderici2016variable,toderici2017full,balle2016end,balle2017end,balle2018variational,rippel2017real,minnen2018joint,li2018learning,mentzer2018conditional,johnston2018improved,tschannen2018deep,theis2017lossy,akbari2019dsslic}.
Codecs in this category fail to produce reconstructions that look good to humans in low bitrates despite their strong R-D performance, as observed in previous work ~\cite{agustsson2019generative}.
The underlying reason of this mismatched performance is that distortion, including the more recent ``perceptual'' distortions~\cite{wang2003multiscale,johnson2016perceptual,zhang2018the}, is fundamentally at odds with perception quality especially in low bitrate settings~\cite{blau2018the}.
\cite{agustsson2019generative} focuses on R-P while ignoring distortion completely.
The reconstructions look appealing at the first glance but have low fidelity.
This suggests that when designing codecs, we should simultaneously consider both perception and distortion for better overall quality.
\cite{tschannen2018deep,santurkar2018generative} focus on joint perception-distortion optimization, but only on thumbnail images.}

\ch{To the best of our knowledge, we are the first to focus on jointly optimizing perception quality and distortion on full-resolution images.
We quantify perception quality directly through a user study that has the largest scale among existing studies in terms of the number of codecs tested, the range of bitrates covered, and the number of users involved.}

\ch{We also demonstrate a new approach for optimizing perception quality.
Existing works typically add a quantification term for perception in the form of some GAN loss to the overall objective function~\cite{tschannen2018deep,santurkar2018generative,agustsson2019generative}.
We show that sending semantics to the decoder naturally improves perception quality.}

\ch{In addition to perception and distortion, we propose to evaluate codec performance along a third ``accuracy'' axis.
This new performance axis is becoming increasingly important due to the wide use of vision algorithms on compressed content.
Note that apart from a clear tradeoff between rate and ``accuracy'' (Fig. \ref{afig14}), we do not claim that a tradeoff necessarily exists among “accuracy” and perception or distortion.}

\paragraph{Semantic Image Synthesis}
Semantic image synthesis methods aim at synthesizing photorealistic and semantically consistent images from given semantic descriptions such as text, sketches, and semantic segmentation maps~\cite{xu2018attngan,zhang2017stackgan,zhang2019stackgan,zhang2018photographic,hong2018inferring,qiao2019mirrorgan,reed2016generative,zhu2017unpaired,wang2018high,park2019semantic,chen2017photographic,qi2018semi}.
This problem can be modeled and subsequently solved with a conditional GAN as follows.
Given semantics \(\mathbf{x}_{\text{sem}}\), one would like a generator \(G\) that can generate an image \(G(\mathbf{x}_{\text{sem}})\) that has the given semantics and is distributed according to the distribution of some real image \(\mathbf{x}\).
This goal can be achieved by training \(G\) (and a discriminator \(D\)) with the following conditional GAN objective function:
\begin{equation}
    \label{sem_CGAN_loss}
    L_{\text{GAN}}
    = \mathop{\mathbb{E}} f\left(D\left(\mathbf{x},\,\mathbf{x}_{\text{sem}}\right)\right)
    + \mathop{\mathbb{E}} g\left(D\left(G\left(\mathbf{x}_{\text{sem}}\right),\,\mathbf{x}_{\text{sem}}\right)\right),
\end{equation}
where \(f\) and \(g\) depend on the GAN formulation used.

Recent methods have been able to synthesize high-definition photographic images with pixel-level semantic accuracy from semantic segmentation maps using conditional GANs~\cite{zhu2017unpaired,wang2018high,park2019semantic,chen2017photographic,qi2018semi}.
Instance-wise low-level visual features are sometimes used to enhance the realism of the synthesized images~\cite{wang2018high}.
For semantic image synthesis, there is no ``original image'' and the evaluation standards are completely different: The generated images are typically evaluated in terms of perception quality and semantic consistency.
In this work, we consider rate, perception quality, performance of downstream vision algorithms, and distortion with respect to the original image.

\section{Building Semantically-Enhanced Codecs}
\label{sec1}
\subsubsection{The Main Pipeline}
\ch{Denote the input image as \(\mathbf{x}\in X\) and its semantics \(\mathbf{x}_\text{sem}\in X_\text{sem}\), where \(X\) is a space of image tensors with shape \(H\times W\times C\) and \(X_\text{sem}\) a space of semantics.
The semantics can be obtained computationally with another component that may be jointly optimized with our compression pipeline.}

\ch{\textbf{Backbone codec \(c\):} 
As the core of the pipeline, we need a backbone image codec \(c = d\circ e\) that encodes and decodes the image input, where \(e: X\to Z\) is an encoder, \(d:Z\to X\) a decoder, \(\circ\) denotes function composition, and \(e(\mathbf{x})\in Z\) is a hidden representation of \(\mathbf{x}\) in some feature space \(Z\) that will be transmitted after potential quantization and entropy coding.
For example, this backbone codec can be JPEG or a learned image codec.
And we evaluate different choices of \(c\) in experiments.}

\ch{\textbf{Semantics codec \(s\):} 
Concurrent to the backbone image codec \(c\), we also need a semantics codec \(s = d_s\circ e_s\) that is responsible for encoding and decoding semantics. \(e_s: X_\text{sem}\to Z_\text{sem}\) here is an encoder, \(d_s: Z_\text{sem}\to X_\text{sem}\) a decoder, and \(e_s(\mathbf{x}_\text{sem})\) a hidden representation of \(\mathbf{x}_\text{sem}\). \(Z_\text{sem}\) is a feature space for semantics.
We shall present a non-learned component that serves as \(s\) for encoding and decoding segmentation maps in Section \ref{sem_comp}.}

\ch{\textbf{Fusion network \(F\):}
On top of image codec \(c\) and semantics codec \(s\), we need another fusion network \(F: X\times X_\text{sem}\to X\) that synthesizes the reconstructed image \(\tilde{\mathbf{x}}\) by fusing together the outputs of the two codecs. 
\(F\) should leverage semantics to enhance the output from the backbone codec.
Mathematically, \(F\) maps the tuple \(\left(c(\mathbf{x}), s(\mathbf{x}_\text{sem})\right)\) to \(\tilde{\mathbf{x}}\).
We implement \(F\) using a conditional GAN generator.}

Concisely, the end-to-end pipeline can be described as follows.
\begin{align*}
    \begin{rcases*}
        \mathbf{x}\xmapsto{c}c(\mathbf{x})\\
        \mathbf{x}_\text{sem}\xmapsto{s}s(\mathbf{x}_\text{sem})
    \end{rcases*}
    \left(c(\mathbf{x}), s(\mathbf{x}_\text{sem})\right)\xmapsto{F}\tilde{\mathbf{x}}
\end{align*}
Fig.~\ref{fig1} gives a schematic illustration of this framework.
When \(c\) is a learned codec, it may be included in the joint optimization.
The proposed framework enjoys wide applicability since any existing image codec can be plugged into the pipeline as the backbone \(c\) and be enhanced semantically.

Note that high-level semantics can take many forms.
For example, one can use a class segmentation map, an instance segmentation map, sketches, a set of bounding boxes, text descriptions and so on.
And each form potentially requires an architecturally different \(s\).

\ch{\textbf{Loss Function Components:}
The main components of the training objective function include a distortion term \(L_\text{dist}\), a GAN term \(L_\text{GAN}\), and optionally, a rate term.
The GAN term facilitates the reconstruction of image from a blend of semantics and visuals.
And the distortion term ensures that the reconstruction has high fidelity to the original.
The GAN term can be interpreted as quantifying perception quality as it can be viewed as a divergence measure between the densities of \(\tilde{\mathbf{x}}\) and \(\mathbf{x}\)~\cite{goodfellow2014generative}.}
\chh{We do not include a loss term that optimizes accuracy directly. Instead, through a posteriori evaluation, we demonstrate that the utilization of semantics improves rate-accuracy performance as an additional benefit.}

\ch{When using \(L^1\) loss as the distortion term, we have
\begin{equation}
    L_\text{dist}\left(\tilde{\mathbf{x}}, \mathbf{x}\right) = \left\|\tilde{\mathbf{x}} - \mathbf{x}\right\|_1.
\end{equation}
Some other valid choices for \(L_\text{dist}\) include the \(L^2\) loss (essentially optimizing PSNR directly), or MS-SSIM~\cite{wang2003multiscale}.
The GAN term is based on the standard conditional GAN formulation Eq.~\ref{sem_CGAN_loss}:
\begin{equation}
    \label{sem_CGAN_loss}
    L_{\text{GAN}}\left(\tilde{\mathbf{x}}, \mathbf{x}\right)
    = \mathop{\mathbb{E}} \log\left(D\left(\mathbf{x},\,\mathbf{x}_{\text{sem}}\right)\right)
    + \mathop{\mathbb{E}} \log\left(1 - D\left(\tilde{\mathbf{x}},\,\mathbf{x}_{\text{sem}}\right)\right),
\end{equation}
where \(D\) is the discriminator that maximizes this loss term during training while \(F\) minimizes it.
The expectation is estimated through mini batch sample mean.}

\ch{To stabilize training and enhance visual quality, one typically supplements GAN loss with auxiliary loss terms.
VGG \cite{johnson2016perceptual} loss computes feature space distance and is formulated as  
\begin{equation}
    L_\text{VGG}\left(\tilde{\mathbf{x}}, \mathbf{x}\right) = \sum_{i=1}^N\frac{1}{M_i}\left\|V^{(i)}\left(\mathbf{x}\right) - V^{(i)}\left(\tilde{\mathbf{x}}\right)\right\|_1,
\end{equation}
where \(V^{(i)}\) denotes the \(i\)th layer with \(M_i\) elements in the VGG network with a total of \(N\) layers \cite{simonyan2014very}.
The GAN feature matching loss \cite{wang2018high} is formulated similarly and has been shown to lead to improved quality when used alongside the VGG loss.
\begin{equation}
    L_\text{FM}\left(\tilde{\mathbf{x}}, \mathbf{x}\right) = \sum_{i=1}^{N^{(D)}}\frac{1}{M_i^{(D)}}\left\|D^{(i)}\left(\mathbf{x}\right) - D^{(i)}\left(\tilde{\mathbf{x}}\right)\right\|_1,
\end{equation}
where \chh{\(D^{(i)}\)} is the \(i\)th layer with \(M_i^{(D)}\) elements in the GAN discriminator with a total of \(N^{(D)}\) layers.}
\chh{The feature matching loss differs from the VGG loss mainly in that the former computes distance in the feature space of the GAN's discriminator, whereas the latter does so in that of a pre-trained, off-the-shelf VGG network.}
\chh{We used both VGG loss and GAN feature matching loss for our experiments.}

\chh{To best optimize for joint perception-distortion performance, we empirically found it ideal to train the model in separate stages with each stage involving only some loss terms instead of in one stage with all loss terms.}
The strategy is described in the following subsection.

\subsubsection{A Three-Phase Training Scheme}
\ch{To have \(F\) successfully learn how to synthesize visuals from a combination of semantics and visuals, one may begin with training it to minimize the GAN loss \(L_\text{GAN}\) \chh{supplemented with the VGG loss \(L_\text{VGG}\) and the feature matching loss \(L_\text{FM}\)}, which represents optimizing perceptual quality.
Once \(F\) is capable of generating reasonably realistic visual content, it should then be optimized with respect to a weighted sum of the above loss terms and another distortion loss directly on the reconstructed pixels \(L_\text{dist}\) such that \(\tilde{\mathbf{x}}\) will be close to \(\mathbf{x}\).
In addition, the combination of the distortion loss \(L_\text{dist}\) and the GAN loss \(L_\text{GAN}\) stabilizes training since the former can be interpreted as a prior that prevents GAN from mode collapse~\cite{agustsson2019generative}.
Finally, to mitigate the issue that GAN loss usually dominates distortion loss when used together, leading to hallucination artifacts (indicating that the model is overly optimizing perception)~\cite{agustsson2019generative}, we finish training by optimizing \(F\) with respect to \(L_\text{dist}\) only.
\chh{We later show that training for three stages guides the model to produce better joint perception-distortion quality than training for just the first two. The latter option, as we have observed, led to models that heavily emphasize on perceptual quality but ignore distortion despite careful weighting of the GAN and the distortion loss terms during training.}
If \(c\) is implemented as a learned codec, it should be jointly optimized with \(F\) during training.}

\chh{We summarize the three training phases below, in which we ignore the rate term for simplicity.
\begin{itemize}[label=$\bullet$]
    \setlength\itemsep{0em}
    \item Phase 1: Minimize \(L_\text{GAN} + \lambda_1 L_\text{VGG} + \lambda_2 L_\text{FM}\) for some \(\lambda_1, \lambda_2 > 0\);
    \item Phase 2: Minimize \(L_\text{GAN} + \lambda_1 L_\text{VGG} + \lambda_2 L_\text{FM} + \lambda_3 L_\text{dist}\) for some \(\lambda_3 > 0\);
    \item Phase 3: Minimize \(L_\text{dist}\).
\end{itemize}}

\subsubsection{Compressing the Semantics}
\label{sem_comp}
Instead of implementing the semantics encoder \(s\) with yet another network, we propose a simple but effective non-learned approach to better utilize some characteristics of semantics in the form of a segmentation map.\footnote{\cite{agustsson2019generative} proposed a similar method with limited motivation and details.}
We observe that pixel-wise semantics in the form of a segmentation map usually contains large uniform areas separated by a few class or instance boundaries.
Instead of compressing the map as a bitmap image, which is prone to creating semantically impossible artifacts such as blurred boundaries, we represent it with a set of paths, each being the boundary of a semantic region.
For each path, we additionally store a single number encoding the class or instance identity of the semantic region.
And we only keep the two end points for each line segment in the paths.
The path-based representation can be obtained from the map with a graph traversal algorithm.
Since the boundaries rarely change drastically in natural images, we apply a simple delta encoding on the paths and entropy code the resulting sequences to further reduce the bitrate.
This algorithm is lossless but if desired, one can convert it to a lossy one by smoothing the paths.
On Cityscapes at \(1024\times 512\) resolution, this method (lossless version) reduces the average semantics overhead (class and instance segmentation maps combined) from approximately \(0.30\) bpp (computed using the rendered maps) to approximately \(0.03\) bpp.

\section{Experiments}
In this section, we present the experimental results of our codecs on Cityscapes~\cite{cordts2016the} and ADE20k~\cite{zhou2016semantic,zhou2017scene} both at \(1024\times 512\) resolution.
The former consists mostly of street scenes and the latter generic everyday images.
We test our pipeline using both learned and conventional codecs including JPEG, JPEG 2000, WebP, and BPG as the backbone codec \(c\).
An extensive ablation study will be presented to demonstrate the usefulness of semantics.
Note that we focus our evaluations on low and medium bitrates since distortion and perception diverge most severely in this regime and empirically, existing codecs typically fail to provide visually appealing results, which necessitates a better compression method and also helps demonstrate the effectiveness of semantics for improving compression quality.

In these experiments, we keep our instantiation of the general framework simple for two reasons:
First, we would like to show that even with a simple implementation and little engineering effort, semantic enhancement can bring significant improvements to perception-accuracy-distortion, which would more effectively prove the usefulness of high-level semantics in compression.
Second, this work focuses on evaluating the high-level idea of leveraging semantics instead of pursuing state-of-the-art performance.
Thus, we feel that leaving out some sophisticated engineering used in other learned compression works can help simplify the presentation and make the interpretation of results easier.

\subsection{On Choice of Baselines}
Note that different from most existing works on learned compression, we are not proposing a single new codec.
Rather, we are proposing a generic framework that can be used to enhance any given codec.
As a result, our experimental evaluations focus on demonstrating the \textit{relative} improvements gained from semantic enhancement with respect to the chosen backbone codecs to show the usefulness of semantics.
This contrasts most existing works, where it would make sense to compare one or two newly proposed codecs ``in parallel'' against existing methods.
Therefore, when interpreting our results, each codec should be primarily compared against and only against its semantically-enhanced counterpart.

Also, note that we mostly employ engineered codecs such as BPG instead of learned ones as our backbone codecs due to their speed and memory efficiency.
And we extensively test with all popular engineered lossy codecs.
At the time of this writing, BPG still performs on par with state-of-the-art learned codecs in terms of R-D~\cite{li2020learning}.
Thus, our results on BPG should let us gauge how much improvement our SE pipeline can bring to advanced learned codecs.

\subsection{Experimental Details}
\subsubsection{Dataset}
For Cityscapes, we trained the models on the densely-annotated training set of \(2975\) images.
For evaluation, we randomly sampled \(10\) images from each of the three cities in the validation set to create a test set.
We used the ground truth class segmentation map and instance boundary map instead of learned ones as the semantics to simplify the set-up and focus more on compression.
Nevertheless, we provide results from using learned semantics at test time to demonstrate how our pipeline would perform with the currently available segmentation technology.
In these cases, the class (instance) segmentation maps were extracted by a trained DeepLab v3+~\cite{chen2018encoder} (Mask R-CNN~\cite{he2017mask}).
One may also train the compression pipeline with learned semantics or jointly train the segmentation network with the codec and we leave this as a future work.

For ADE20k, we trained the models on the full training set of \(20210\) images.
For evaluation, we randomly sampled one image from each leaf folder in the dataset directory, creating a total of \(682\) images.
We then randomly split this set into a validation set and a test set.
During training, we used the best hyperparameters from Cityscapes and saved the best models according to their validation performance on our ADE20k validation set.
The images were all reshaped to \(1024\times 512\).

Note that the images in ADE20k have already been compressed by JPEG and therefore are no longer suitable for a rigorous quantitative analysis for compression performance.
We mainly show qualitative results as proof of concept that our method works well both for specialized images such as street scenes and in the wild for any generic image.

\subsubsection{Architecture}
We implemented \(F\) using a generator network with the same architecture as in~\cite{wang2018high}.
\ch{The decoded class segmentation map was one-hot encoded to form a \(k\)-channel 3D tensor with height and width the same as the original image, where \(k\) is the number of classes and is \(35\) for Cityscapes and \(151\) for ADE20k.
To form the input of \(F\), this segmentation map tensor was channel-concatenated to the decoded RGB image and another single-channel instance segmentation edge map.}

For \(c\), we tested with JPEG, JPEG 2000, WebP, BPG, and an autoencoder-based learned codec.
BPG was implemented using the official \textit{libbpg}~\footnote{\url{https://bellard.org/bpg/}}.
JPEG, JPEG 2000, and WebP were implemented with \textit{Pillow}~\footnote{\url{https://github.com/python-pillow/Pillow}}.
For the autoencoder, the encoder (decoder) consists of \(4\) convolutional (transposed convolutional) layers, each downsampling (upsampling) the image by \(2\).
We added a binarization layer in between to quantize the continuous hidden activations to binary codes, using the method specified in~\cite{toderici2016variable}.
A \chh{Prediction by Partial Matching (PPM)}~\cite{cleary1984data} coder was used on the bitstream produced by the binarizer to further reduce bitrate.
For simplicity, we did not explicitly estimate and minimize the entropy of the bitstream during training.

The semantics was losslessly compressed following descriptions in Section~\ref{sec1}.
A PPM~\cite{cleary1984data} coder was used on the delta encoded boundaries to further reduce bitrate.
On the \(30\)-image test set of Cityscapes, the average semantics bpp overhead is \(0.03\).

\subsubsection{Training}
We used LS-GAN~\cite{mao2017least} as our \(L_\text{GAN}\) formulation.
For the first and second training phases, we supplemented the GAN loss with VGG loss \(L_\text{VGG}\)~\cite{johnson2016perceptual} and GAN feature matching loss \(L_\text{FM}\)~\cite{wang2018high}.
\ch{For pipelines using conventional codecs as \(c\), we trained \(F\) with the \(L^1\) loss as the distortion term.
For pipelines using learned \(c\), we trained \(F\) and \(c\) with the \(L^2\) loss to demonstrate that our framework works well with any generic distortion loss.}
The weightings of the loss terms follow those in \cite{wang2018high}.
Adam~\cite{kingma2015adam} was used as the optimizer and we set the learning rate to \(0.0001\) with no scheduling.
With a batch size of \(2\), we trained the networks on Cityscapes (ADE20k) for \(30\), \(30\), and \(50\) (\(5\), \(5\), and \(7\)) epochs for the three phases, respectively.
Except for the objective function formulation, all settings remain unchanged across the three phases for a given model.
All experiments were performed on an NVIDIA Tesla V100 GPU.
The inference time depends on the underlying backbone image codec and its compression settings.
As a reference, each test image took on average \(1\) second to process for JPEG-SE with the backbone JPEG codec running at quality 15\footnote{This is using \textit{Pillow}'s metrics. All other JPEG settings were kept at \textit{Pillow}'s defaults for this test.}.

\subsubsection{Evaluation}
To quantify R-P performance, we conducted a comprehensive user study on Amazon Mechanical Turk (MTurk).
For each pair of original and SE codecs with comparable bpp values, we created a questionnaire consisting of a sequence of images for \(30\) distinct, randomly-chosen workers by randomly sampling \(10\) images out of our \(30\)-image Cityscapes test set, producing a total of \(300\) responses per comparison.
For each image in a questionnaire, we presented the original and reconstructions from the two codecs \ch{(baseline codec and its SE counterpart)} with relative orders of the latter two randomized.
The task for the worker was to choose between the compressed images which one they thought was a better compressed version of the original with the only standard being their subjective opinion.
As quality control, three sanity checks were added to random positions in each questionnaire using \(3\) more randomly chosen test images, in which one of the compressed images was produced by a lossless codec.
Failing to choose the lossless codec in any of the sanity checks will cause a response to be rejected and the job reposted to other workers.
We also rejected responses finished within \(60\) seconds to further filter out workers that picked images quickly at random and happened to have passed all three sanity checks by luck.

To measure R-A performance, we used bounding-box object detection~\footnote{We did not choose segmentation because our pipeline involves transmitting segmentation maps to the receiver's end. Thus, there would be no need to perform segmentation again.} as an example to show that the SE codecs produce reconstructions that not only look better to human viewers but also enable machines to achieve better performance on high-level vision tasks.
For this study, we used a trained Faster R-CNN~\cite{ren2015faster} from~\cite{chen2019mmdetection} for bounding-box object detection on the codec reconstructions.
Detection performance is reported using AP@IoU=.50:.05:.95, the primary challenge metric of COCO 2019~\cite{lin2014microsoft}.

We used LPIPS~\cite{zhang2018the}, MS-SSIM~\cite{wang2003multiscale}, and PSNR as the distortion metrics to quantify R-D performance.
LPIPS measures the distance between two images using the deep features learned by a trained classification network.
We used the authors' official implementation (v0.1) with default settings, which uses an AlexNet~\cite{krizhevsky2012imagenet} as the trained network.
Note that although LPIPS was named a ``perceptual'' metric by the authors, it is still a pairwise metric that does not perfectly align with the perception quality~\cite{blau2018the}.
Nevertheless, the authors in~\cite{zhang2018the} showed that LPIPS aligns with human perception much better than the more traditional distortion losses.

\begin{figure*}[h]
    \subfloat{
        \includegraphics[width=.64\columnwidth]{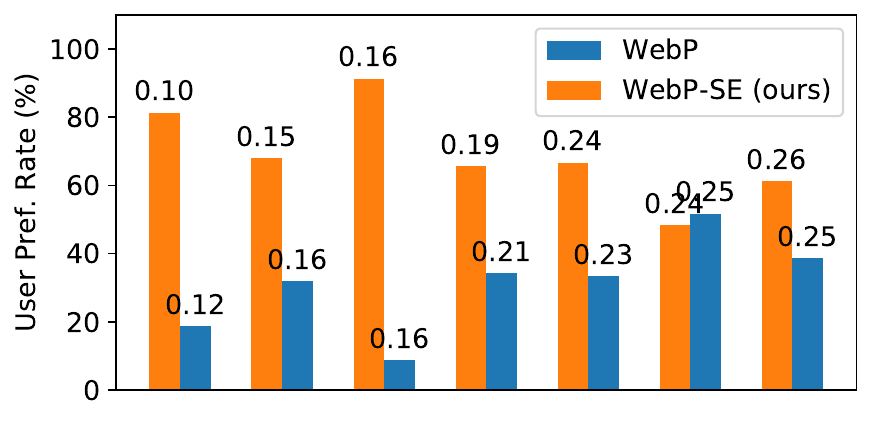}
    }
    \subfloat{
        \includegraphics[width=.64\columnwidth]{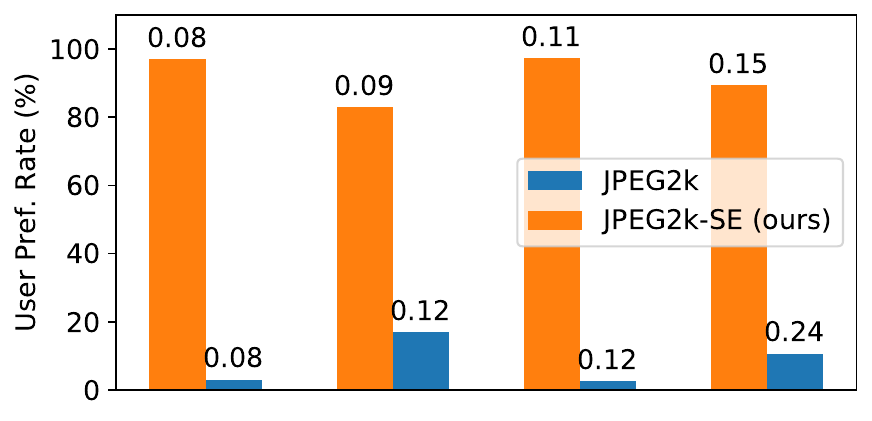}
    }
    \subfloat{
        \includegraphics[width=.64\columnwidth]{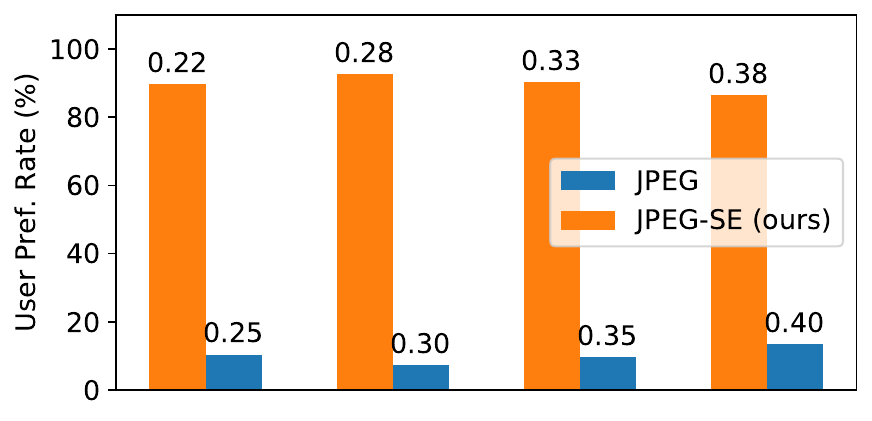}
    }
    \vspace{-13pt}
    \subfloat{
        \includegraphics[width=.64\columnwidth]{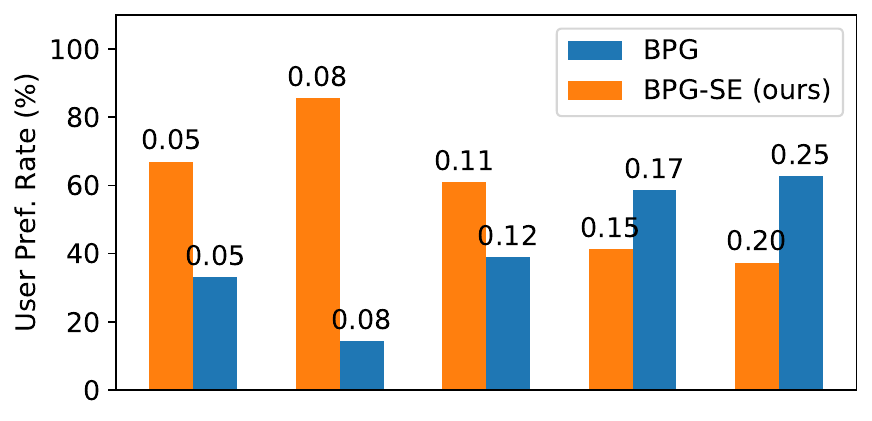}
    }
    \subfloat{
        \includegraphics[width=.64\columnwidth]{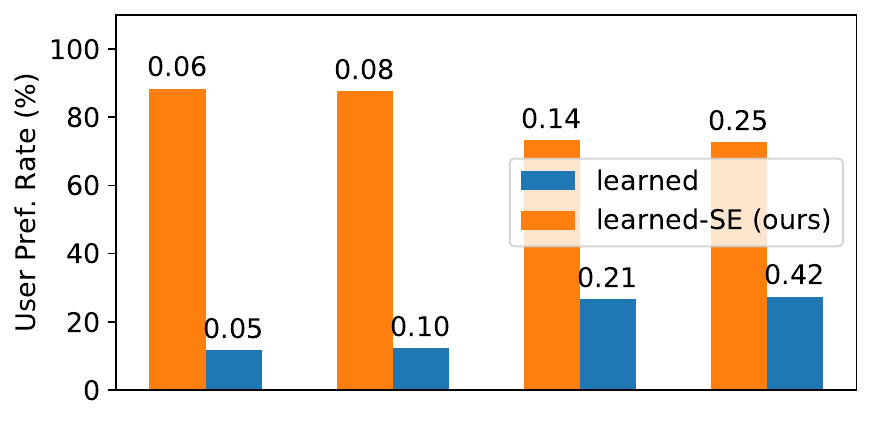}
    }
    \subfloat{
        \includegraphics[width=.64\columnwidth]{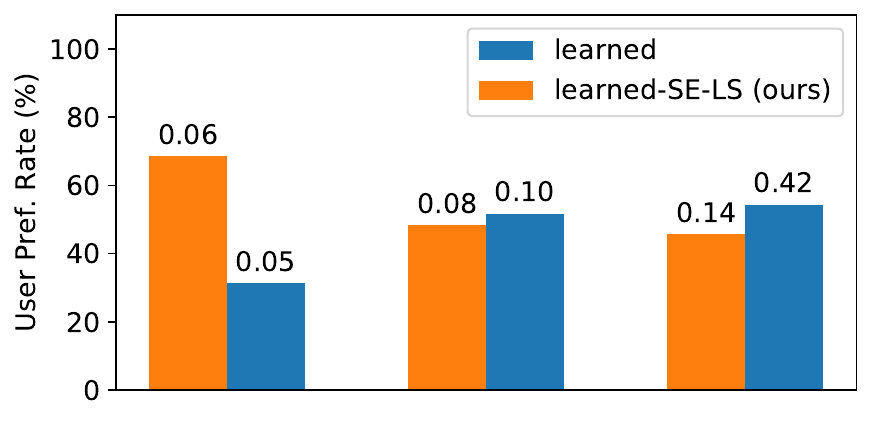}
    }
    \caption{
        Rate-perception performance on Cityscapes.
        Each pair of neighboring bars represent two codecs compared by human raters.
        The height of each bar shows the \% of times this codec is preferred among \(300\) distinct user responses.
        Text annotations are the bpp values.
        \ch{The number of bars can differ across subfigures depending on the number of models trained.}
        For this and all results in the paper, the bpp overhead from semantics has \textit{already} been included in the bpp values of the SE codecs.
        ``LS'' refers to using learned semantics extracted by off-the-shelf segmentation networks at test time.
        The SE models are almost always preferred, often by a considerable margin.
    }
    \label{fig2c}
\end{figure*}

\subsection{Main Results}
In this section, we present the main results of our study.
Following the trend in learned image compression community~\cite{clic2020}, we consider human ratings to be the most important metric among perception, accuracy, and distortion and therefore present R-P results first.
Quantitative evaluations will be performed on Cityscapes and qualitative results will be provided for both Cityscapes and ADE20k.
We refer the readers to the supplementary materials for the qualitative results on ADE20k.

\subsubsection{Rate-Perception}
Fig.~\ref{fig2c} quantitatively shows that the SE codecs produce more visually appealing reconstructions compared to the originals.
Indeed, the SE codecs are almost always preferred by the human raters, often by a considerable margin.
From the visualizations in Fig.~\ref{fig3}, it is clear that the SE codecs ``understood'' the images.
To be specific, the original codecs typically failed to respect boundaries between semantic regions, producing image patches that lack realism.
This can be seen in between the tires and the road surface, among other places.
Indeed, we know that tires are not the same as the road surface in real-world, yet the non-SE codecs do not ``understand'' this.
Even worse, some non-SE codecs produced severe artifacts, which are essentially low-level features that are not possible in real scenes.
In comparison, even though the SE codecs could not get all the details correct due to the low bpp allowances, they preserved well the semantic boundaries and they rarely produced low-level features that are incoherent or unnatural.

\subsubsection{Robustness Against Noise on Low-Level Visuals}
\begin{figure} 
    \centering
    \subfloat{
        \rotatebox{90}{\scriptsize{(a) JPEG-SE (1.51 bpp)}}
        \includegraphics[trim={25cm 7cm 3cm 1cm},clip,width=.45\columnwidth]{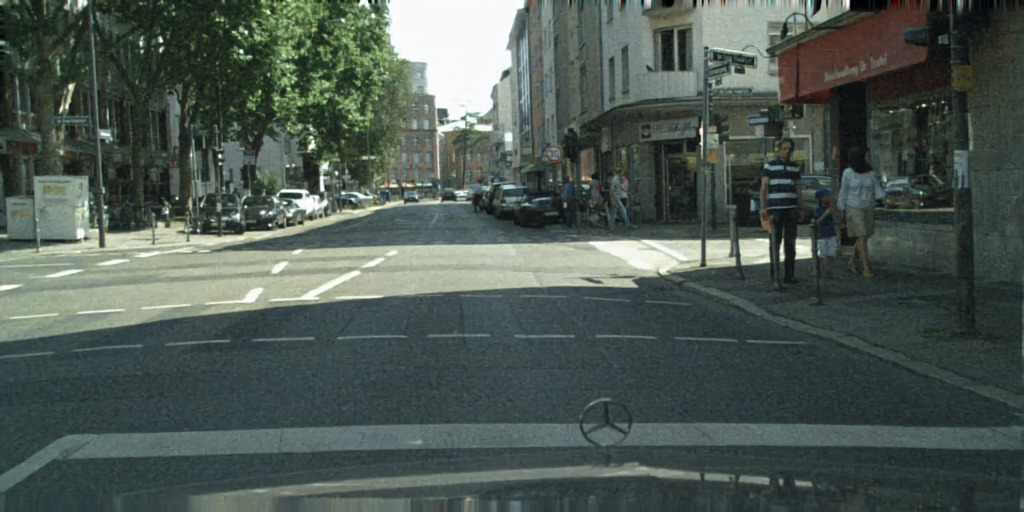}
    }
    \subfloat{
        \includegraphics[trim={25cm 7cm 3cm 1cm},clip,width=.45\columnwidth]{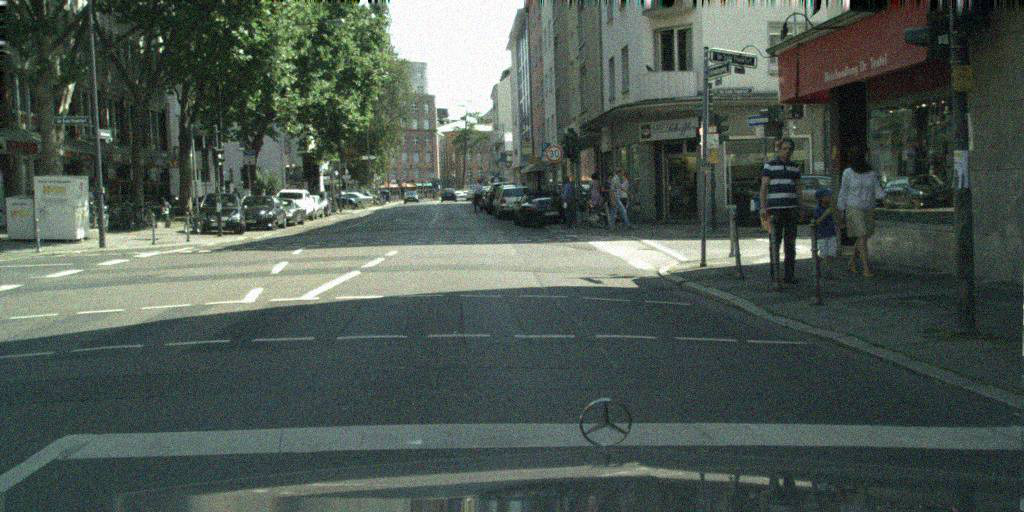}
        \rotatebox{90}{\scriptsize{(b) JPEG (1.51 bpp)}}
    }
    \vspace{-6pt}
    \subfloat{
        \rotatebox{90}{\scriptsize{(c) Noisy Original}}
        \includegraphics[trim={25cm 7cm 3cm 1cm},clip,width=.45\columnwidth]{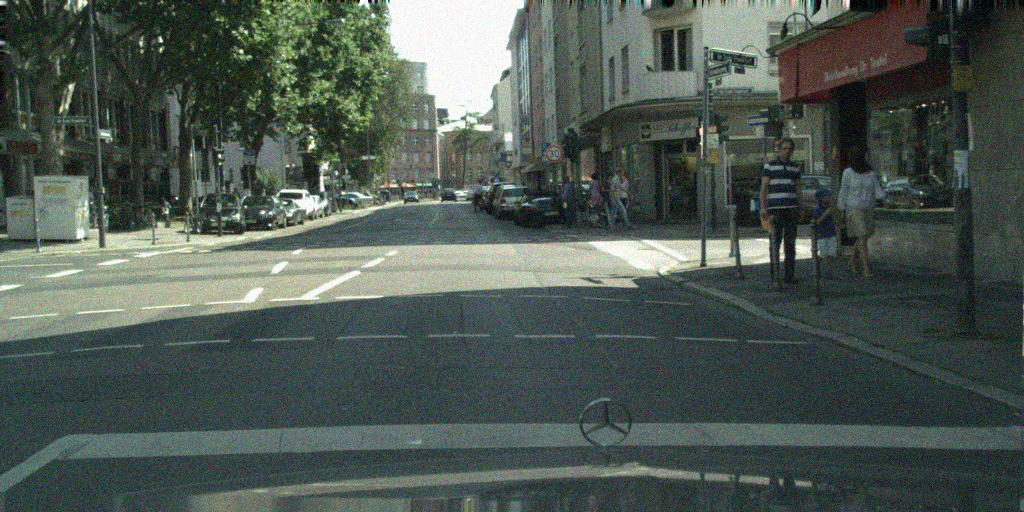}
    }
    \subfloat{
        \includegraphics[trim={25cm 7cm 3cm 1cm},clip,width=.45\columnwidth]{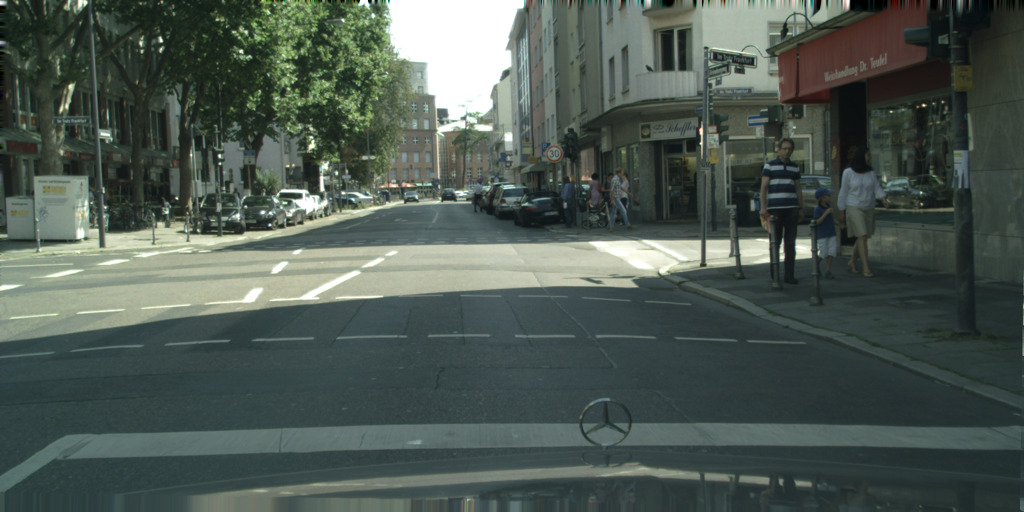}
        \rotatebox{90}{\scriptsize{(d) Original}}
    }
    \caption{
        Their high-level understanding of the scenes robustifies the SE codecs against noise on the low-level visual features without additional training on noisy images.
    }
    \label{fig6}
\end{figure}

We have argued and demonstrated that the high-level knowledge gained via training helps the SE codecs produce more natural and perceptually appealing reconstructions.
Intuitively, it should therefore also help the SE codecs distinguish between valid low-level features and artifacts for any given semantics.
For undesired artifacts such as noise, the SE codecs should be able to filter them out in the reconstruction.
As proof of concept, we added Gaussian noise to a test image to be compressed.
And from Fig.~\ref{fig6}, we can see that the SE codec indeed filtered out the added noise at test time whereas the original codec tried to reconstruct noise as well, which, in most scenarios, is undesirable.

It is worth pointing out that the SE model can achieve this denoising effect without having been trained with noisy examples.
In contrast, existing denoising pipelines require noisy images for training~\cite{lehtinen2018noise,mildenhall2018burst,chen2018learning}.
To obtain noisy images, one either explicitly device a synthetic noise model, which makes the generation process cheap but does not necessarily produce realistic noisy images~\cite{mildenhall2018burst}.
Alternatively, one captures real noisy images and sometimes the corresponding ground truth images from the real world, which is certainly a nontrivial procedure~\cite{chen2018learning}.

Our work potentially opens up new directions for image denoising research, in which one does not need at all or no longer need as many noisy images for training to achieve the same level of denoising effect.
This contrasts a branch of works aiming at removing the need for noisy-clean pairs during training~\cite{lehtinen2018noise} since our framework only requires clean-clean pairs.

\subsubsection{Rate-Accuracy}

\begin{figure}[h]
    \centering
    \includegraphics[width=.7\columnwidth]{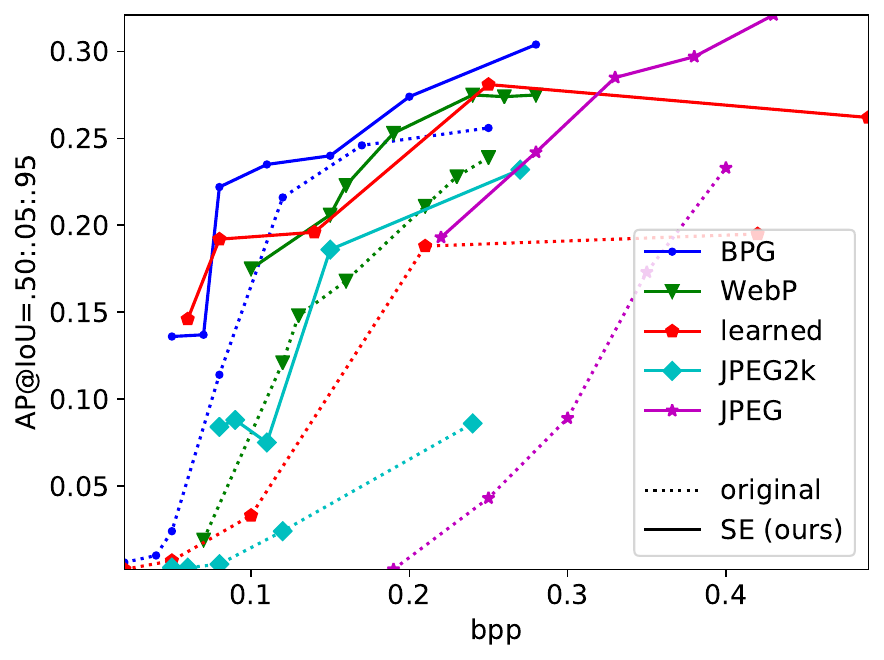}
    \caption{
        Rate-accuracy performance on Cityscapes.
        Each AP value was calculated with the same off-the-shelf bounding-box object detector operating on the codec's reconstructions at that bpp.
        The higher the AP, the more accurate the detection.
        For all codecs, the SE model enables the downstream object detector to detect much more accurately at any given bpp.
    }
    \label{afig14}
\end{figure}

From Fig.~\ref{afig14}, we see that for all of the five codecs, the object detector performs much better on the SE codecs' outputs.
In practice, this means that one can now transmit fewer bits to the receiver's end for the machine agent to achieve a certain level of performance, that is, the agent can make a better decision faster.
For the ever-growing set of machine-based vision tasks including autonomous driving and face recognition, this observation potentially makes the SE codecs more suitable than the existing ones.
And considering how rapidly the industry is embracing automated algorithms for vision tasks and how ubiquitous these tasks are, these results are highly relevant and the SE codecs are much better geared toward an AI-powered future.
\chh{We emphasize that the formulation of our training objective does not directly optimize for rate-accuracy. The improved rate-accuracy performance should be seen as a byproduct from using semantics.}

\subsubsection{Rate-Distortion}
\begin{figure*}[h!]
    \centering
    \subfloat[][LPIPS\(\downarrow\)~\cite{zhang2018the}.]{
        \includegraphics[width=.66\columnwidth]{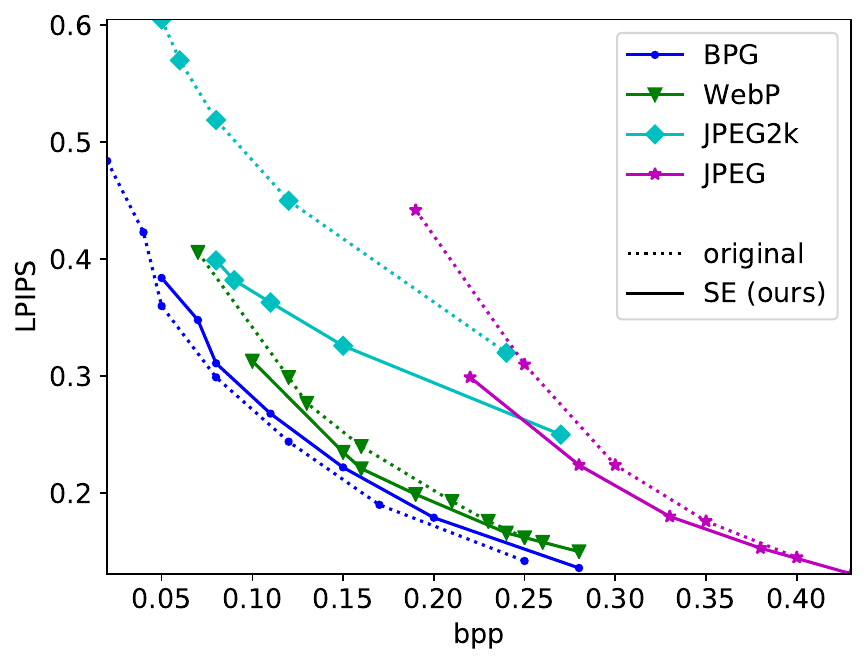}
        \label{fig2d}
    }
    \subfloat[][MS-SSIM\(\uparrow\)~\cite{wang2003multiscale}.]{
        \includegraphics[width=.69\columnwidth]{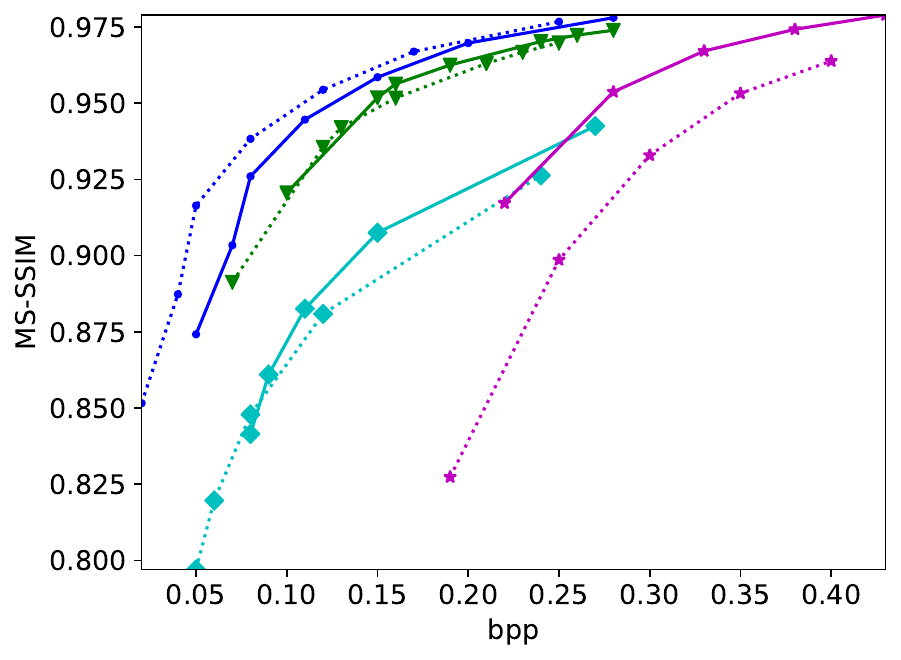}
        \label{fig2a}
    }
    \subfloat[][PSNR (dB)\(\uparrow\).]{
        \includegraphics[width=.66\columnwidth]{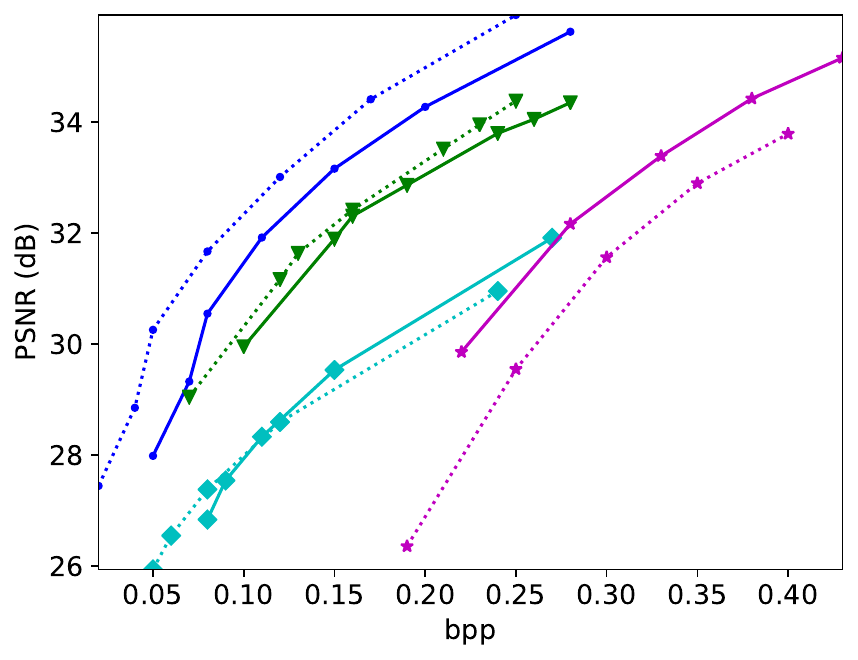}
        \label{fig2b}
    }
    \caption{
        Rate-distortion performance on Cityscapes.
        \(\uparrow\) (\(\downarrow\)) means the higher (lower) the better.
        Among the three distortion metrics used, PSNR, the most traditional one, has been known to be inconclusive in quantifying compression quality~\cite{wang2003multiscale,zhang2018the}.
        LPIPS~\cite{zhang2018the} and MS-SSIM~\cite{wang2003multiscale} have been shown to align much better with human perception.
        The SE models achieve better or comparable rate-distortion performance, especially when characterized by LPIPS and MS-SSIM.
    }
    \label{fig2}
\end{figure*}

\begin{figure}[h!]
    \centering
    \includegraphics[width=.7\columnwidth]{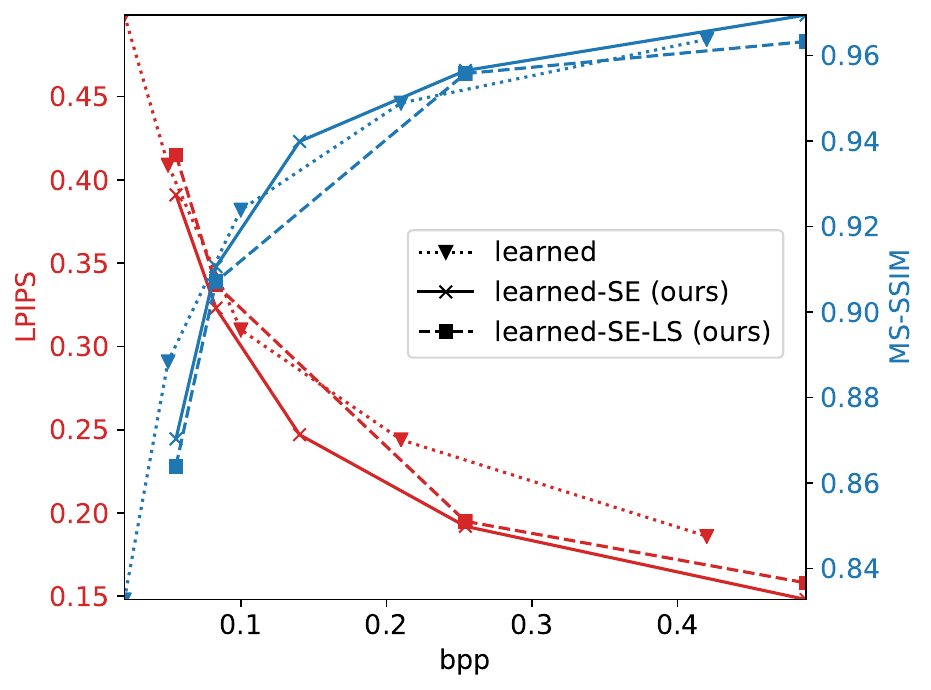}
    \caption{
        Rate-distortion performance on Cityscapes.
        ``LS'' refers to using learned semantics extracted by off-the-shelf segmentation networks at test time.
    }
    \label{fig_learned_rd}
\end{figure}

From Fig.~\ref{fig2}, \ref{fig_learned_rd}, and~\ref{fig3}, we see that for all codecs, the SE models achieve comparable or better distortion at most bpp values.
On both LPIPS and MS-SSIM, which are the distortion metrics that align more closely with human perception among the three~\cite{wang2003multiscale,zhang2018the}, the SE codecs outperform the originals in all cases except for BPG.

Note that to achieve their superior performance in terms of R-P, the SE codecs likely had to compromise R-D to some extent due to the known tension between distortion and perception~\cite{blau2018the,blau2019rethinking}.
Further, since compressing semantics is not the main focus of this work, we always losslessly compressed semantics using the aforementioned simple method, creating a constant bpp overhead of 0.03 that becomes significant in some of the extreme low-bitrate settings in which the evaluated BPG-SE models mainly operated, 
where it may suffice to use lossy compression instead and reduce rate.

Further, as a side note, all SE codecs outperform the originals in nearly all comparisons with the semantics overhead ignored and we refer the readers to the supplementary materials for more details.
This represents a performance upper bound of our method that can hopefully be approached with advances in compression techniques tailored for semantics.
This, however, is out of the scope of this paper and we hope our work can serve as evidence of the usefulness of semantics and draw more attention to this previously overlooked topic of compressing semantics.

For the models using a learned codec as \(c\), we also provide results from using learned semantics at test time.
These SE models still obtained performance comparable to those using ground truth semantics despite the imperfect segmentation.
Naturally, we expect that future advances in semantic segmentation will help close this performance gap.

\subsubsection{\chh{Comparing With Related Work}}
\chh{In \cite{agustsson2019generative}, the authors proposed a version of their GAN-based codec that leveraged semantics (named ``selective generative compression'' (SC)).
Among existing works, this perhaps resembles our work the most.
Although \cite{agustsson2019generative} only used semantics for bitrate control in a restricted, human-in-the-loop compression use case, their proposed pipeline is actually similar to ours.
In terms of training objective, however, they only optimized for rate-perception performance, whereas we focus on jointly optimizing rate-perception-distortion.
Another major difference is that they only focused on one specific codec instantiation, whereas we propose a generic framework that can be used to enhance any codec with semantics.}

\begin{figure*}[h]
    \centering
    \subfloat[\scriptsize{\chh{Left to right: \cite{agustsson2019generative} (0.18 bpp), original, BPG (0.16 bpp), BPG-SE (0.17 bpp)}}]{
        \label{fig34a}
        \includegraphics[width=0.5\textwidth]{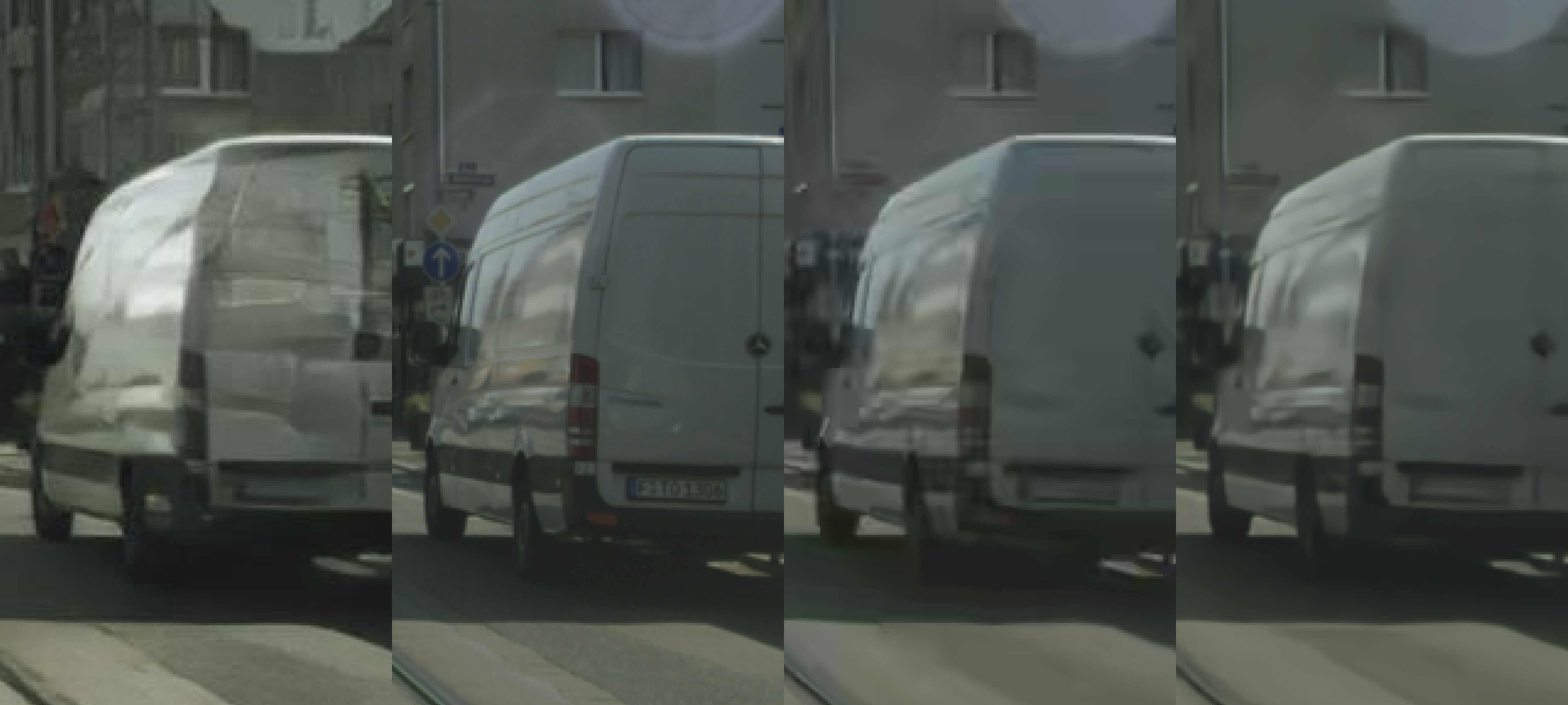}
    }
    \subfloat[\scriptsize{\chh{Left to right: \cite{agustsson2019generative} (0.33 bpp), original, BPG (0.24 bpp), BPG-SE (0.24 bpp)}}]{
        \includegraphics[width=0.5\textwidth]{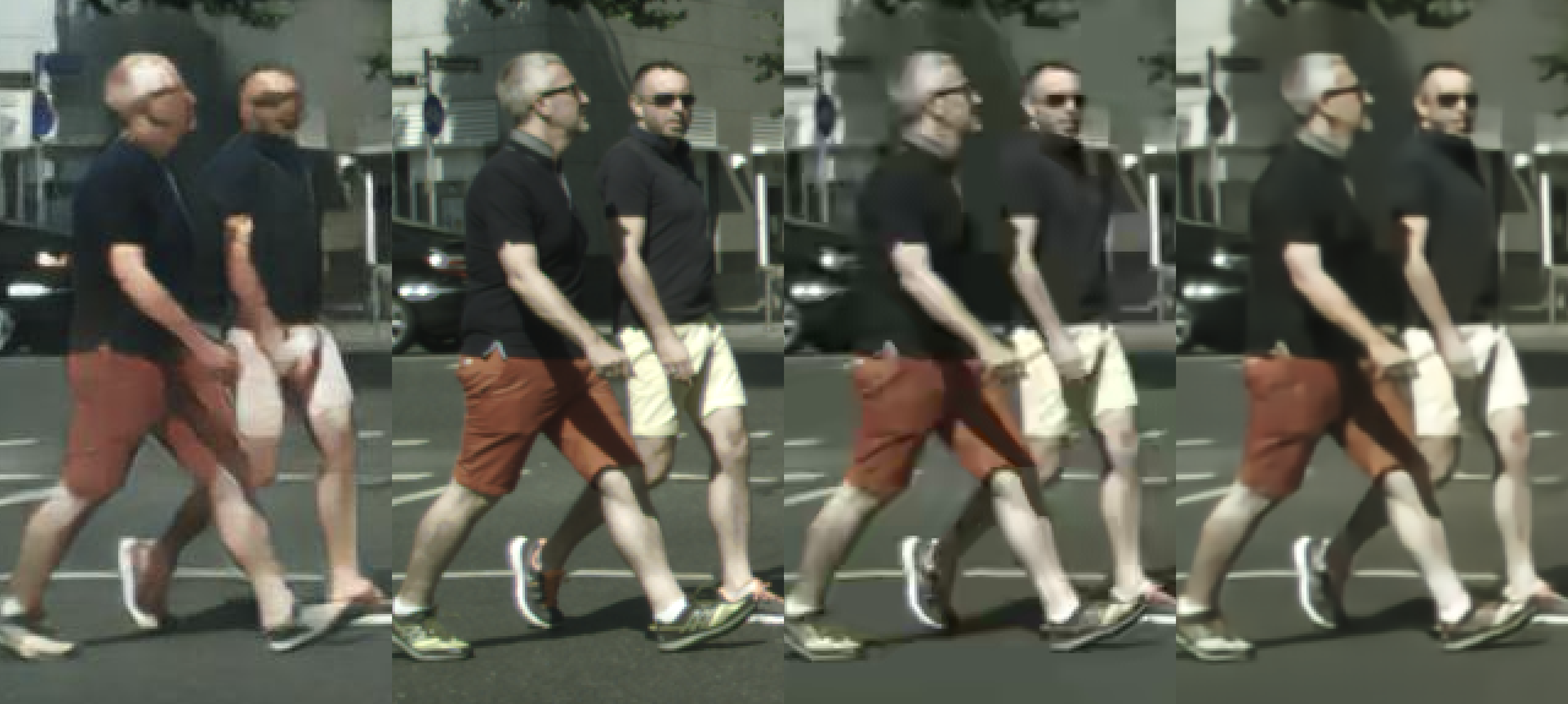}
        \label{fig34b}
    }
    \caption{
    \chh{Comparing \cite{agustsson2019generative} against our BPG-SE. While both \cite{agustsson2019generative} and BPG-SE generate more visually pleasing results than the BPG baseline (cleaner object boundaries, less artifacts, etc.), our BPG-SE additionally preserves high image fidelity. Full images and more comparisons are in the supplementary materials.}
    }
    \label{fig34}
\end{figure*}
\begin{table}[h!]
    \caption{
        \chh{Comparing against \cite{agustsson2019generative}: Average rate-distortion performance on the 3 test images provided by \cite{agustsson2019generative}.
        \(\uparrow\) (\(\downarrow\)) means the higher (lower) the better.
        Our BPG-SE performs much better than \cite{agustsson2019generative} in terms of distortion metrics.}
    }
    \begin{center}
        \begin{tabular}{c|cccc}
            \hline
            Codec & bpp & LPIPS\(\downarrow\) & MS-SSIM\(\uparrow\) & PSNR (dB)\(\uparrow\)\\
            \hline
            \cite{agustsson2019generative} & 0.23 & 0.203 & 0.8634 & 22.50\\ 
            BPG-SE & 0.20 & 0.196 & 0.9654 & 31.56\\ 
            \hline
        \end{tabular}
    \end{center}
    \label{comp_w_6_table}
\end{table}

\chh{In Fig. \ref{fig34}, we present qualitative comparisons of one of our codecs (BPG-SE) against theirs.
Results from a baseline codec are also illustrated as references.
Both \cite{agustsson2019generative} and our BPG-SE are able to generate more visually appealing reconstructions than the stock BPG baseline.
Indeed, \cite{agustsson2019generative} and BPG-SE produce cleaner object boundaries and less compression artifacts.
However, due to their disregard for image fidelity, result from \cite{agustsson2019generative} can look quite different from original.
In contrast, our training objective can simultaneously lead to good perceptual quality as well as small distortion with respect to the original.
Table \ref{comp_w_6_table} further quantifies the advantage of our method in terms of image fidelity.
}

\chh{Note that since \cite{agustsson2019generative} did not disclose their trained models or code, we could only conduct comparisons with the test images they demonstrated in their paper.
With that said, their semantic-aware codec (the SC codec in their paper) can be viewed as a special case of our SE codecs but trained with only Phase 1 and 2.
With this connection established, our ablation study on the importance of Phase 3 then serves as another set of large-scale comparisons between our approach and \cite{agustsson2019generative}.
This ablation study can be found in Sec. \ref{minus_phase3}. 
}

\subsection{Ablation Study}
\subsubsection{The Importance of Semantics}
To rigorously validate the usefulness of semantics, we performed an extensive ablation study using various codecs.
First, we fed dummy all-zero semantics to a trained JP2-SE model during test time.
Comparing Fig.~\ref{fig3f} with~\ref{fig3h} and also the first row of Table~\ref{ablation}, we conclude that the removal of semantic information at test time significantly worsened performance and in particular, caused the codec to behave similarly to the original JP2 and produce blurred semantic boundaries.
However, it may be that the same quality improvement can be achieved using any learned post-processing module without semantics had we removed semantics during training such that the model adapts to its absence at test time.
To rule out this possibility, we trained one SE model for each engineered backbone codec from scratch with the semantics channels zeroed out throughout.
The GAN loss was left unchanged during training and these models did not use semantics at test time.
From Fig.~\ref{fig4b}, and \ref{fig4d}, it can be seen from the silhouette of the car that without semantics, the codec returned to producing reconstructions with blurred semantic boundaries and unnatural details.
Rows 2-5 of Table~\ref{ablation} further quantitatively verifies that the use of semantics is indispensable for the improved compression performance of SE codecs.

\subsubsection{Contribution From GAN}
\begin{figure*}[h] 
    \centering
    \subfloat[\scriptsize{JPEG (0.22 bpp)}]{
        \label{fig4a}
        \includegraphics[trim={25cm 0 0 0},clip,width=.38\columnwidth]{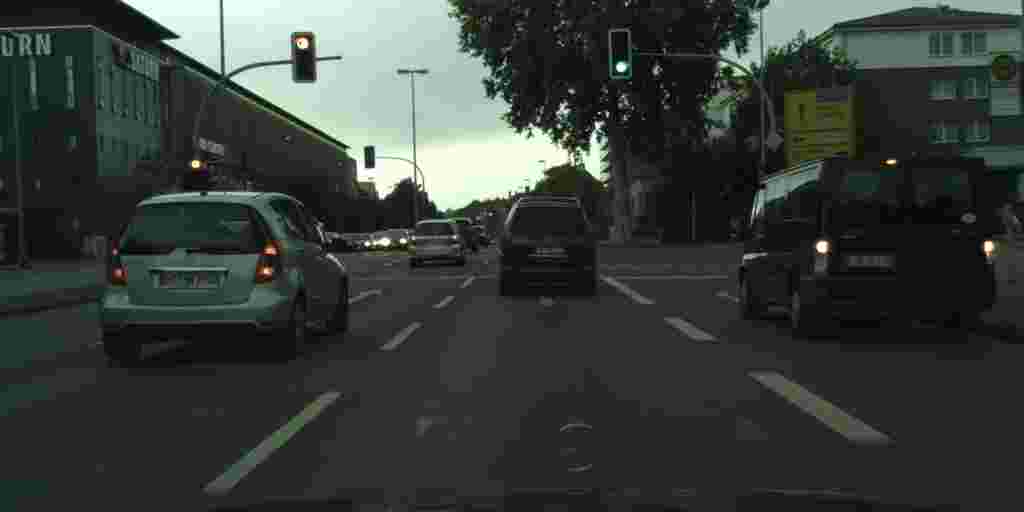}
    }
    \subfloat[\scriptsize{JPEG-SE (0.21 bpp)}]{
        \label{fig4b}
        \includegraphics[trim={25cm 0 0 0},clip,width=.38\columnwidth]{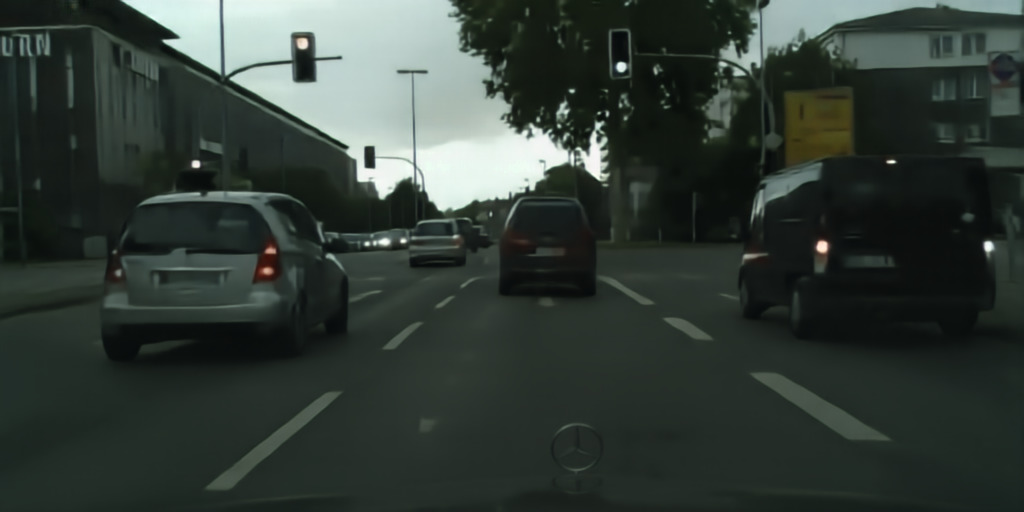}
    }
    \subfloat[\scriptsize{JPEG-SE (-sg; tt)}]{
        \label{fig4c}
        \includegraphics[trim={25cm 0 0 0},clip,width=.38\columnwidth]{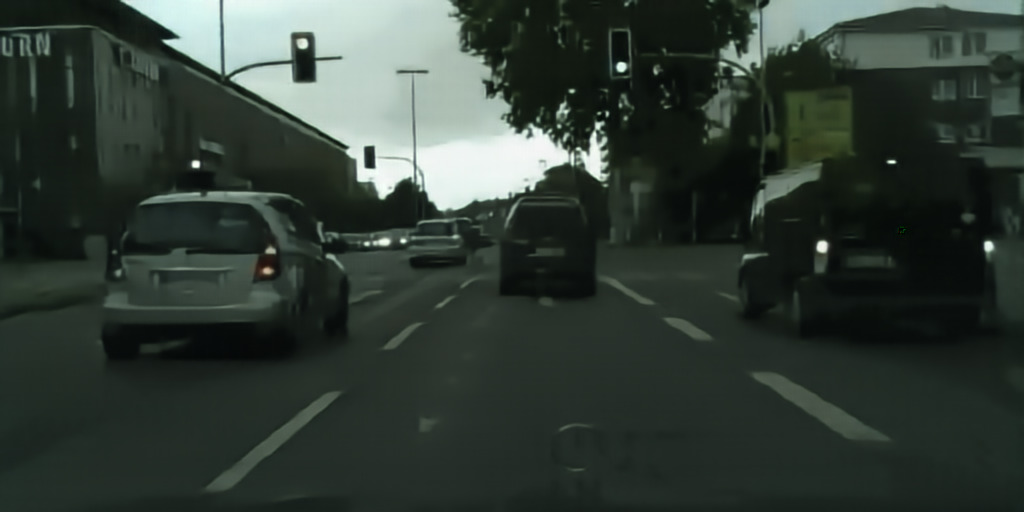}
    }
    \subfloat[\scriptsize{JPEG-SE (-s; tt)}]{
        \label{fig4d}
        \includegraphics[trim={25cm 0 0 0},clip,width=.38\columnwidth]{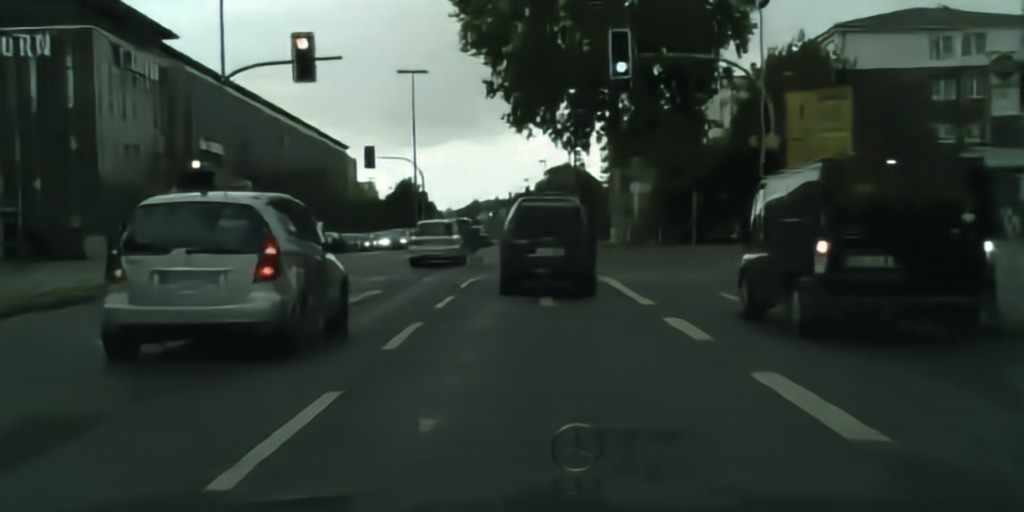}
    }
    \subfloat[\scriptsize{Original}]{
        \includegraphics[trim={25cm 0 0 0},clip,width=.38\columnwidth]{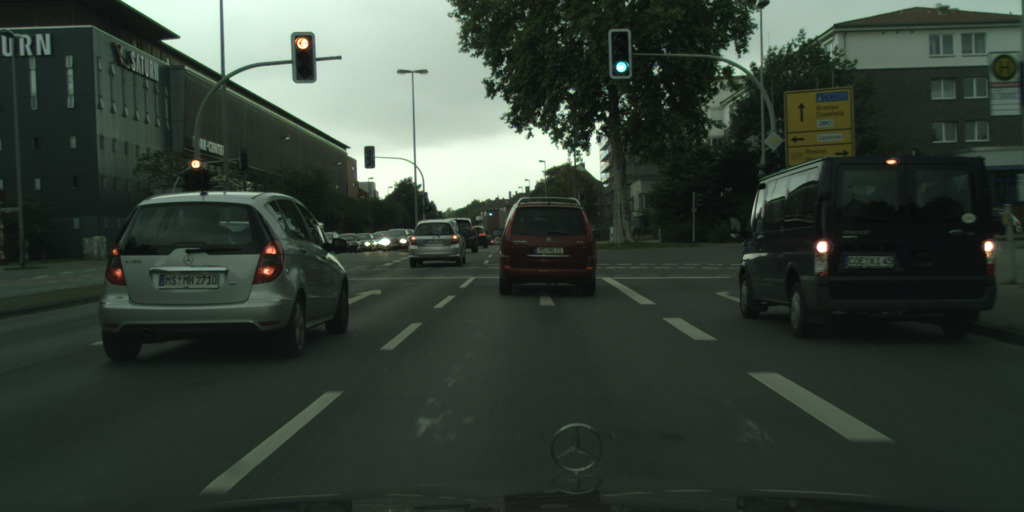}
    }
    \caption{
        Ablation study: ``-sg;tt'' means the model has been trained and tested without semantics or GAN;
        ``-s;tt'' means the model has been trained and tested without semantics.
        The effect of semantics cannot be attained with purely low-level visuals even with the help of GAN, as \ref{fig4b} and \ref{fig4d} suggest.
        \ref{fig4c} and \ref{fig4d} show that GAN can help improve perception quality through creating more vibrant colors without severely increasing distortion or adding hallucination artifacts.
        Full images are provided in the supplementary materials.
    }
    \label{fig4}
\end{figure*}

The GAN loss in our framework helps improve perception quality since GAN loss can be interpreted as a divergence measure between the original density and that of the reconstruction.
On the other hand, GAN is known to create hallucination artifacts even in the presence of a distortion loss~\cite{agustsson2019generative} and we proposed a distortion-loss-only training phase to mitigate this issue.
The question, however, is whether our third training phase washes off all the improvement in perception quality introduced by GAN.
To prove that our training scheme enables GAN to contribute to perception quality without introducing severe hallucination artifacts, we trained a JPEG-SE model with the GAN component (hence also semantics) removed throughout and compare it with the same model trained with only the semantics removed.
From Fig.~\ref{fig4c} and \ref{fig4d}, it is evident that GAN helped produce much more vibrant colors than what could be achieved with a vanilla learned post-processing module.
And from the second and sixth rows of Table~\ref{ablation}, removing GAN even caused a performance drop in terms of distortion and accuracy.
Note that the user preference rates on the two rows were both computed with respect to the SE baseline and therefore cannot be used to conclude that the removal of GAN did not significantly affect the perceptual quality.

\subsubsection{The Three-Phase Training Scheme}
\label{minus_phase3}
\begin{figure}[h] 
    \centering
    \subfloat{
        \label{fig5d}
        \rotatebox{90}{\scriptsize{(a) JP2-SE (0.08 bpp)}}
        \includegraphics[trim={0 1cm 25cm 6cm},clip,width=.45\columnwidth]{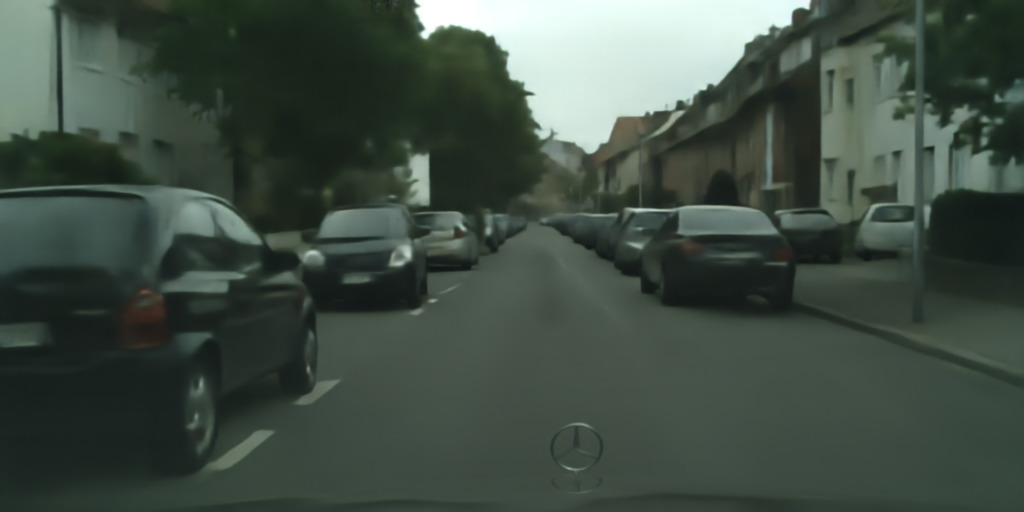}
    }
    \subfloat{
        \label{fig5c}
        \includegraphics[trim={0 1cm 25cm 6cm},clip,width=.45\columnwidth]{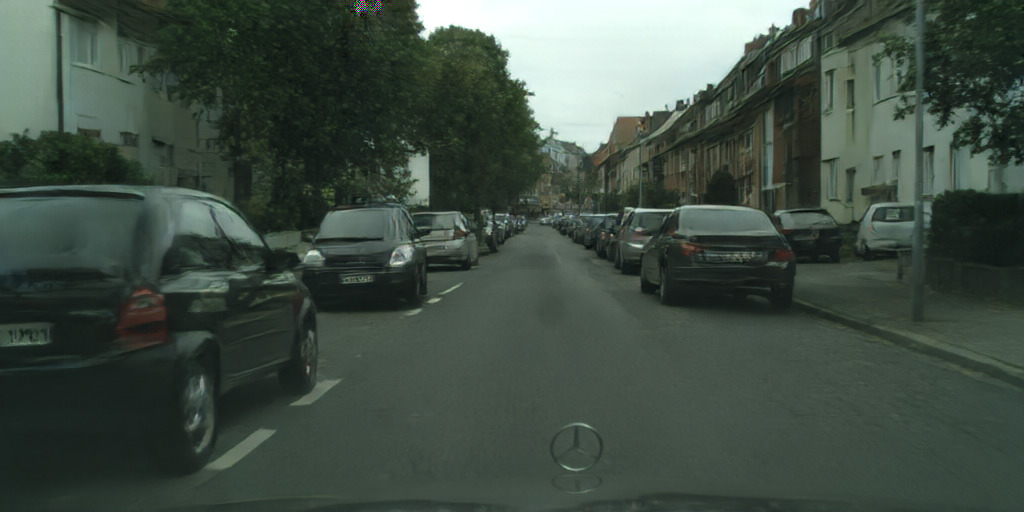}
        \rotatebox{90}{\scriptsize{(b) JP2-SE (-Phase 3)}}
    }
    \vspace{-6pt}
    \subfloat{
        \rotatebox{90}{\scriptsize{(c) JP2 (0.08 bpp)}}
        \includegraphics[trim={0 1cm 25cm 6cm},clip,width=.45\columnwidth]{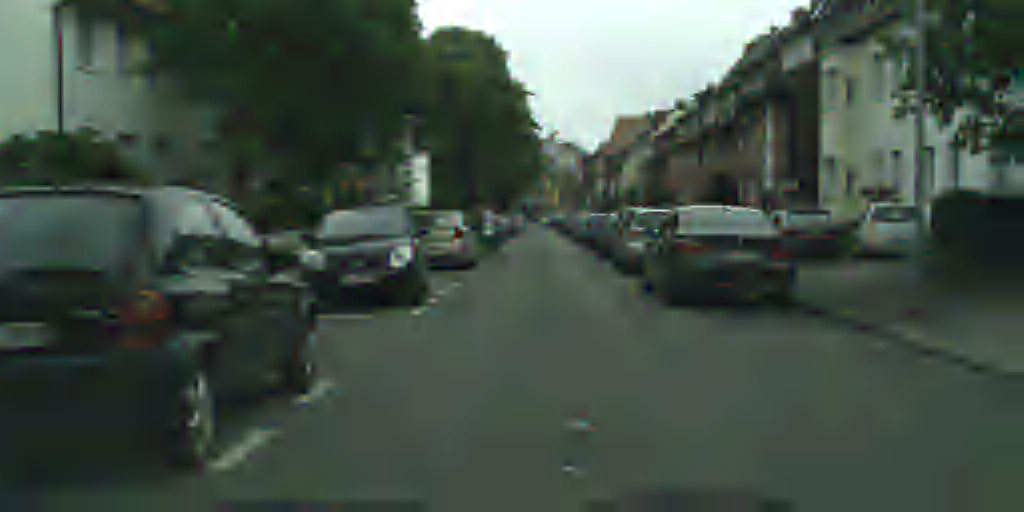}
    }
    \subfloat{
        \includegraphics[trim={0 1cm 25cm 6cm},clip,width=.45\columnwidth]{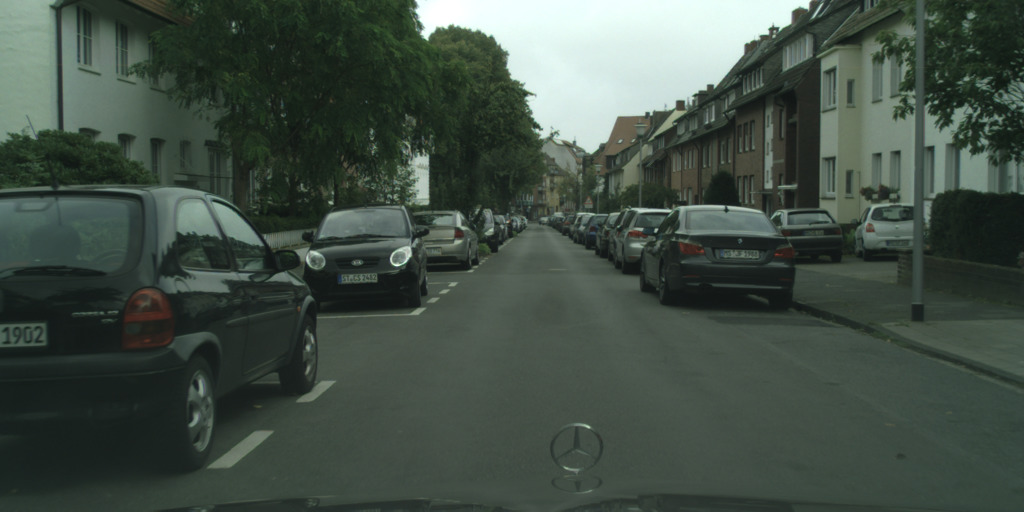}
        \rotatebox{90}{\scriptsize{(d) Original}}
    }
    \caption{
        Ablation study: ``-Phase 3'' means the model has been trained without Phase 3.
        Without the refinement from the distortion-loss-only Phase 3, the reconstructions suffer heavily from hallucination artifacts and high distortion, indicating that the model has been overly optimizing for perceptual quality.
        Full images are in the supplementary materials.
    }
    \label{fig5}
\end{figure}

Fig.~\ref{fig5c} makes it clear that the models trained without Phase 3 produced reconstructions with heavy hallucination artifacts.
In comparison, the distortion-loss-only Phase 3 effectively improved the fidelity of the reconstructed images.
This, however, came at a price of decreased perception quality and detection accuracy (the last two rows of Table~\ref{ablation}), which is expected since in the evaluated bitrates, distortion and perception quality diverge severely~\cite{blau2018the}.

\begin{table*}[h!]
    \caption{
        Ablation study: Each table entry corresponds to bpp/\red{\% of times this codec is preferred when compared against the full SE model among \(300\) distinct user responses}/\blue{AP@IoU=.50:.05:.95\(\uparrow\)}/\magenta{LPIPS\(\downarrow\)}/\green{MS-SSIM\(\uparrow\)} on the Cityscapes test set.
        \(\uparrow\) (\(\downarrow\)) means the higher (lower) the better.
        Notations: ``-s'' means ``no semantics'' and ``-sg'' ``no semantics or GAN''.
        The stage(s) at which the ablation was performed is specified afterwards with ``tt'' meaning ``during both training and test'' and ``test'' meaning ``only during test''.
        ``-Phase 3'' means the model has been trained without Phase 3.
        The full SE model has the best overall R-PAD performance in each original-SE-SE (ablated) triplet.
    }
    \begin{center}
        \begin{tabular}{l|cccc}
            \hline
            Codec & Original & SE & SE (Ablated Models) & Ablation\\
            \hline
            JP2 & .12/\red{2.67}/\blue{.024}/\magenta{.450}/\green{.8809} & .11/\red{-}/\blue{.075}/\magenta{.363}/\green{.8826} & -/\red{13.23}/\blue{.018}/\magenta{.414}/\green{.8416} & -s; test\\
            \hline
            JPEG & .25/\red{10.33}/\blue{.043}/\magenta{.310}/\green{.8986} & .22/\red{-}/\blue{.193}/\magenta{.299}/\green{.9172} & -/\red{20.00}/\blue{.081}/\magenta{.318}/\green{.9138} & \multirow{4}{*}{-s; tt}\\
            JP2 & .24/\red{10.67}/\blue{.086}/\magenta{.320}/\green{.9263} & .15/\red{-}/\blue{.186}/\magenta{.326}/\green{.9075} & -/\red{7.33}/\blue{.069}/\magenta{.364}/\green{.9002} & \\
            BPG & .08/\red{14.33}/\blue{.114}/\magenta{.299}/\green{.9383} & .08/\red{-}/\blue{.222}/\magenta{.311}/\green{.9260} & -/\red{25.33}/\blue{.093}/\magenta{.334}/\green{.9210} & \\
            WebP & .16/\red{8.67}/\blue{.168}/\magenta{.240}/\green{.9517} & .16/\red{-}/\blue{.223}/\magenta{.221}/\green{.9563} & -/\red{29.67}/\blue{.213}/\magenta{.238}/\green{.9517} & \\
            \hline
            JPEG & .25/\red{10.33}/\blue{.043}/\magenta{.310}/\green{.8986} & .22/\red{-}/\blue{.193}/\magenta{.299}/\green{.9172} & -/\red{19.67}/\blue{.073}/\magenta{.332}/\green{.9094} & -sg; tt\\
            \hline
            JP2 & .08/\red{2.00}/\blue{.005}/\magenta{.519}/\green{.8479} & .08/\red{-}/\blue{.084}/\magenta{.399}/\green{.8415} & -/\red{96.90}/\blue{.242}/\magenta{.251}/\green{.7989} & \multirow{2}{*}{-Phase 3} \\
            WebP & .16/\red{8.67}/\blue{.168}/\magenta{.240}/\green{.9517} & .16/\red{-}/\blue{.223}/\magenta{.221}/\green{.9563} & -/\red{95.00}/\blue{.335}/\magenta{.120}/\green{.9444} & \\
            \hline
        \end{tabular}
    \end{center}
    \label{ablation}
\end{table*}

\section{Conclusion}
In this paper, we proposed a generic framework that can be used to semantically enhance any image codec.
Our work is the first to study the efficacy of high-level semantics as a fundamental supplement to low-level visual features in the general compression setting.
\chh{Moreover, in contrast to how existing works mostly focus only on rate-distortion, we are the first to consider the joint optimization of perception quality and distortion on full-resolution images. We also demonstrate improved accuracy of downstream vision algorithm.}

Despite its simplicity, an instantiation of our framework effectively improved the perception-accuracy-distortion performance of existing learned and engineered codecs by enabling them to leverage high-level semantics in the form of segmentation maps.
And extensive ablation study was performed to validate that the use of semantics is the key to the improved performance.

\bibliographystyle{IEEEtran}
\bibliography{main}

\renewcommand{\thesection}{\Alph{section}}
\setcounter{section}{0}
\clearpage
\section*{Supplementary Materials for ``JPD-SE: High-Level Semantics for Joint Perception-Distortion Enhancement in Image Compression''}

\section{Full Table for R-D Performance}
See Table~\ref{full_table} for the complete results of the R-D performance of all the models that we have trained and tested.
We put each SE codec side-by-side against its original to clearly demonstrate how high-level semantics affected codec performance.
This provides more insights in addition to the R-D curves in Fig.~\ref{fig2}.

In this table, the SE codecs achieve better distortion performance in nearly all comparisons.
This comes at the price of an extra \(0.03\) bpp overhead from semantics which necessarily worsens the R-D performance.
Although, note that this overhead is based on our primitive lossless semantics compression method described in Section~\ref{sec1}.
Given that compressing semantics has not been a major research focus in the past and that few works on this topic exist, we believe that more efficient methods can be developed in the future and we hope that our work can draw attention to this largely unexplored area, which would help the SE codecs approach the R-D performance upper bound given in Table~\ref{full_table}.

\begin{sidewaystable*}[]
    \scriptsize
    \centering
    \begin{tabular}{@{}clllllllllllll@{}}
        \toprule
        \multicolumn{1}{l}{} & & \multicolumn{2}{c}{JPEG} & \multicolumn{2}{c}{JP2} & \multicolumn{2}{c}{WebP} & \multicolumn{2}{c}{BPG} & \multicolumn{2}{c}{learned} & \multicolumn{2}{c}{learned (LS)}\\
        \midrule
        \multicolumn{1}{l}{} & & \multicolumn{1}{c}{Original} & \multicolumn{1}{c}{SE (Ours)} & \multicolumn{1}{c}{Original} & \multicolumn{1}{c}{SE (Ours)} & \multicolumn{1}{c}{Original} & \multicolumn{1}{c}{SE (Ours)} & \multicolumn{1}{c}{Original} & \multicolumn{1}{c}{SE (Ours)} & \multicolumn{1}{c}{Original} & \multicolumn{1}{c}{SE (Ours)} & \multicolumn{1}{c}{Original} & \multicolumn{1}{c}{SE (Ours)} \\
        \midrule
        & bpp & 0.19 & --- & 0.05 & --- & 0.07 & --- & 0.02 & --- & 0.02 & 0.03 & 0.02 & 0.03 \\
        & LPIPS & 0.442  & \textbf{0.299}    & 0.605  & \textbf{0.399}    & 0.406  & \textbf{0.313}    & 0.484  & \textbf{0.384}    & 0.499  & \textbf{0.391}    & 0.499  & \textbf{0.415}    \\
        & MS-SSIM  & 0.8274  & \textbf{0.9172}    & 0.7971  & \textbf{0.8415}    & 0.8913  & \textbf{0.9206}    & 0.8516  & \textbf{0.8742}    & 0.8327  & \textbf{0.8703}    & 0.8327  & \textbf{0.8639}    \\
        & PSNR  & 26.36 & \textbf{29.86}   & 25.94 & \textbf{26.84}   & 29.05 & \textbf{29.96}   & 27.45 & \textbf{27.99}   & 25.11 & \textbf{26.20}   & 25.11 & \textbf{25.88}   \\
        \midrule

        & bpp & 0.25 & --- & 0.06 & --- & 0.12 & --- & 0.04 & --- & 0.05 & 0.05 & 0.05 & 0.05 \\
        & LPIPS & 0.310  & \textbf{0.224}    & 0.570  & \textbf{0.382}    & 0.299  & \textbf{0.235}    & 0.423  & \textbf{0.348}    & 0.409  & \textbf{0.323}    & 0.409  & \textbf{0.337}    \\
        & MS-SSIM  & 0.8986  & \textbf{0.9537}    & 0.8197  & \textbf{0.8610}    & 0.9356  & \textbf{0.9518}    & 0.8873  & \textbf{0.9034}    & 0.8884  & \textbf{0.9106}    & 0.8884  & \textbf{0.9071}    \\
        & PSNR  & 29.55 & \textbf{32.17}   & 26.55 & \textbf{27.54}   & 31.17 & \textbf{31.90}   & 28.85 & \textbf{29.33}   & 26.87 & \textbf{27.84}   & 26.87 & \textbf{27.57}   \\

        \midrule
        & bpp & 0.30 & --- & 0.08 & --- & 0.13 & --- & 0.05 & --- & 0.10 & 0.11 & 0.21 & 0.22  \\
        & LPIPS & 0.224  & \textbf{0.180}    & 0.519  & \textbf{0.363}    & 0.277  & \textbf{0.221}    & 0.360  & \textbf{0.311}    & 0.310  & \textbf{0.247}    & 0.244  & \textbf{0.195}\\
        & MS-SSIM  & 0.9329  & \textbf{0.9671}    & 0.8479  & \textbf{0.8826}    & 0.9420  & \textbf{0.9563}    & 0.9164  & \textbf{0.9260}    & 0.9239  & \textbf{0.9399}    & 0.9489  & \textbf{0.9558} \\
        & PSNR  & 31.57 & \textbf{33.39}   & 27.38 & \textbf{28.33}   & 31.64 & \textbf{32.31}   & 30.26 & \textbf{30.55}   & 28.54 & \textbf{29.39}   & 29.80 & \textbf{30.27}  \\

        \midrule
        & bpp & 0.35 & --- & 0.12 & --- & 0.16 & --- & 0.08 & --- & 0.21 & 0.22 & 0.43 & 0.46 \\
        & LPIPS & 0.176  & \textbf{0.153}    & 0.450  & \textbf{0.326}    & 0.240  & \textbf{0.199}    & 0.299  & \textbf{0.268}    & 0.244  & \textbf{0.192}    & 0.186  & \textbf{0.158}   \\
        & MS-SSIM  & 0.9532  & \textbf{0.9742}    & 0.8809  & \textbf{0.9075}    & 0.9517  & \textbf{0.9625}    & 0.9383  & \textbf{0.9446}    & 0.9489  & \textbf{0.9565}    & \textbf{0.9638}    & 0.9633    \\
        & PSNR  & 32.90 & \textbf{34.43}   & 28.60 & \textbf{29.54}   & 32.42 & \textbf{32.87}   & 31.67 & \textbf{31.92}   & 29.80 & \textbf{30.35}   & \textbf{30.95} & 30.49   \\
        \midrule
        & bpp & 0.40 & --- & 0.24 & --- & 0.21 & --- & 0.12 & --- & 0.43 & 0.46 & & \\
        & LPIPS & 0.145  & \textbf{0.131}    & 0.320  & \textbf{0.250}    & 0.193  & \textbf{0.166}    & 0.244  & \textbf{0.222}    & 0.186  & \textbf{0.148}    & & \\
        & MS-SSIM  & 0.9638  & \textbf{0.9789}    & 0.9263  & \textbf{0.9425}    & 0.9630  & \textbf{0.9702}    & 0.9544  & \textbf{0.9585}    & 0.9638  & \textbf{0.9695}    & & \\
        & PSNR  & 33.79 & \textbf{35.16}   & 30.96 & \textbf{31.92}   & 33.52 & \textbf{33.80}   & 33.01 & \textbf{33.16}   & 30.95 & \textbf{31.97}   & & \\
        \midrule
        & bpp & & & & & 0.23 & --- & 0.17 & --- & & & & \\
        & LPIPS &     &     &     &     & 0.176  & \textbf{0.158}    & 0.190  & \textbf{0.179}    &     &     &     &     \\
        & MS-SSIM  &    &    &    &    & 0.9667  & \textbf{0.9723}    & 0.9669  & \textbf{0.9697}    &    &    &    &    \\
        & PSNR  &     &     &     &     & 33.95 & \textbf{34.05}   & \textbf{34.41}   & 34.28 &     &     &     &     \\
        \midrule
        & bpp & & & & & 0.25 & --- & 0.25 & --- & & & & \\
        & LPIPS &    &    &    &    & 0.162  & \textbf{0.150}    & 0.142    & \textbf{0.136}  &    &    &    &    \\
        & MS-SSIM  &    &    &    &    & 0.9698  & \textbf{0.9739}    & 0.9767  & \textbf{0.9780}    &    &    &    &    \\
        & PSNR  &     &     &     &     & \textbf{34.38} & 34.35   & \textbf{35.93}   & 35.63 &     &     &     &     \\
        \bottomrule
    \end{tabular}
    \caption{
        Full table for R-D performance of all models.
        The bpp of an SE codec equals that of its backbone \(c\) plus an extra \(0.03\) bpp overhead from semantics.
        The bpp values from the visuals differ for learned and learned-SE because we retrained a new backbone codec \(c\) for the latter.
        In learned (LS), the SE models used learned semantics extracted by off-the-shelf segmentation networks at test time.
        The better in each original-SE pair is marked in bold.
        The SE model almost always achieves superior distortion performance.
    }
    \label{full_table}
\end{sidewaystable*}

\section{Full Images of Visualizations in Main Text}
\begin{figure*}[h] 
    \thisfloatpagestyle{empty}
    \vskip -.7in
    \centering
    \subfloat{
        \hspace{-.17in}
        \rotatebox{90}{\scriptsize{(a) JPEG (0.24 bpp)}}
        \includegraphics[width=1\columnwidth]{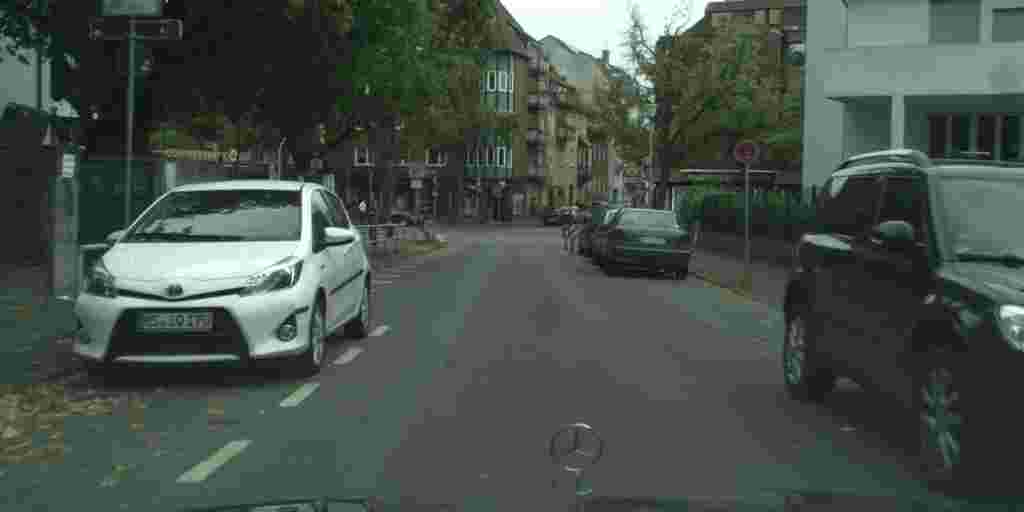}
    }
    \subfloat{
        \hspace{-.095in}
        \includegraphics[width=1\columnwidth]{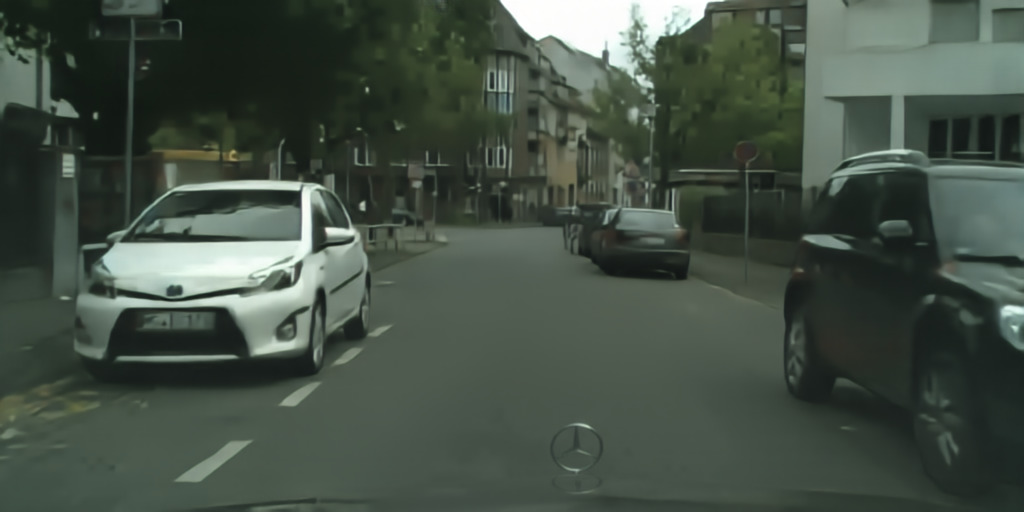}
        \rotatebox{90}{\scriptsize{(b) JPEG-SE (0.21 bpp)}}
    }
    \vspace{-11pt}
    \subfloat{
        \hspace{-.17in}
        \rotatebox{90}{\scriptsize{(c) BPG (0.05 bpp)}}
        \includegraphics[width=1\columnwidth]{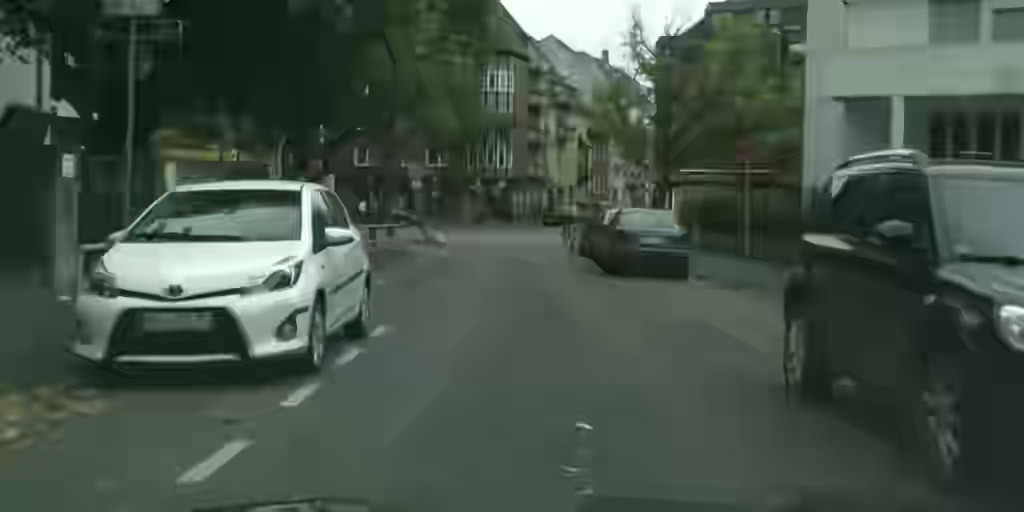}
    }
    \subfloat{
        \hspace{-.095in}
        \includegraphics[width=1\columnwidth]{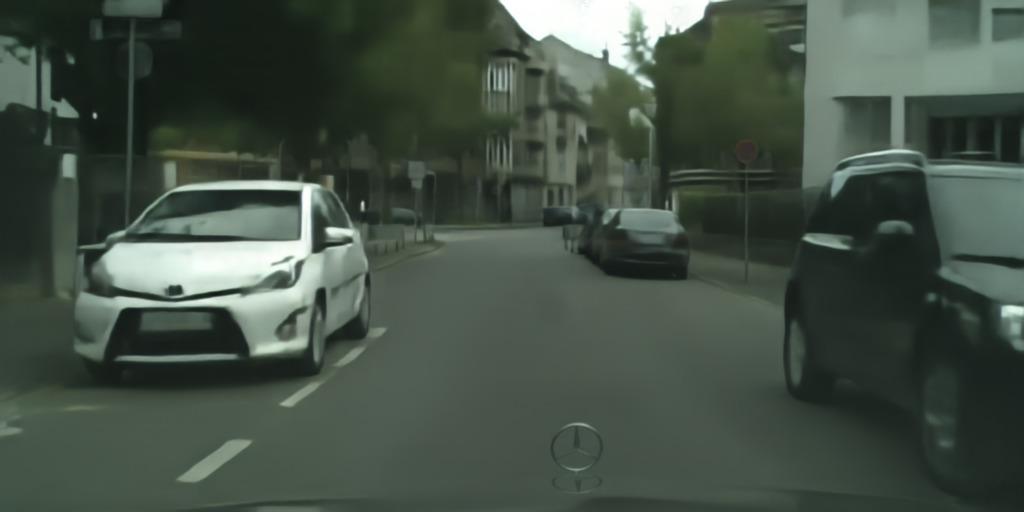}
        \rotatebox{90}{\scriptsize{(d) BPG-SE (0.05 bpp)}}
    }
    \vspace{-11pt}
    \subfloat{
        \hspace{-.17in}
        \rotatebox{90}{\scriptsize{(e) WebP (0.11 bpp)}}
        \includegraphics[width=1\columnwidth]{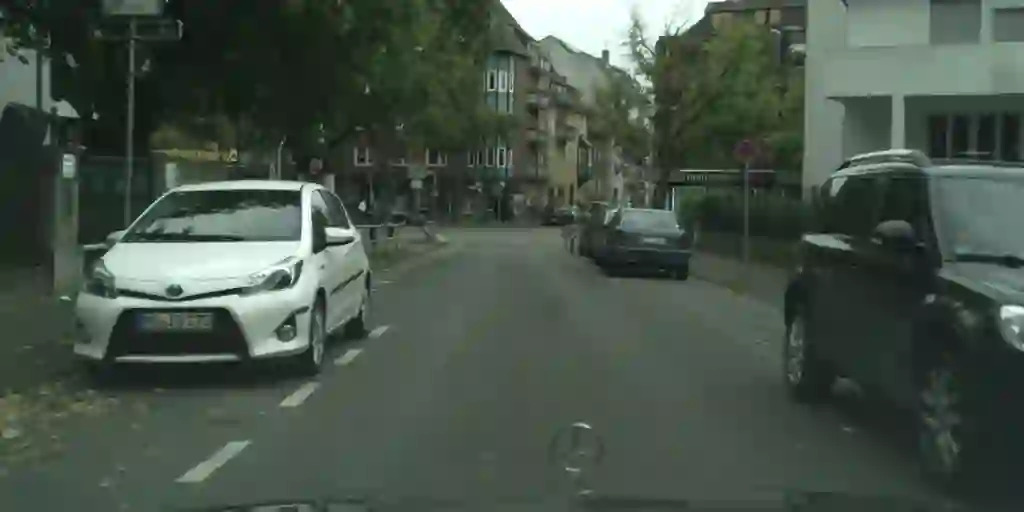}
    }
    \subfloat{
        \hspace{-.095in}
        \includegraphics[width=1\columnwidth]{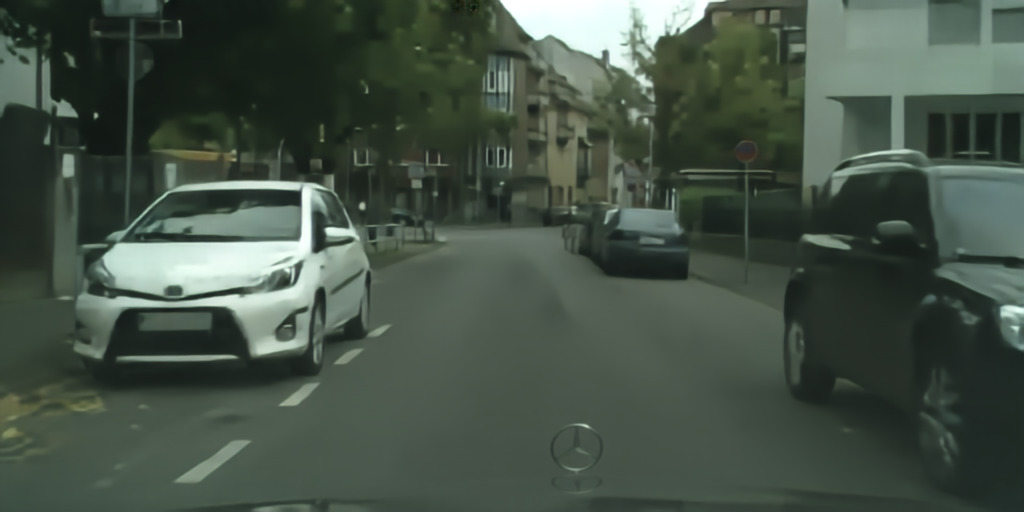}
        \rotatebox{90}{\scriptsize{(f) WebP-SE (0.10 bpp)}}
    }
    \vspace{-11pt}
    \subfloat{
        \hspace{-.17in}
        \rotatebox{90}{\scriptsize{(g) learned (0.21 bpp)}}
        \includegraphics[width=1\columnwidth]{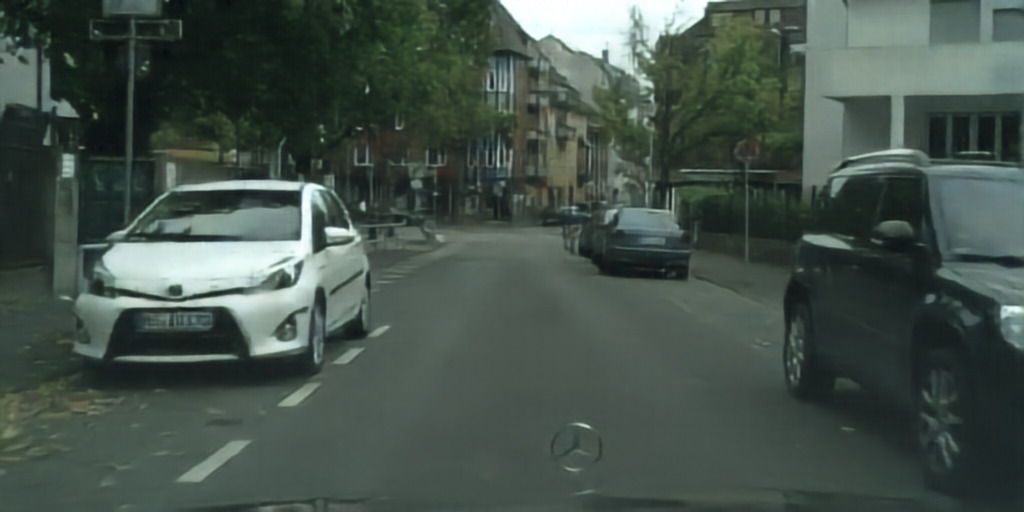}
    }
    \subfloat{
        \hspace{-.095in}
        \includegraphics[width=1\columnwidth]{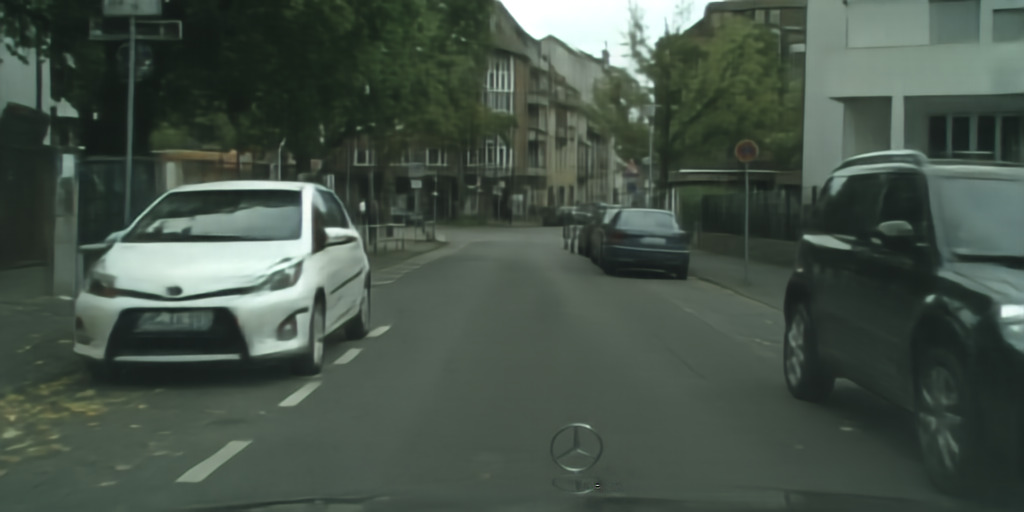}
        \rotatebox{90}{\scriptsize{(h) learned-SE (0.14 bpp)}}
    }
    \vspace{-11pt}
    \subfloat{
        \hspace{-.17in}
        \rotatebox{90}{\scriptsize{(i) JP2 (0.12 bpp)}}
        \includegraphics[width=1\columnwidth]{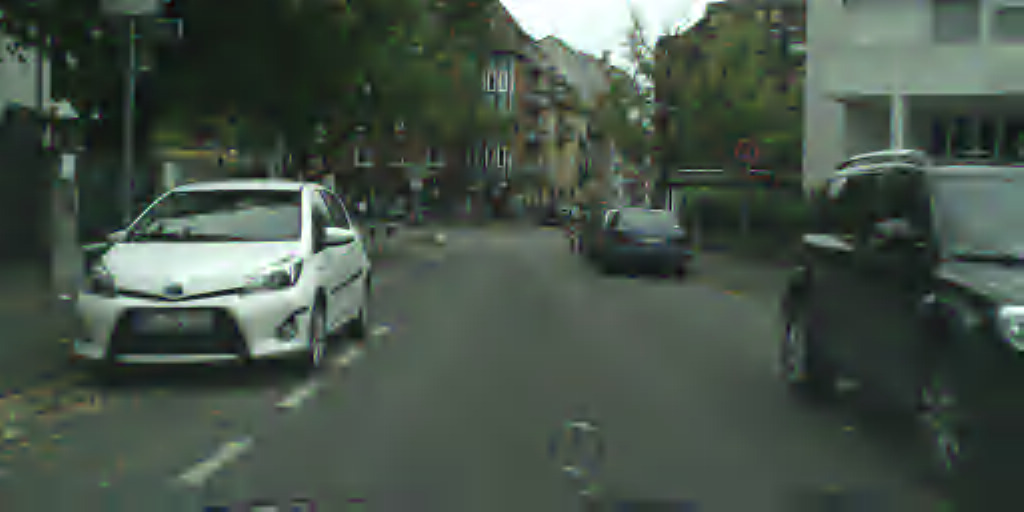}
        \label{afig3g}
    }
    \subfloat{
        \hspace{-.095in}
        \includegraphics[width=1\columnwidth]{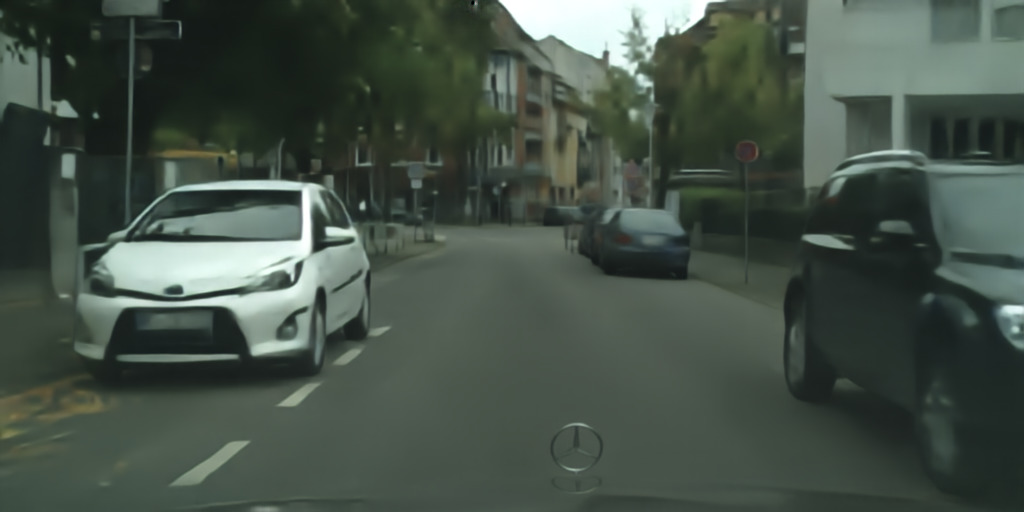}
        \rotatebox{90}{\scriptsize{(j) JP2-SE (0.11 bpp)}}
        \label{afig3h}
    }
    \vspace{-11pt}
    \subfloat{
        \hspace{-.17in}
        \rotatebox{90}{\scriptsize{(k) Original}}
        \includegraphics[width=1\columnwidth]{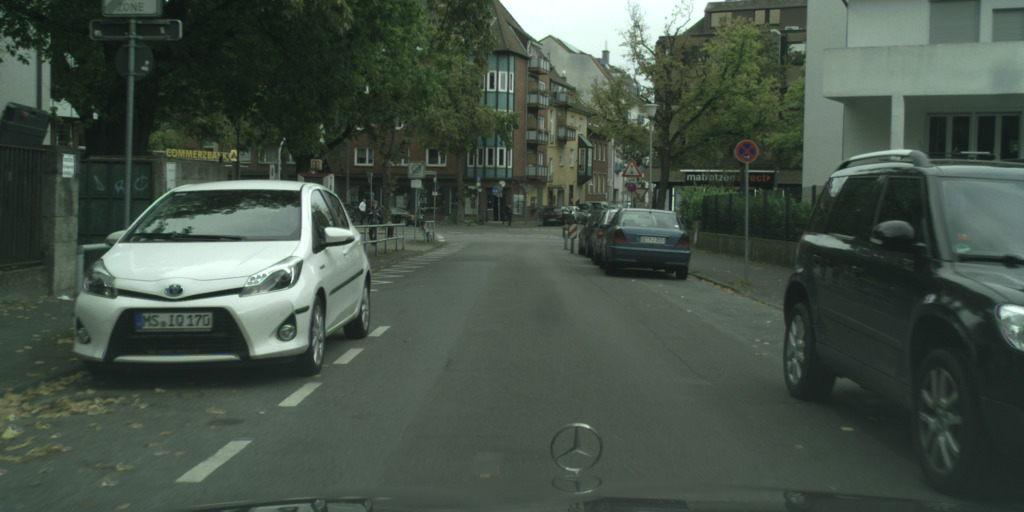}
    }
    \subfloat{
        \hspace{-.095in}
        \includegraphics[width=1\columnwidth]{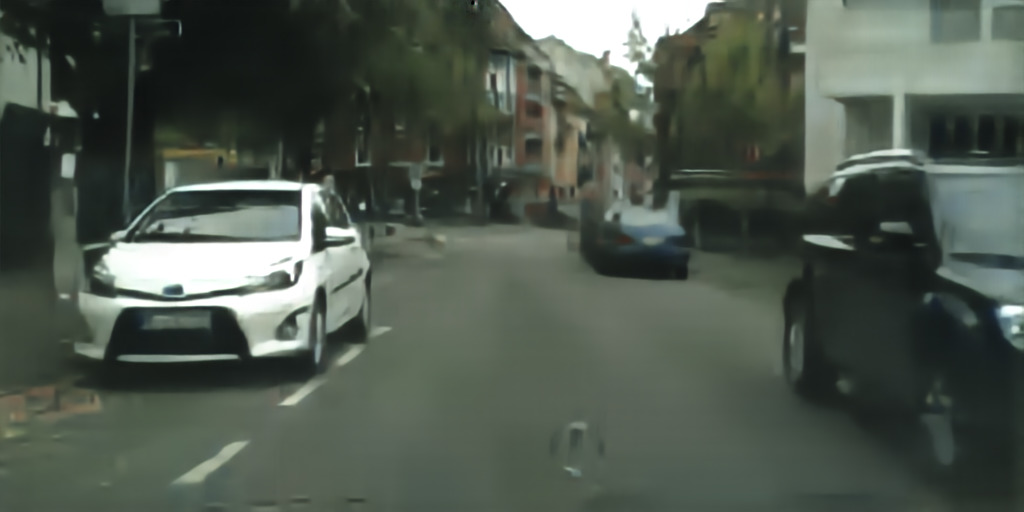}
        \rotatebox{90}{\scriptsize{(l) JP2-SE (-s; test)}}
        \label{afig3f}
    }
    \caption{
        Full images for Fig.~\ref{fig3}
    }
    \label{afig3}
\end{figure*}

See Fig.~\ref{afig3}, \ref{afig34}, \ref{afig4}, and \ref{afig5} for the full images of Fig.~\ref{fig3}, \ref{fig34}, \ref{fig4}, and \ref{fig5}, respectively.

\begin{figure*}
    \centering
    \includegraphics[width=1\columnwidth]{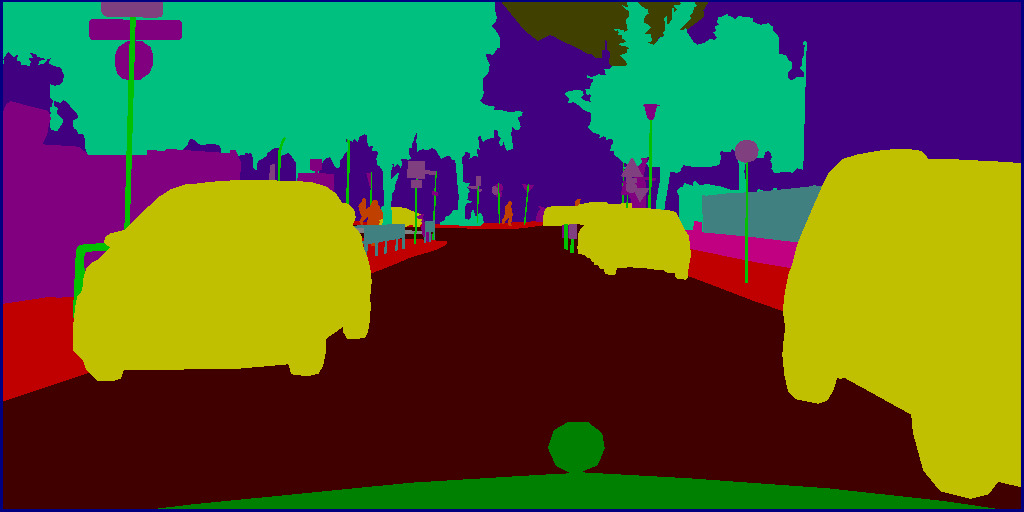}
    \caption{
        Semantics (class seg. map) for Fig.~\ref{afig3}.
    }
\end{figure*}
\begin{figure*}[h]
    \thisfloatpagestyle{empty}
    \vskip -.7in
    \centering
    \subfloat{
        \hspace{-.17in}
        \rotatebox{90}{\scriptsize{\chh{(a) Original}}}
        \includegraphics[width=1\columnwidth]{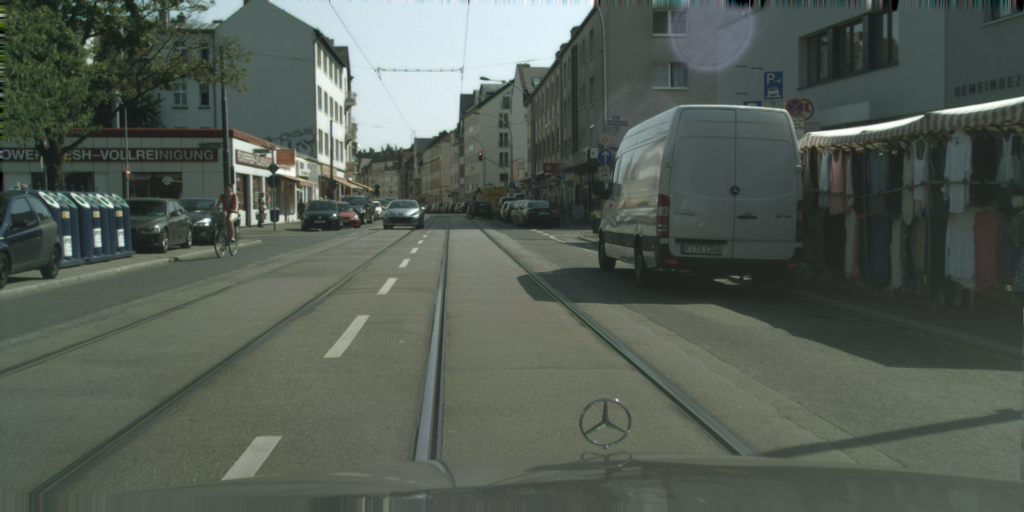}
    }
    \subfloat{
        \hspace{-.095in}
        \includegraphics[width=1\columnwidth]{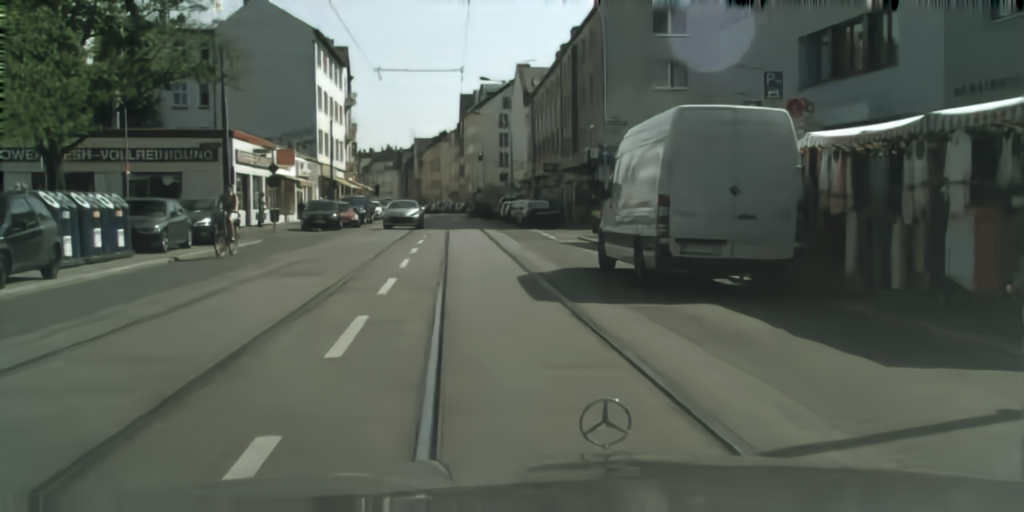}
        \rotatebox{90}{\scriptsize{\chh{(b) BPG-SE (0.17 bpp)}}}
    }
    \vspace{-11pt}
    \subfloat{
        \hspace{-.17in}
        \rotatebox{90}{\scriptsize{\chh{(c) \cite{agustsson2019generative} (0.18 bpp)}}}
        \includegraphics[width=1\columnwidth]{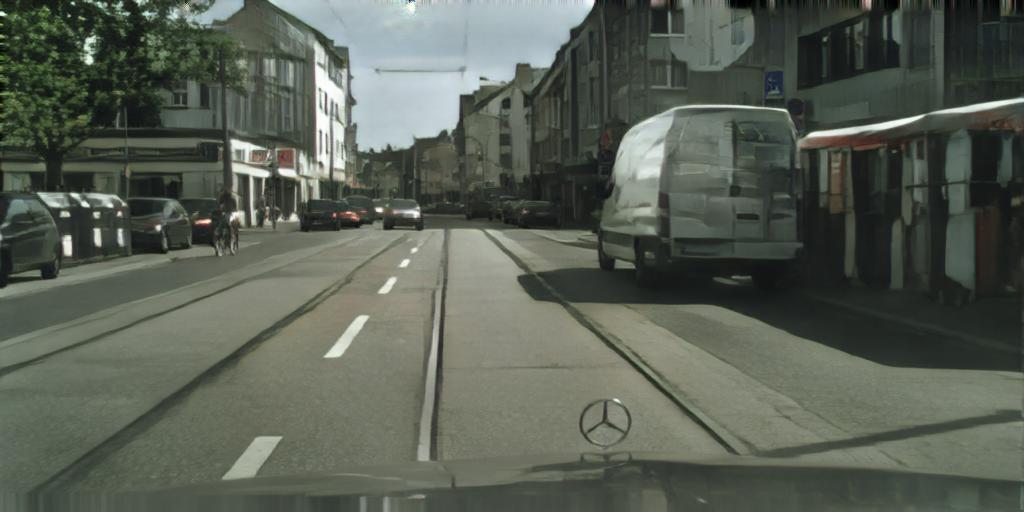}
    }
    \subfloat{
        \hspace{-.095in}
        \includegraphics[width=1\columnwidth]{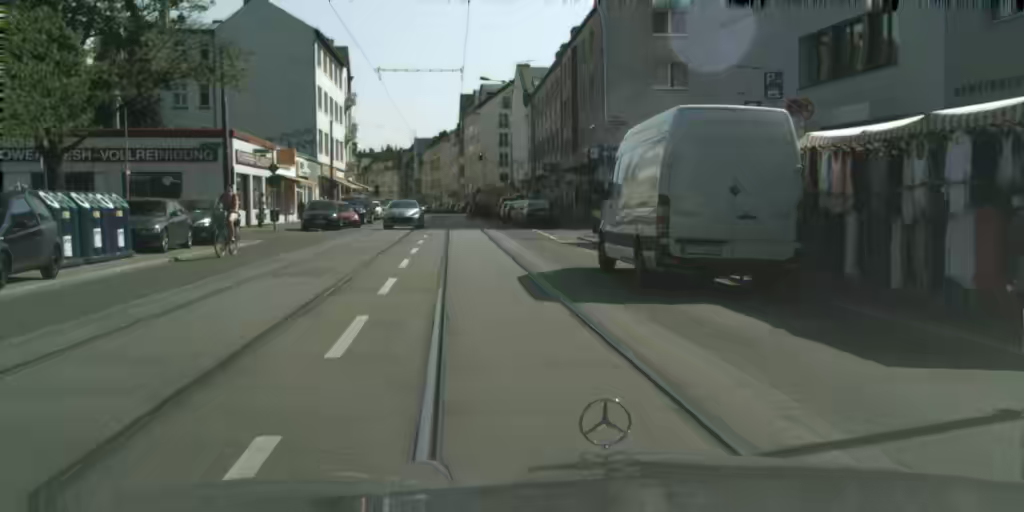}
        \rotatebox{90}{\scriptsize{\chh{(d) BPG (0.16 bpp)}}}
    }
    \vspace{-11pt}
    \subfloat{
        \hspace{-.17in}
        \rotatebox{90}{\scriptsize{\chh{(e) Original}}}
        \includegraphics[width=1\columnwidth]{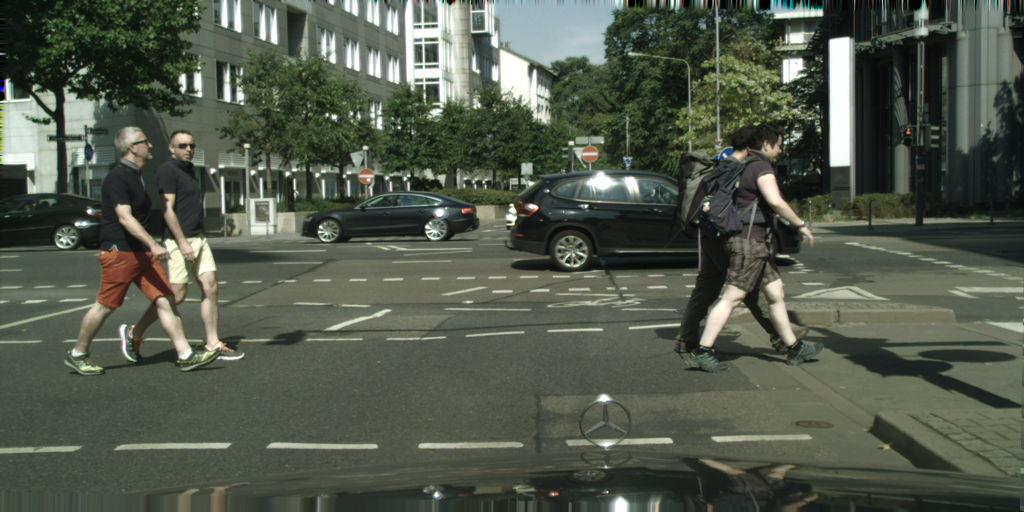}
    }
    \subfloat{
        \hspace{-.095in}
        \includegraphics[width=1\columnwidth]{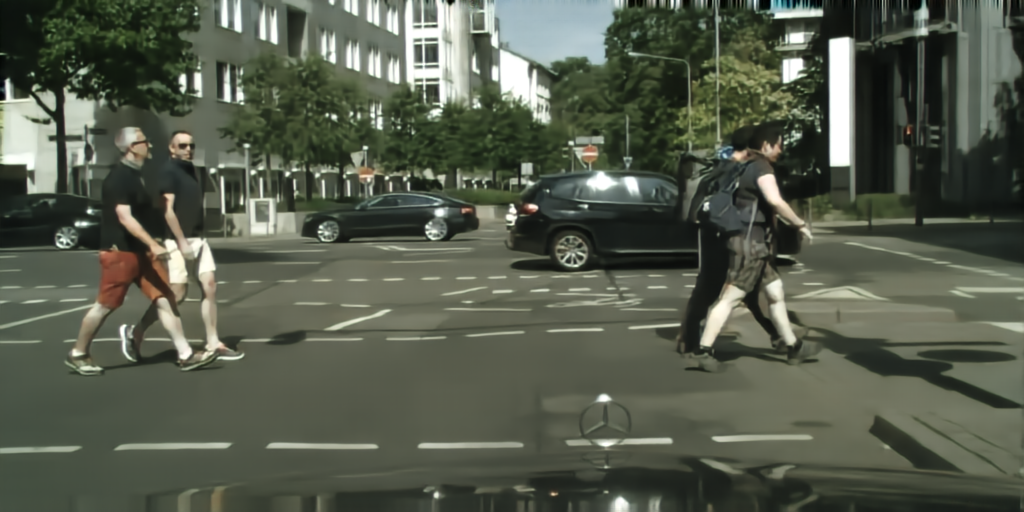}
        \rotatebox{90}{\scriptsize{\chh{(f) BPG-SE (0.24 bpp)}}}
    }
    \vspace{-11pt}
    \subfloat{
        \hspace{-.17in}
        \rotatebox{90}{\scriptsize{\chh{(g) \cite{agustsson2019generative} (0.33 bpp)}}}
        \includegraphics[width=1\columnwidth]{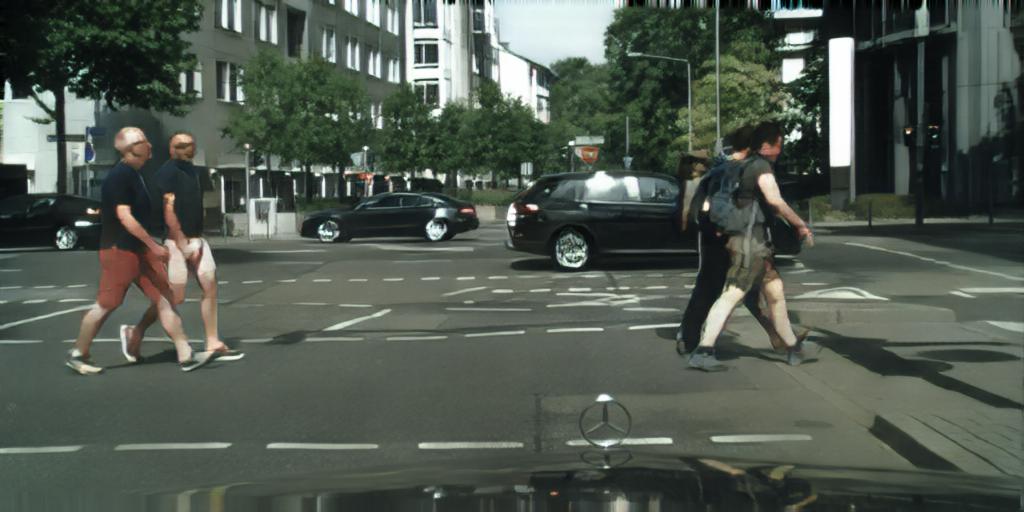}
    }
    \subfloat{
        \hspace{-.095in}
        \includegraphics[width=1\columnwidth]{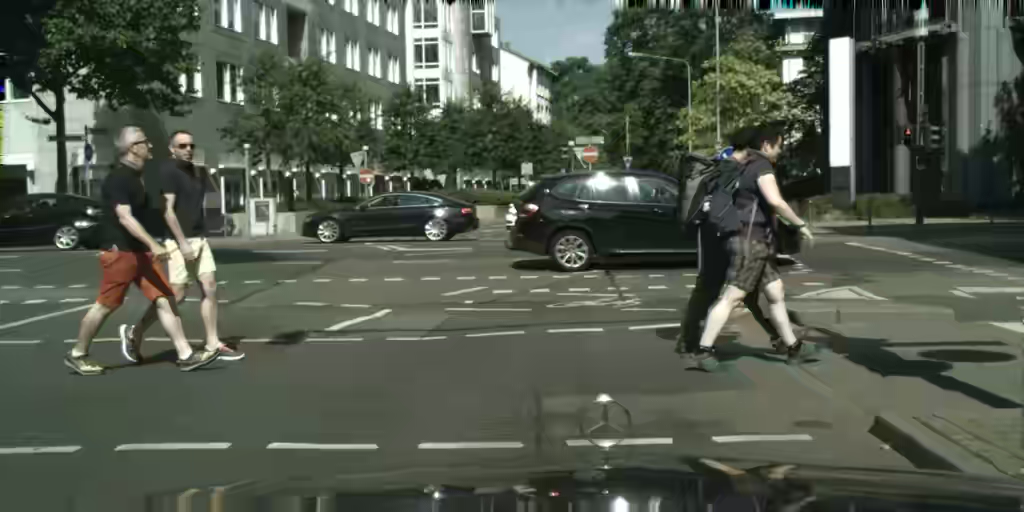}
        \rotatebox{90}{\scriptsize{\chh{(h) BPG (0.24 bpp)}}}
    }
    \vspace{-11pt}
    \subfloat{
        \hspace{-.17in}
        \rotatebox{90}{\scriptsize{\chh{(i) Original}}}
        \includegraphics[width=1\columnwidth]{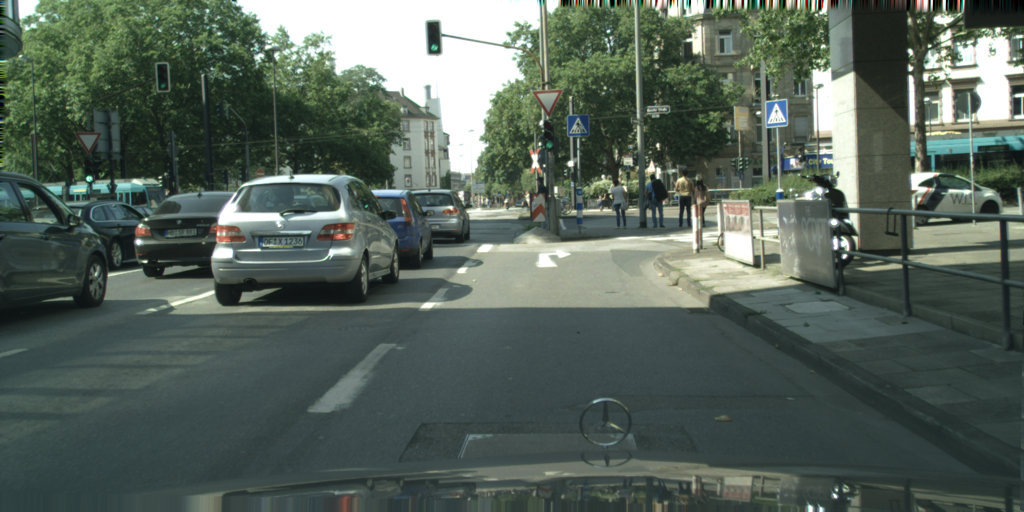}
    }
    \subfloat{
        \hspace{-.095in}
        \includegraphics[width=1\columnwidth]{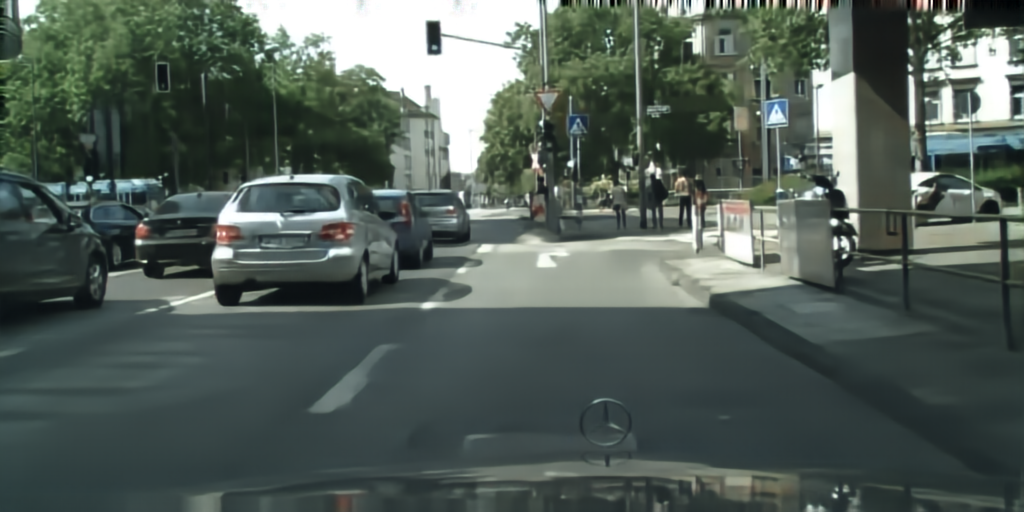}
        \rotatebox{90}{\scriptsize{\chh{(j) BPG-SE (0.19 bpp)}}}
    }
    \vspace{-11pt}
    \subfloat{
        \hspace{-.17in}
        \rotatebox{90}{\scriptsize{\chh{(k) \cite{agustsson2019generative} (0.19 bpp)}}}
        \includegraphics[width=1\columnwidth]{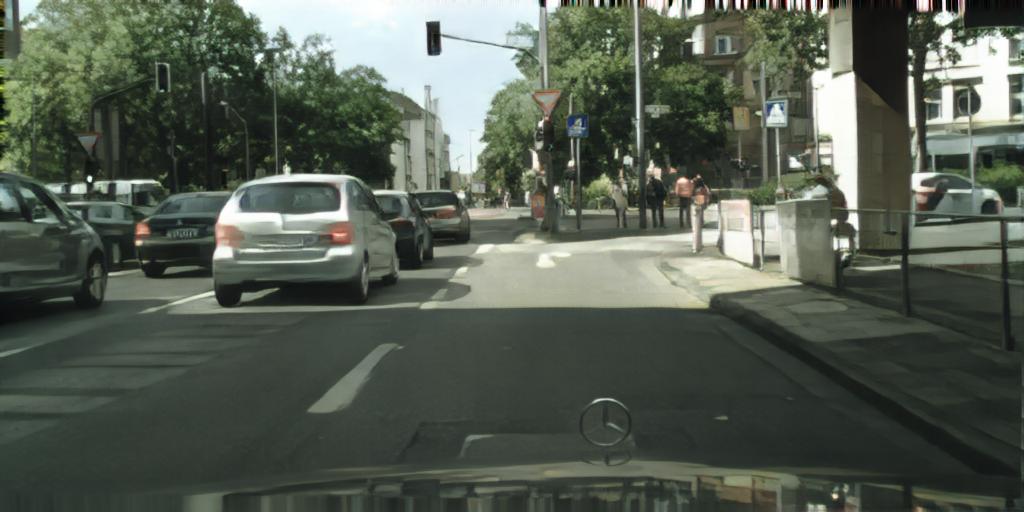}
    }
    \subfloat{
        \hspace{-.095in}
        \includegraphics[width=1\columnwidth]{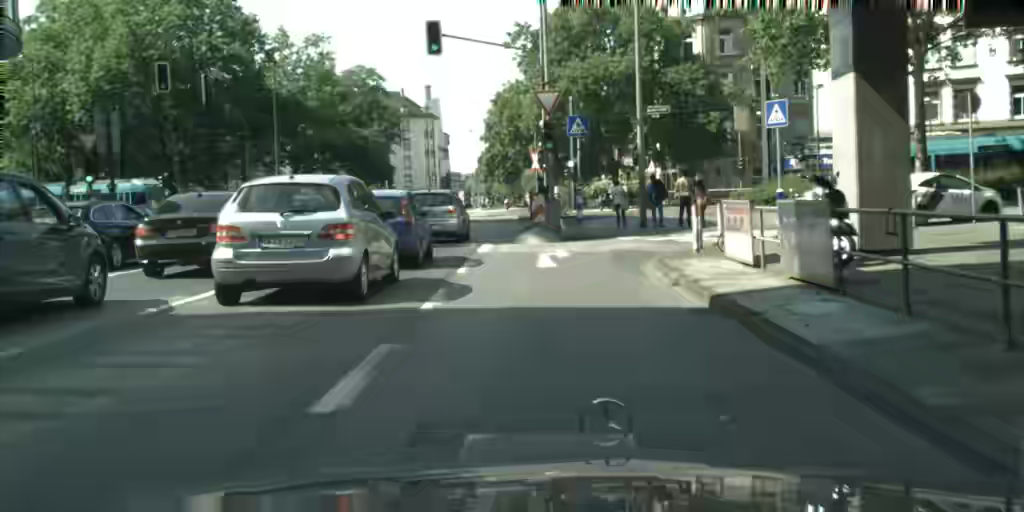}
        \rotatebox{90}{\scriptsize{\chh{(l) BPG (0.19 bpp)}}}
    }
    \caption{
        \chh{Full images for Fig.~\ref{fig34} and more comparisons with \cite{agustsson2019generative}.}
    }
    \label{afig34}
\end{figure*}
\begin{figure*} 
    \centering
    \subfloat{
        \label{afig4a}
        \hspace{-.17in}
        \rotatebox{90}{\scriptsize{(a) JPEG (0.22 bpp)}}
        \includegraphics[width=1\columnwidth]{./figures/jpgq10_3}
    }
    \subfloat{
        \label{afig4b}
        \hspace{-.095in}
        \includegraphics[width=1\columnwidth]{./figures/jpgq5_sehs_3}
        \rotatebox{90}{\scriptsize{(b) JPEG-SE (0.21 bpp)}}
    }
    \vspace{-11pt}
    \subfloat{
        \label{afig4c}
        \hspace{-.17in}
        \rotatebox{90}{\scriptsize{(c) JPEG-SE (-sg; tt)}}
        \includegraphics[width=1\columnwidth]{./figures/jpgq5_sehs_3_wo_sem_wo_gan_at_traintest}
    }
    \subfloat{
        \label{afig4d}
        \hspace{-.095in}
        \includegraphics[width=1\columnwidth]{./figures/jpgq5_sehs_3_wo_sem_at_traintest}
        \rotatebox{90}{\scriptsize{(d) JPEG-SE (-s; tt)}}
    }
    \vspace{-11pt}
    \subfloat{
        \hspace{-.17in}
        \rotatebox{90}{\scriptsize{(e) Original}}
        \includegraphics[width=1\columnwidth]{./figures/3}
    }
    \subfloat{
        \hspace{-.095in}
        \includegraphics[width=1\columnwidth]{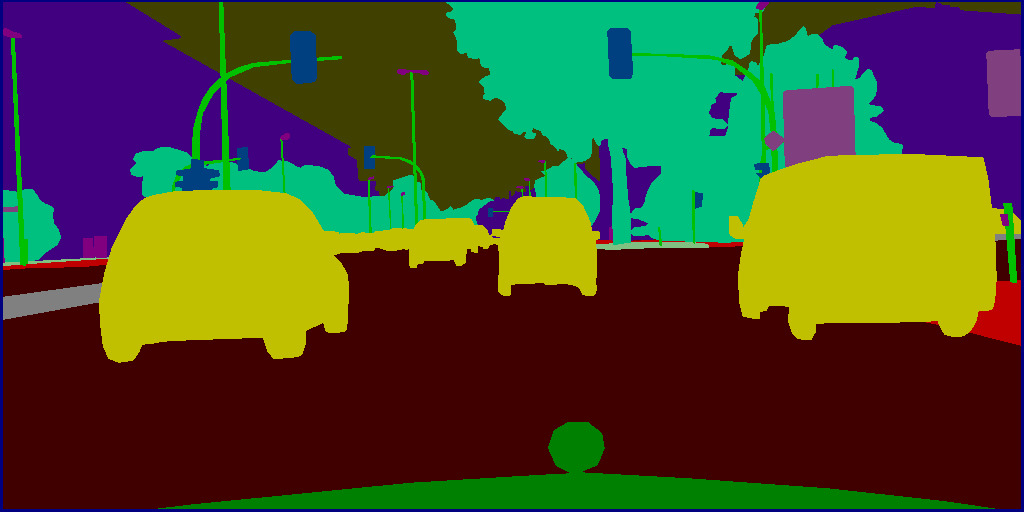}
        \rotatebox{90}{\scriptsize{(f) Semantics (class seg. map)}}
    }
    \caption{
        Full images for Fig.~\ref{fig4}.
    }
    \label{afig4}
\end{figure*}
\begin{figure*} 
    \centering
    \subfloat{
        \hspace{-.17in}
        \rotatebox{90}{\scriptsize{(a) JP2-SE (0.08 bpp)}}
        \includegraphics[width=1\columnwidth]{./figures/j2kq500_sehs_5}
    }
    \subfloat{
        \hspace{-.095in}
        \includegraphics[width=1\columnwidth]{./figures/j2kq300_5}
        \rotatebox{90}{\scriptsize{(b) JP2 (0.08 bpp)}}
    }
    \vspace{-11pt}
    \subfloat{
        \hspace{-.17in}
        \rotatebox{90}{\scriptsize{(c) JP2-SE (-Phase 3)}}
        \includegraphics[width=1\columnwidth]{./figures/j2kq500_sehs_5_twotrainingphasesonly}
    }
    \subfloat{
        \hspace{-.095in}
        \includegraphics[width=1\columnwidth]{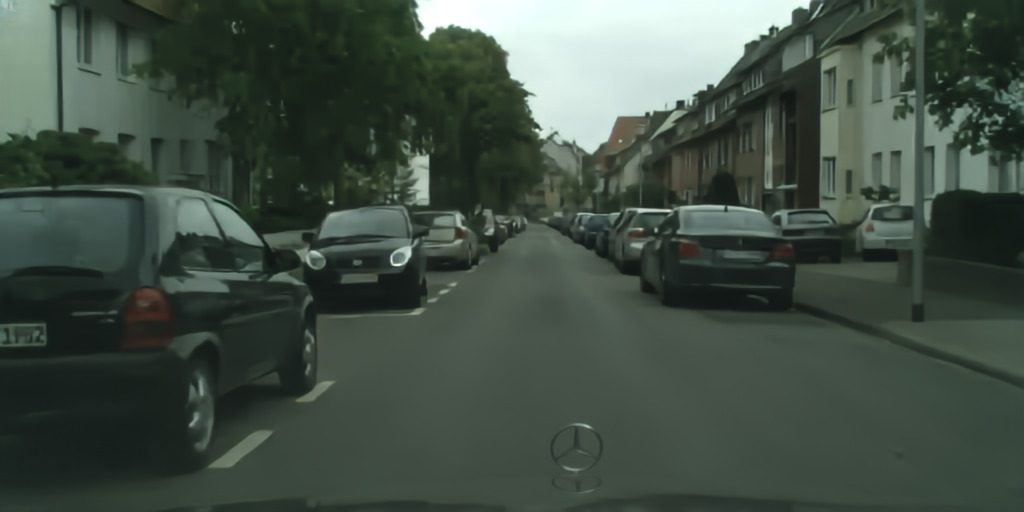}
        \rotatebox{90}{\scriptsize{(d) WebP-SE (0.14 bpp)}}
    }
    \vspace{-11pt}
    \subfloat{
        \hspace{-.17in}
        \rotatebox{90}{\scriptsize{(e) WebP (0.13 bpp)}}
        \includegraphics[width=1\columnwidth]{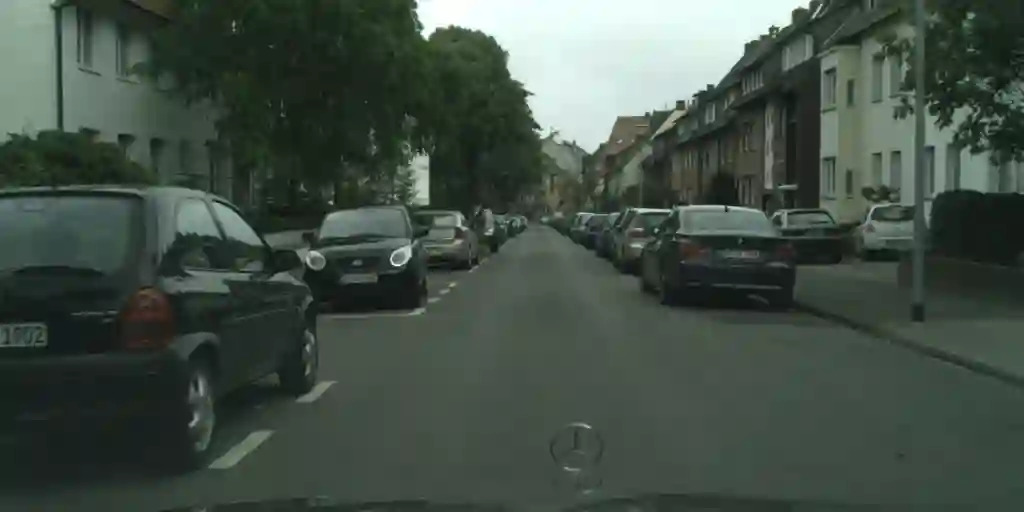}
    }
    \subfloat{
        \hspace{-.095in}
        \includegraphics[width=1\columnwidth]{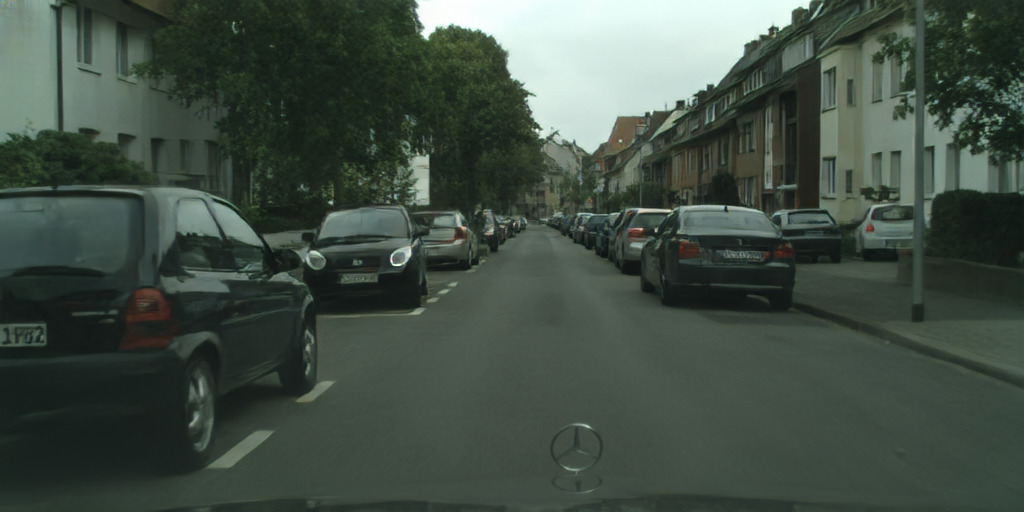}
        \rotatebox{90}{\scriptsize{(f) WebP-SE (-Phase 3)}}
    }
    \vspace{-11pt}
    \subfloat{
        \hspace{-.17in}
        \rotatebox{90}{\scriptsize{(g) Original}}
        \includegraphics[width=1\columnwidth]{./figures/5}
    }
    \subfloat{
        \hspace{-.095in}
        \includegraphics[width=1\columnwidth]{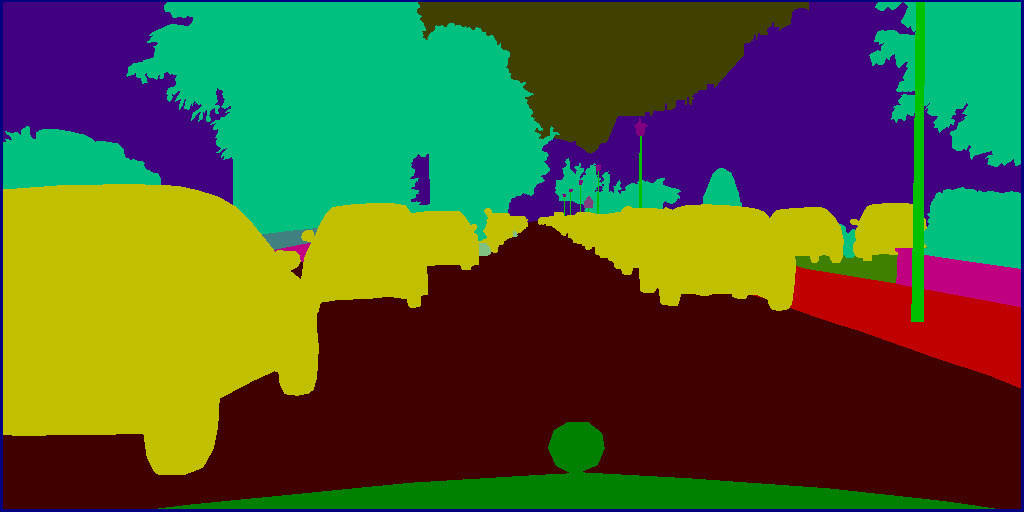}
        \rotatebox{90}{\scriptsize{(h) Semantics (class seg. map)}}
    }
    \caption{
        Full images for Fig.~\ref{fig5}.
        Additionally, we also provide visualization for a WebP model trained without Phase 3, the quantitative performance of which we reported in Table~\ref{ablation}.
    }
    \label{afig5}
\end{figure*}

\section{More Visualizations for Cityscapes}
\begin{figure*}[h]
    \thisfloatpagestyle{empty}
    \vskip -.7in
    \centering
    \subfloat{
        \hspace{-.17in}
        \rotatebox{90}{\scriptsize{(a) JPEG (0.28 bpp)}}
        \includegraphics[width=1.\columnwidth]{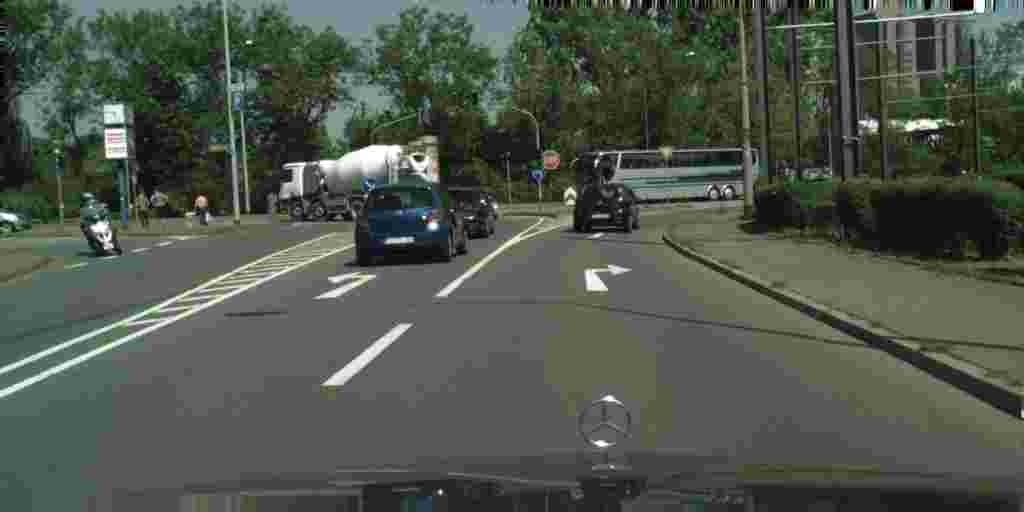}
    }
    \subfloat{
        \hspace{-.095in}
        \includegraphics[width=1.\columnwidth]{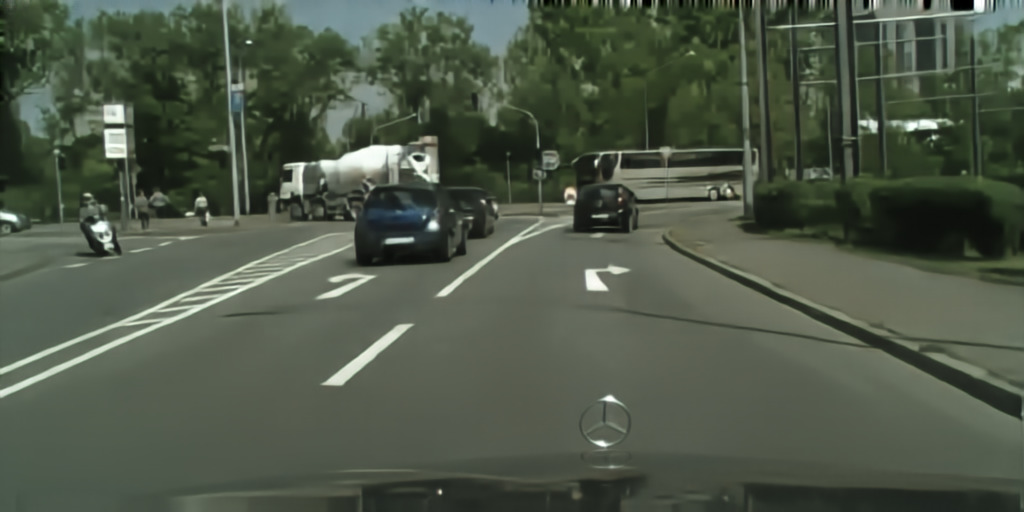}
        \rotatebox{90}{\scriptsize{(b) JPEG-SE (0.25 bpp)}}
    }
    \vspace{-11pt}
    \subfloat{
        \hspace{-.17in}
        \rotatebox{90}{\scriptsize{(c) BPG (0.07 bpp)}}
        \includegraphics[width=1.\columnwidth]{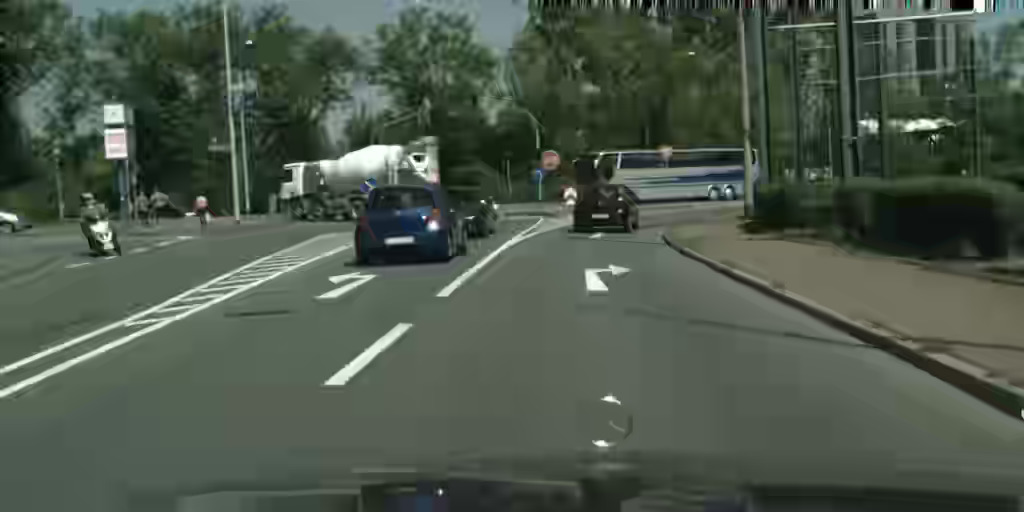}
    }
    \subfloat{
        \hspace{-.095in}
        \includegraphics[width=1.\columnwidth]{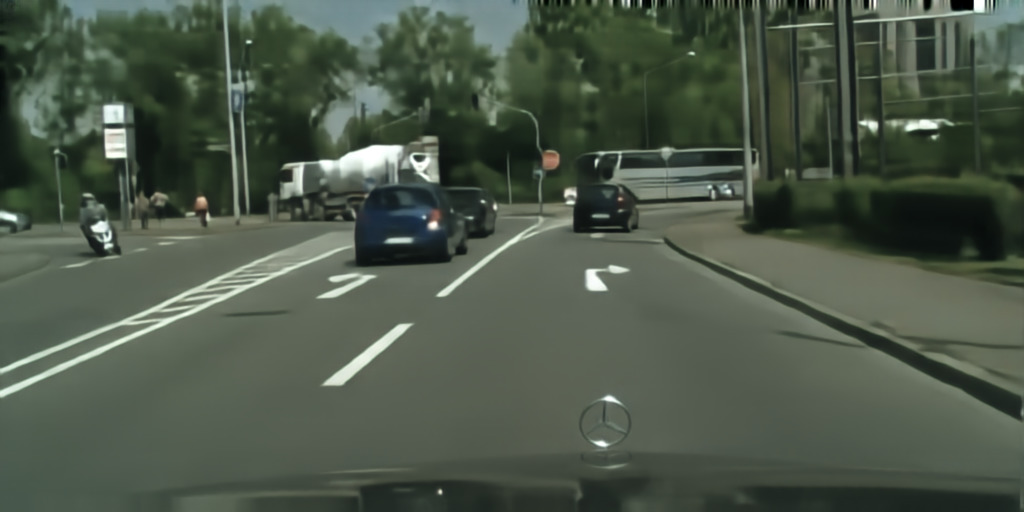}
        \rotatebox{90}{\scriptsize{(d) BPG-SE (0.08 bpp)}} %
    }
    \vspace{-11pt}
    \subfloat{
        \hspace{-.17in}
        \rotatebox{90}{\scriptsize{(e) WebP (0.15 bpp)}}
        \includegraphics[width=1.\columnwidth]{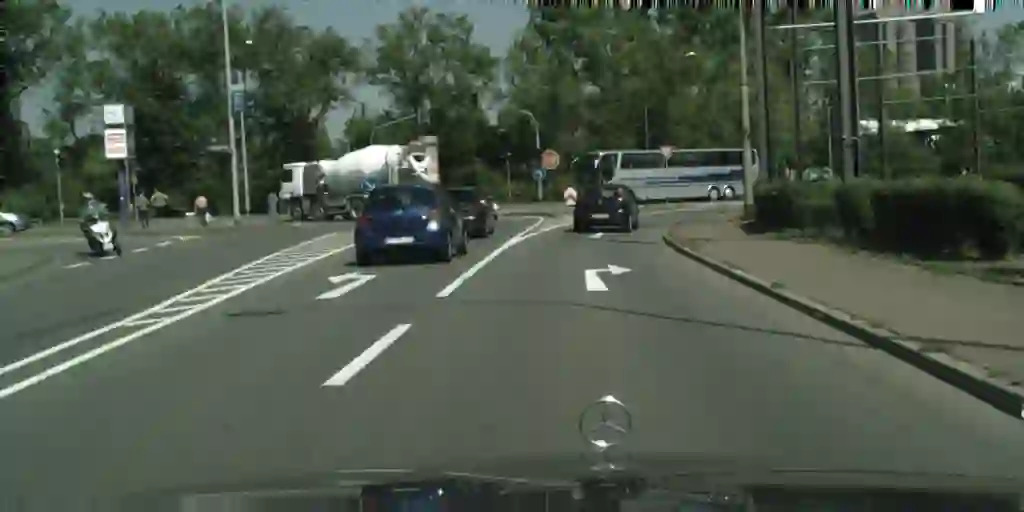}
    }
    \subfloat{
        \hspace{-.095in}
        \includegraphics[width=1.\columnwidth]{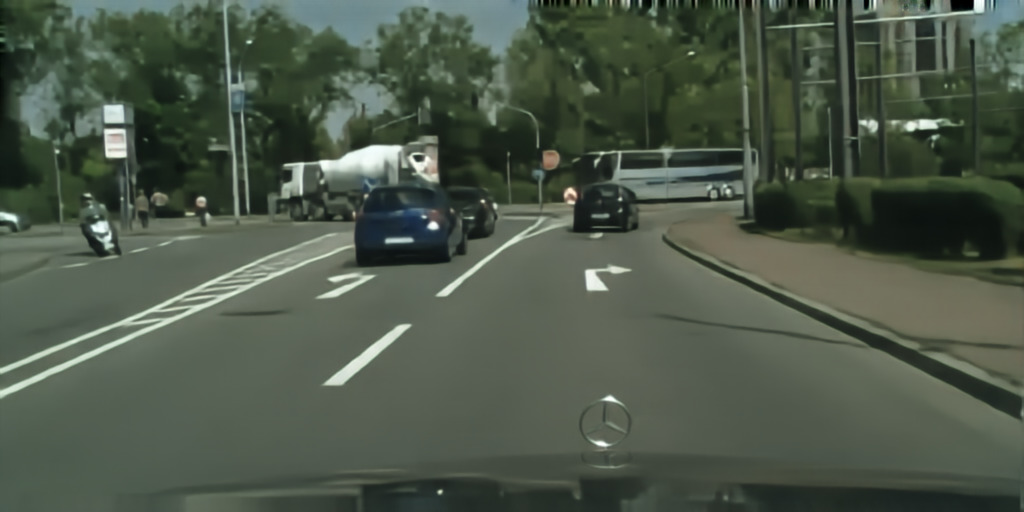}
        \rotatebox{90}{\scriptsize{(f) WebP-SE (0.13 bpp)}}
    }
    \vspace{-11pt}
    \subfloat{
        \hspace{-.17in}
        \rotatebox{90}{\scriptsize{(g) learned (0.21 bpp)}}
        \includegraphics[width=1.\columnwidth]{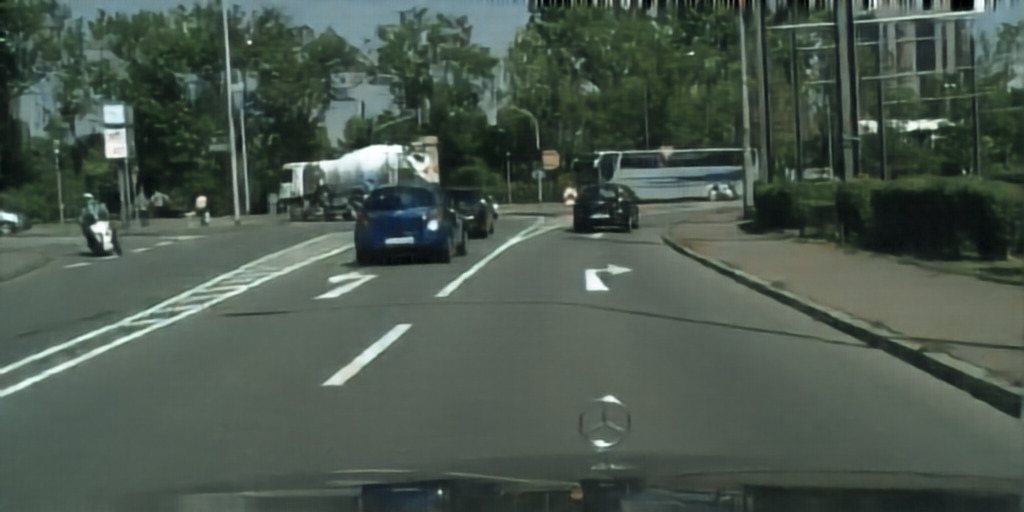}
    }
    \subfloat{
        \hspace{-.095in}
        \includegraphics[width=1.\columnwidth]{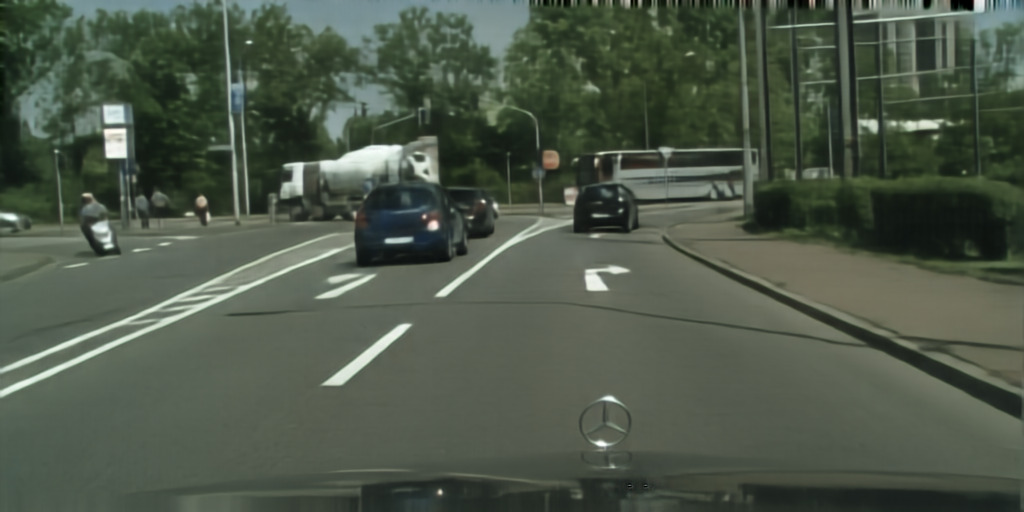}
        \rotatebox{90}{\scriptsize{(h) learned-SE (0.15 bpp)}}
    }
    \vspace{-11pt}
    \subfloat{
        \hspace{-.17in}
        \rotatebox{90}{\scriptsize{(i) JP2 (0.12 bpp)}}
        \includegraphics[width=1.\columnwidth]{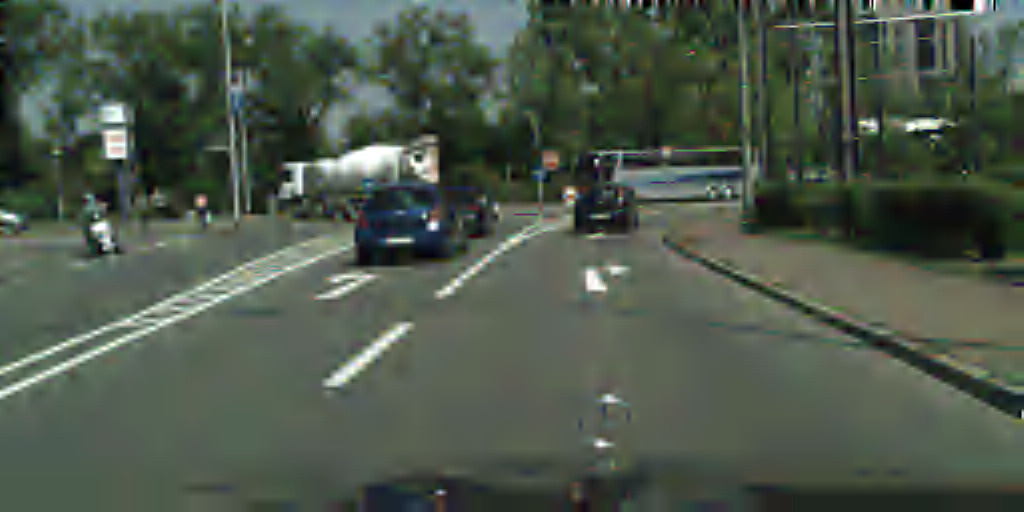}
    }
    \subfloat{
        \hspace{-.095in}
        \includegraphics[width=1.\columnwidth]{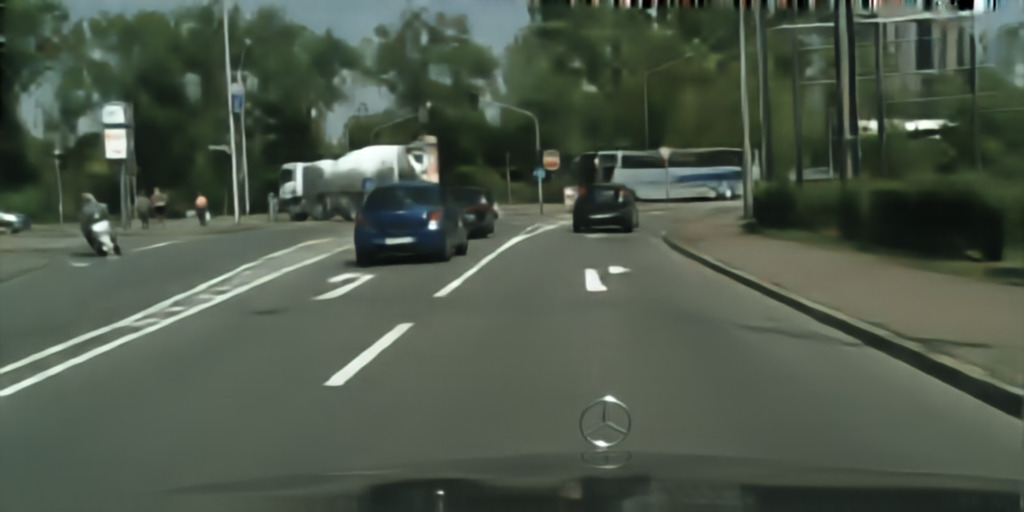}
        \rotatebox{90}{\scriptsize{(j) JP2-SE (0.12 bpp)}}
    }
    \vspace{-11pt}
    \subfloat{
        \hspace{-.17in}
        \rotatebox{90}{\scriptsize{(k) Original}}
        \includegraphics[width=1.\columnwidth]{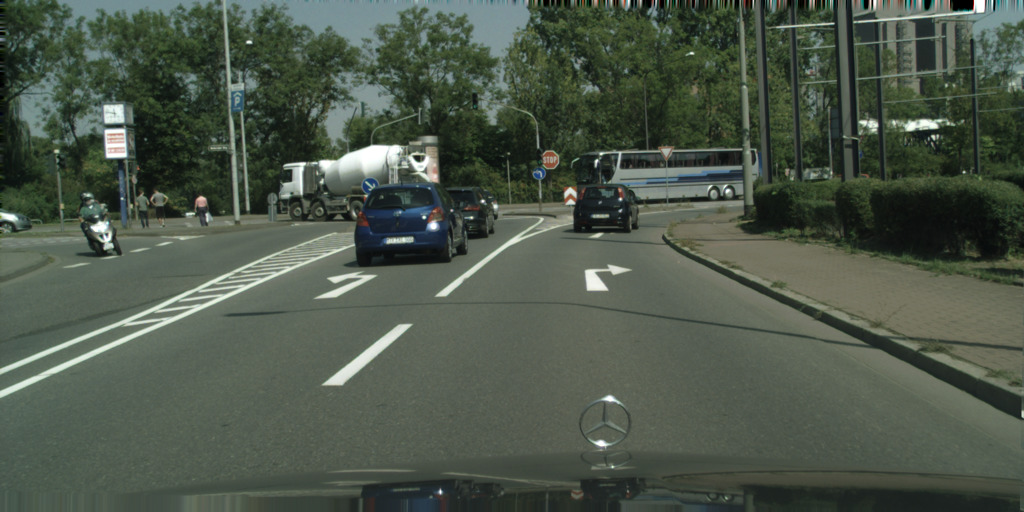}
    }
    \subfloat{
        \hspace{-.095in}
        \includegraphics[width=1.\columnwidth]{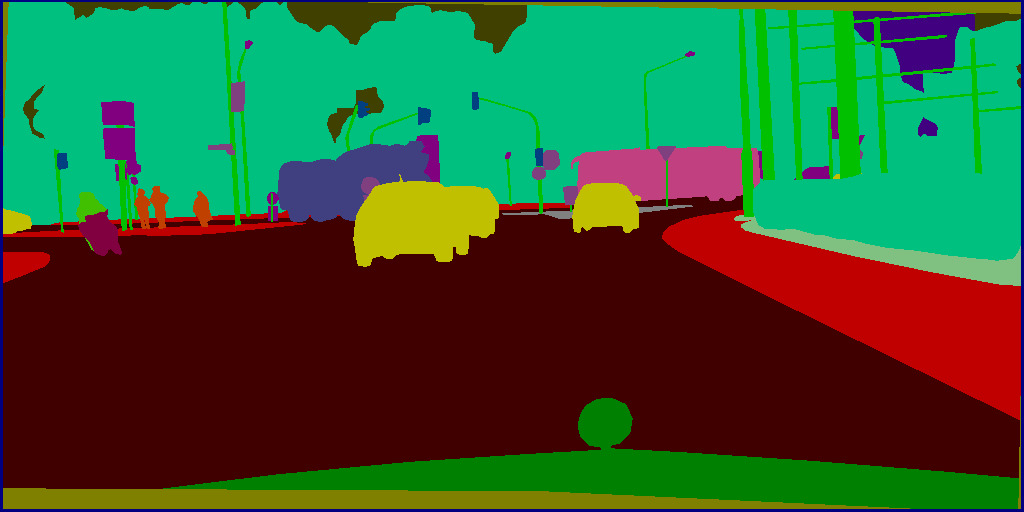}
        \rotatebox{90}{\scriptsize{(l) Semantics (class seg. map)}}
    }
    \caption{
        More visualizations for Cityscapes.
        These results mainly illustrate codec performance in low bitrate.
    }
    \label{afig6}
\end{figure*}

\begin{figure*}[h]
    \thisfloatpagestyle{empty}
    \vskip -.7in
    \centering
    \subfloat{
        \hspace{-.17in}
        \rotatebox{90}{\scriptsize{(a) JPEG (0.22 bpp)}}
        \includegraphics[width=1.\columnwidth]{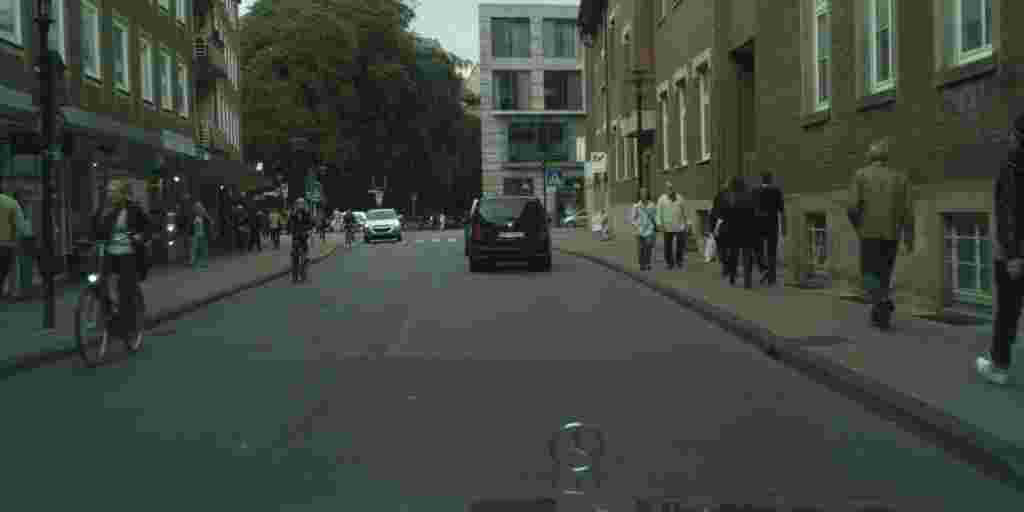}
    }
    \subfloat{
        \hspace{-.095in}
        \includegraphics[width=1.\columnwidth]{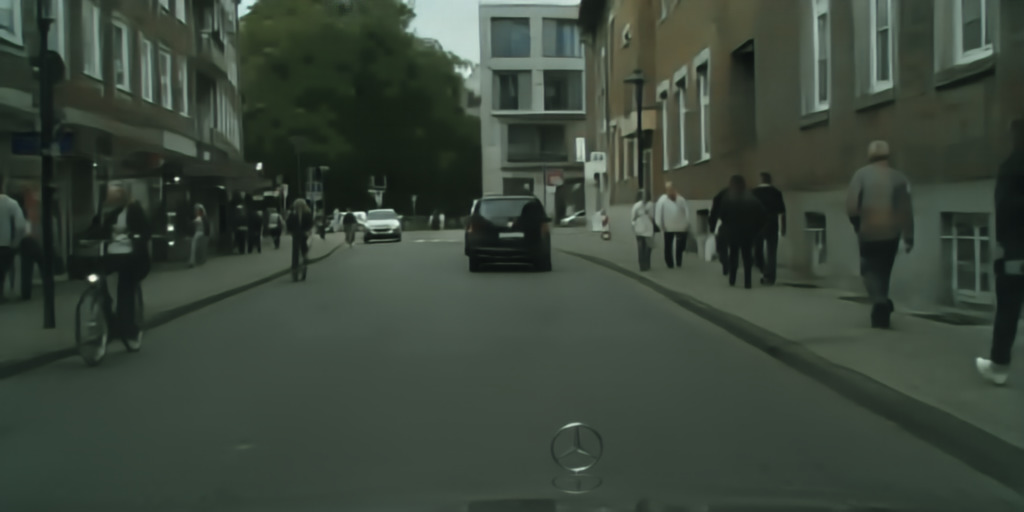}
        \rotatebox{90}{\scriptsize{(b) JPEG-SE (0.21 bpp)}}
    }
    \vspace{-11pt}
    \subfloat{
        \hspace{-.17in}
        \rotatebox{90}{\scriptsize{(c) BPG (0.06 bpp)}}
        \includegraphics[width=1.\columnwidth]{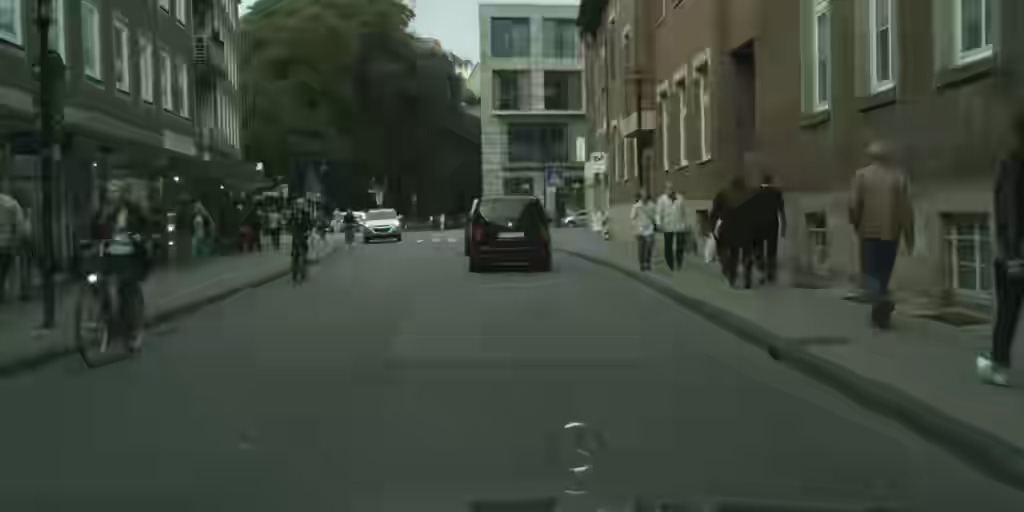}
    }
    \subfloat{
        \hspace{-.095in}
        \includegraphics[width=1.\columnwidth]{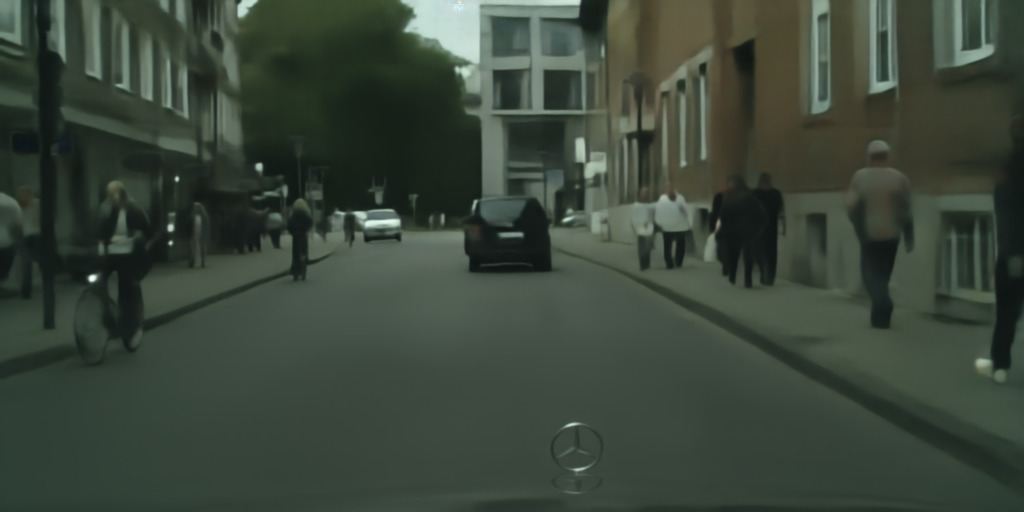}
        \rotatebox{90}{\scriptsize{(d) BPG-SE (0.06 bpp)}}
    }
    \vspace{-11pt}
    \subfloat{
        \hspace{-.17in}
        \rotatebox{90}{\scriptsize{(e) WebP (0.10 bpp)}}
        \includegraphics[width=1.\columnwidth]{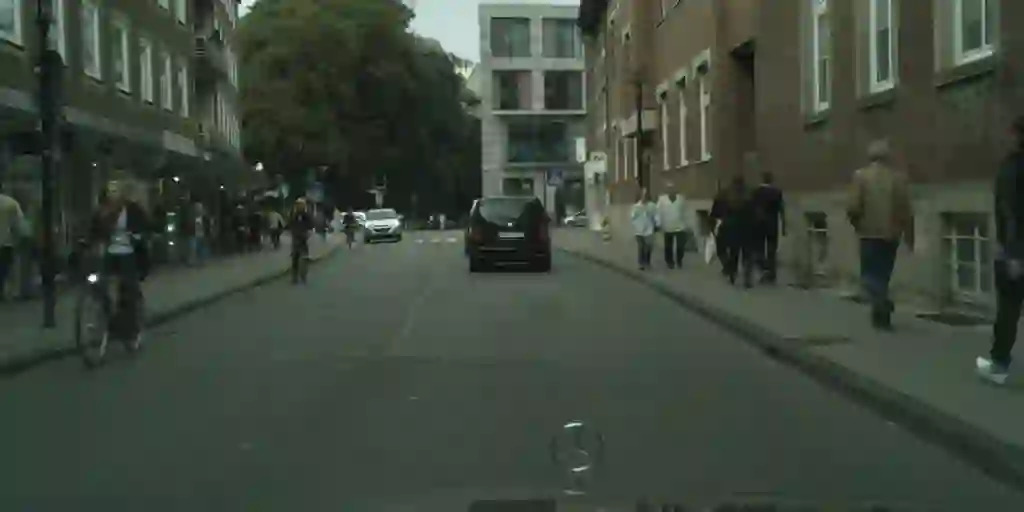}
    }
    \subfloat{
        \hspace{-.095in}
        \includegraphics[width=1.\columnwidth]{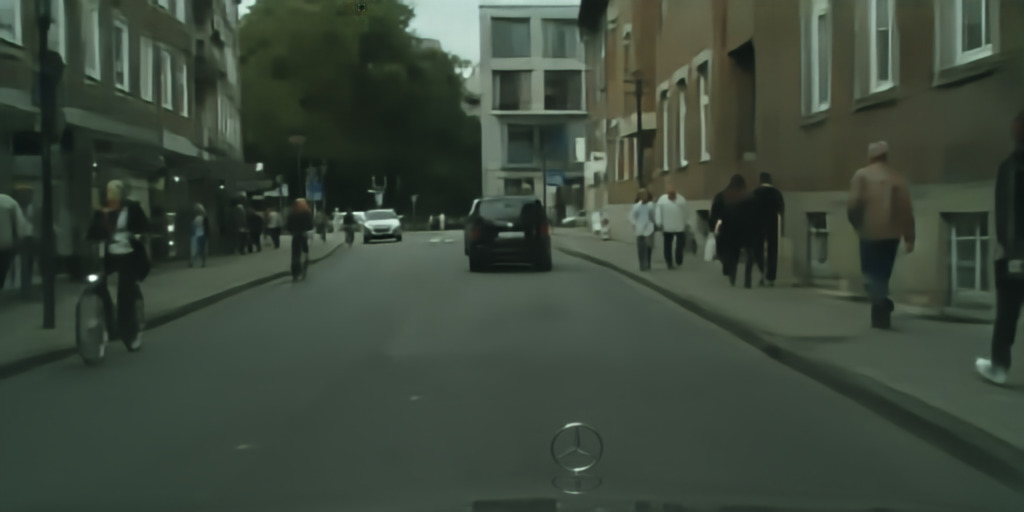}
        \rotatebox{90}{\scriptsize{(f) WebP-SE (0.10 bpp)}}
    }
    \vspace{-11pt}
    \subfloat{
        \hspace{-.17in}
        \rotatebox{90}{\scriptsize{(g) learned (0.21 bpp)}}
        \includegraphics[width=1.\columnwidth]{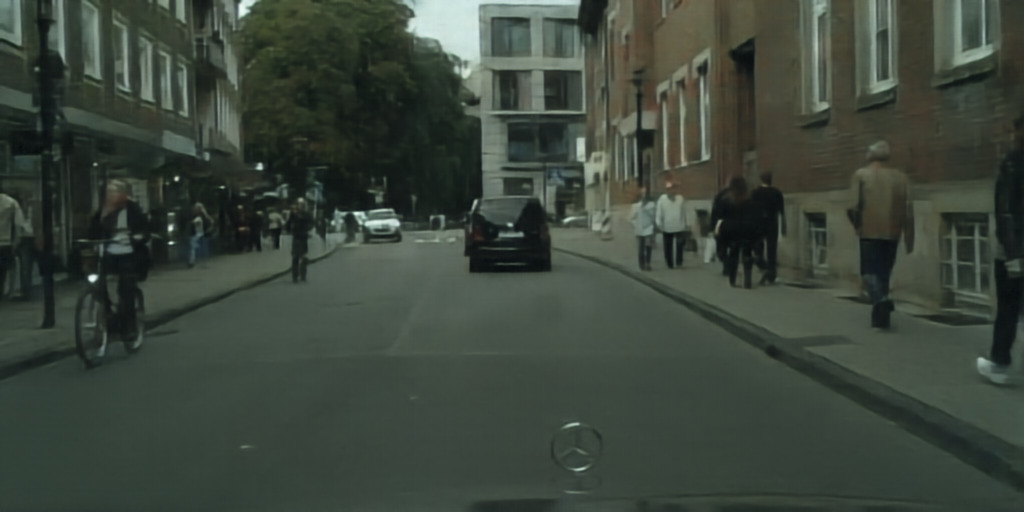}
    }
    \subfloat{
        \hspace{-.095in}
        \includegraphics[width=1.\columnwidth]{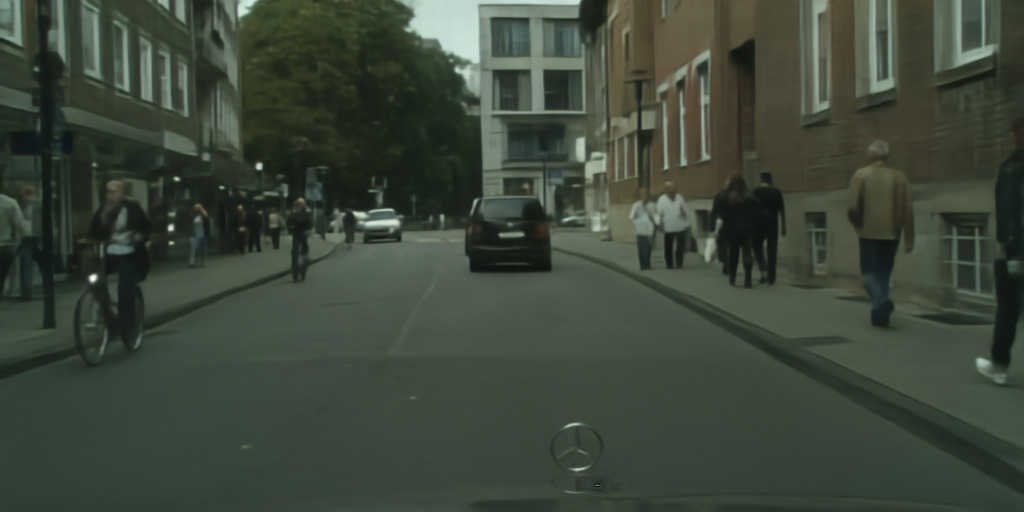}
        \rotatebox{90}{\scriptsize{(h) learned-SE (0.15 bpp)}}
    }
    \vspace{-11pt}
    \subfloat{
        \hspace{-.17in}
        \rotatebox{90}{\scriptsize{(i) JP2 (0.12 bpp)}}
        \includegraphics[width=1.\columnwidth]{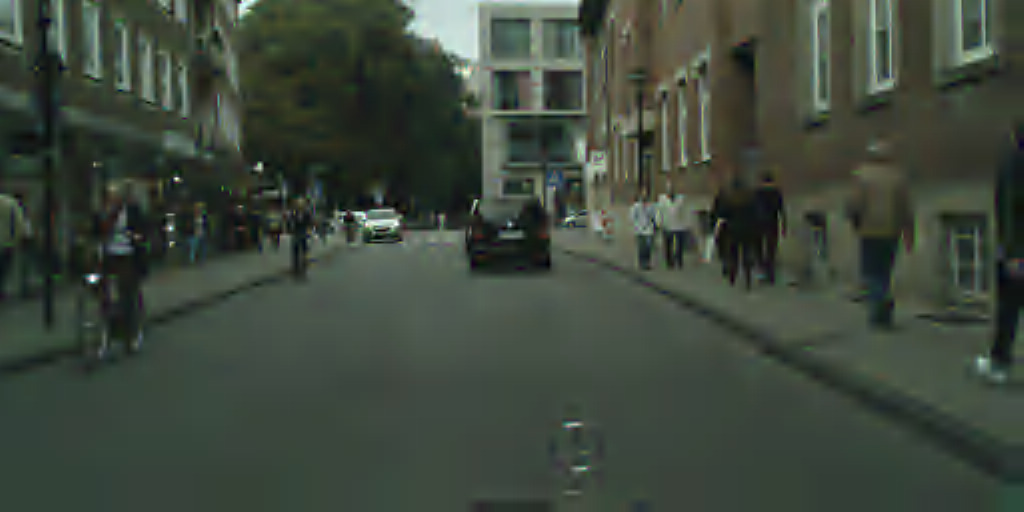}
    }
    \subfloat{
        \hspace{-.095in}
        \includegraphics[width=1.\columnwidth]{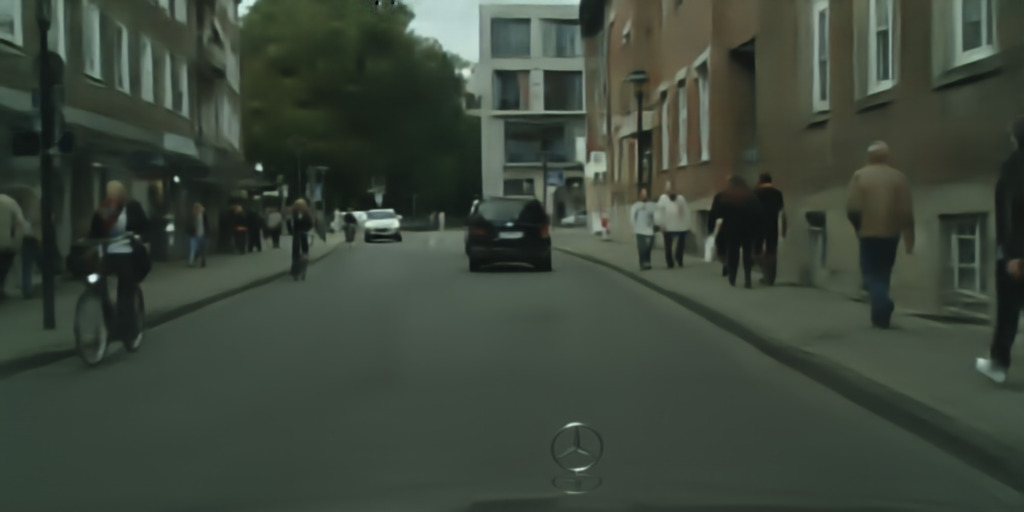}
        \rotatebox{90}{\scriptsize{(j) JP2-SE (0.12 bpp)}}
    }
    \vspace{-11pt}
    \subfloat{
        \hspace{-.17in}
        \rotatebox{90}{\scriptsize{(k) Original}}
        \includegraphics[width=1.\columnwidth]{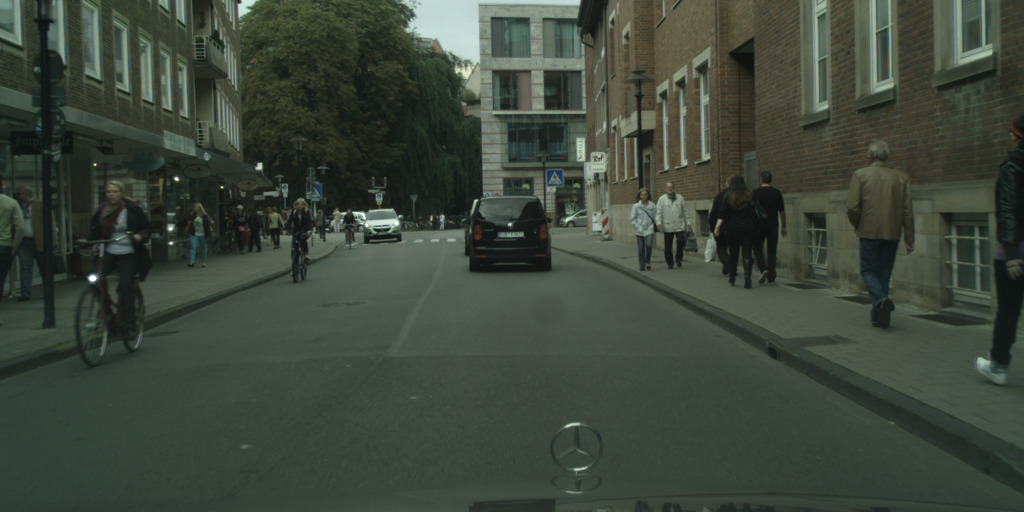}
    }
    \subfloat{
        \hspace{-.095in}
        \includegraphics[width=1.\columnwidth]{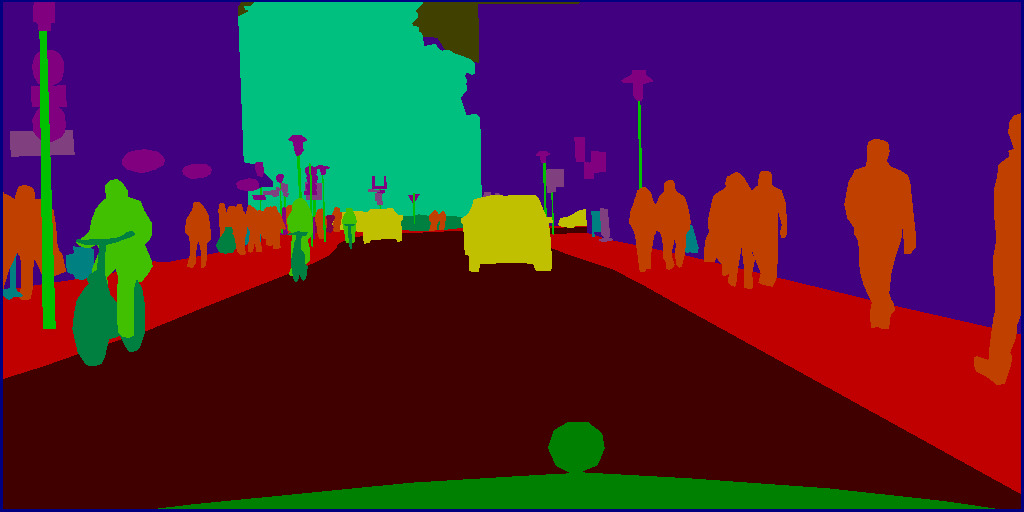}
        \rotatebox{90}{\scriptsize{(l) Semantics (class seg. map)}}
    }
    \caption{
        More visualizations for Cityscapes.
        These results mainly illustrate codec performance in low bitrate.
    }
    \label{afig12}
\end{figure*}

\begin{figure*}[h]
    \thisfloatpagestyle{empty}
    \vskip -.8in
    \centering
    \subfloat{
        \hspace{-.17in}
        \rotatebox{90}{\scriptsize{(a) JPEG (0.35 bpp)}}
        \includegraphics[width=1.\columnwidth]{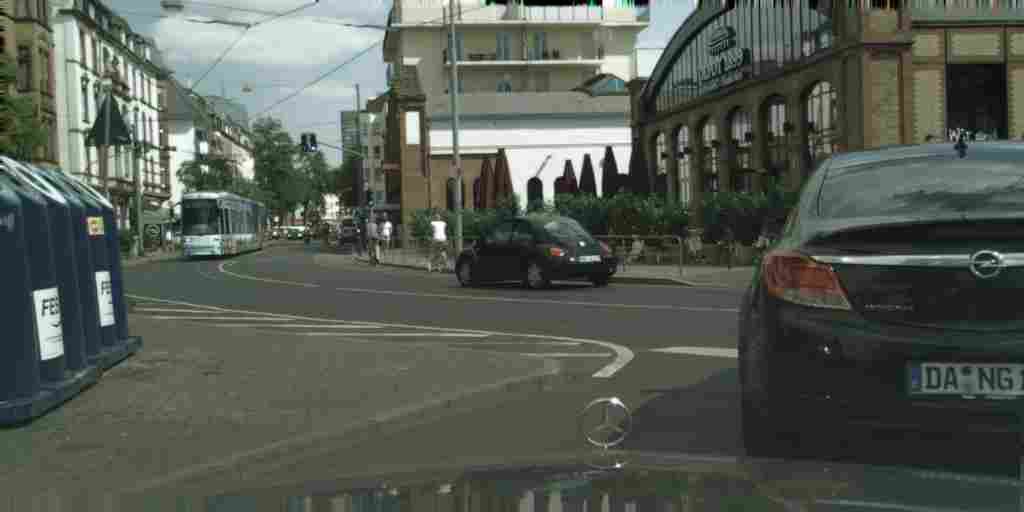}
    }
    \subfloat{
        \hspace{-.095in}
        \includegraphics[width=1.\columnwidth]{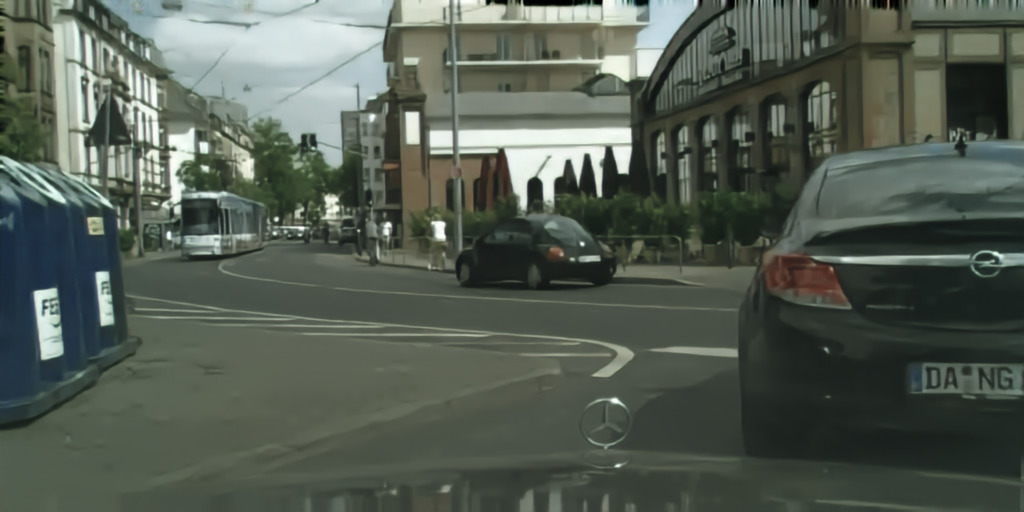}
        \rotatebox{90}{\scriptsize{(b) JPEG-SE (0.32 bpp)}}
    }
    \vspace{-11pt}
    \subfloat{
        \hspace{-.17in}
        \rotatebox{90}{\scriptsize{(c) BPG (0.11 bpp)}}
        \includegraphics[width=1.\columnwidth]{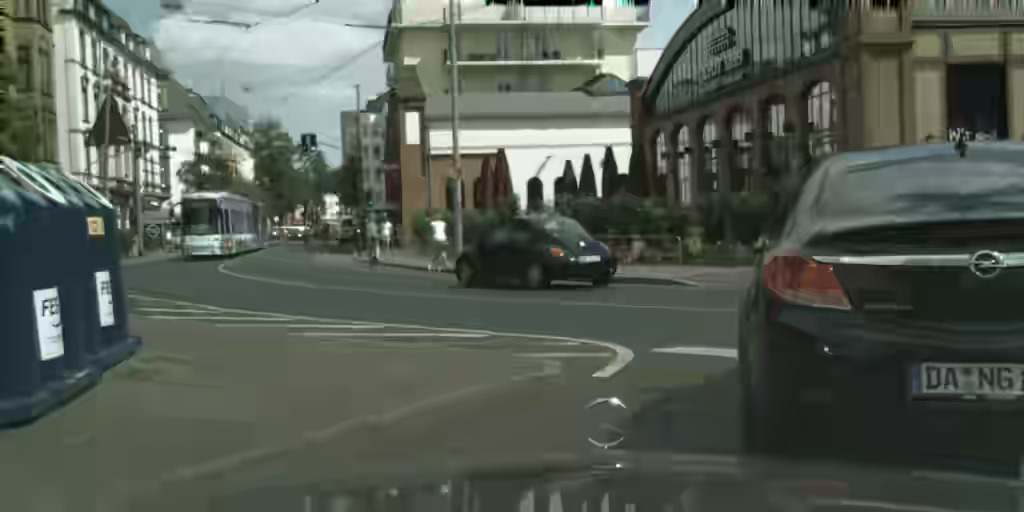}
    }
    \subfloat{
        \hspace{-.095in}
        \includegraphics[width=1.\columnwidth]{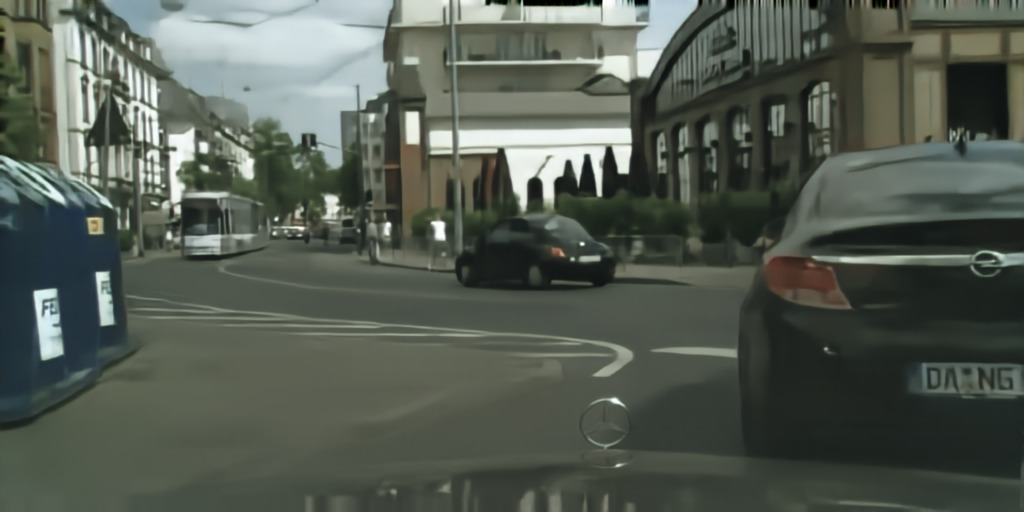}
        \rotatebox{90}{\scriptsize{(d) BPG-SE (0.11 bpp)}}
    }
    \vspace{-11pt}
    \subfloat{
        \hspace{-.17in}
        \rotatebox{90}{\scriptsize{(e) WebP (0.20 bpp)}} 
        \includegraphics[width=1.\columnwidth]{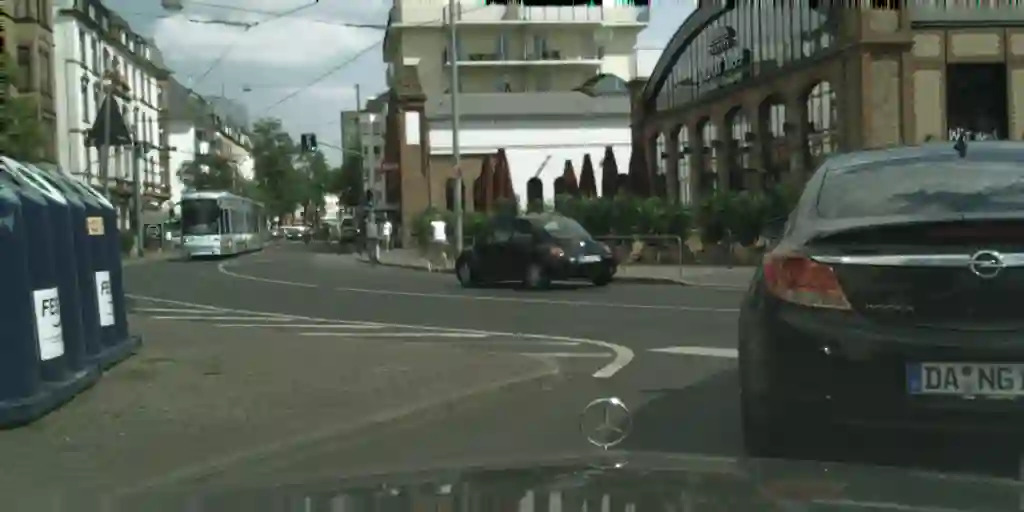}
    }
    \subfloat{
        \hspace{-.095in}
        \includegraphics[width=1.\columnwidth]{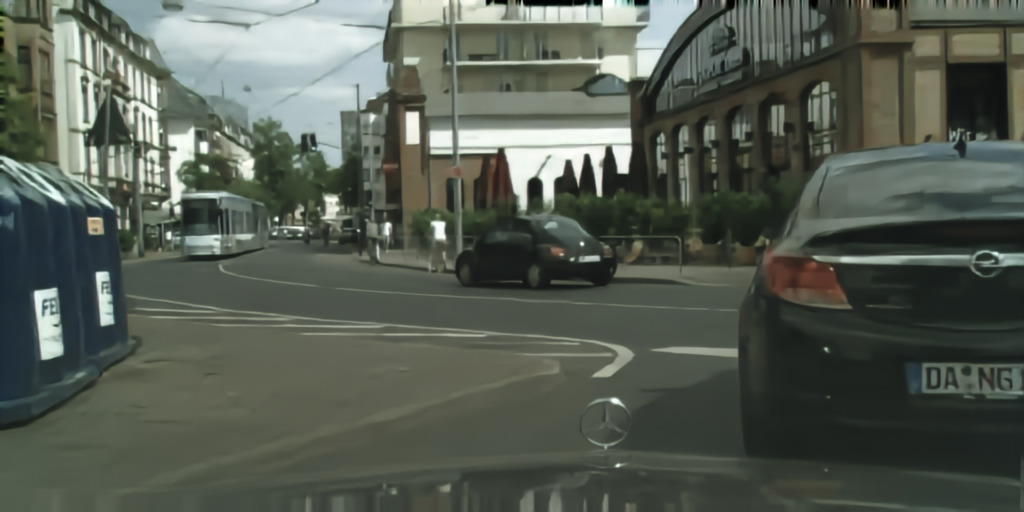}
        \rotatebox{90}{\scriptsize{(f) WebP-SE (0.19 bpp)}}
    }
    \vspace{-11pt}
    \subfloat{
        \hspace{-.17in}
        \rotatebox{90}{\scriptsize{(g) learned (0.21 bpp)}}
        \includegraphics[width=1.\columnwidth]{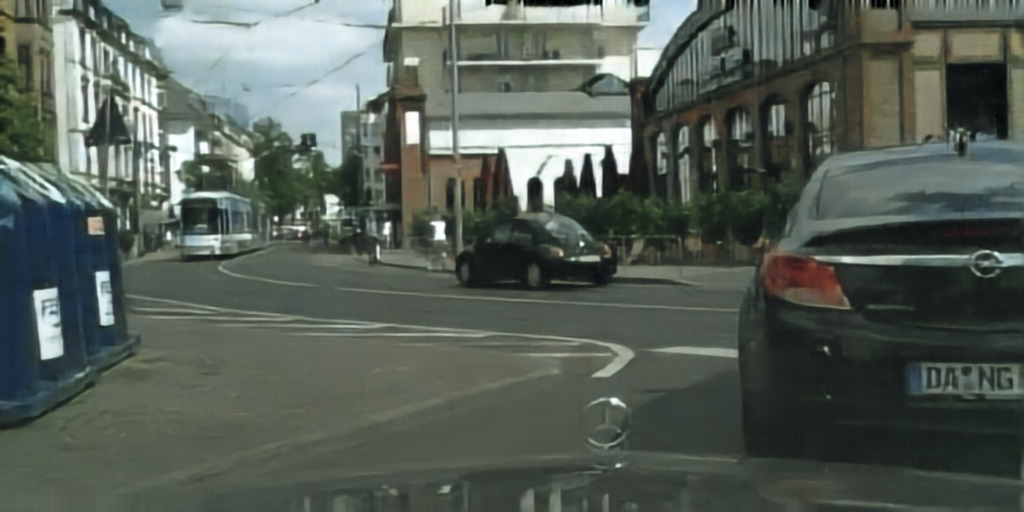}
    }
    \subfloat{
        \hspace{-.095in}
        \includegraphics[width=1.\columnwidth]{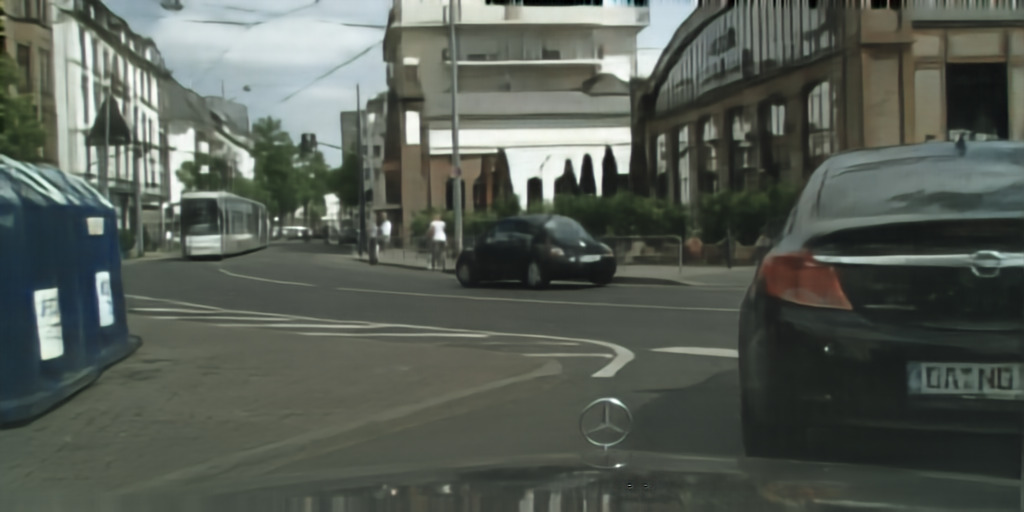}
        \rotatebox{90}{\scriptsize{(h) learned-SE (0.15 bpp)}}
    }
    \vspace{-11pt}
    \subfloat{
        \hspace{-.17in}
        \rotatebox{90}{\scriptsize{(i) JP2 (0.24 bpp)}}
        \includegraphics[width=1.\columnwidth]{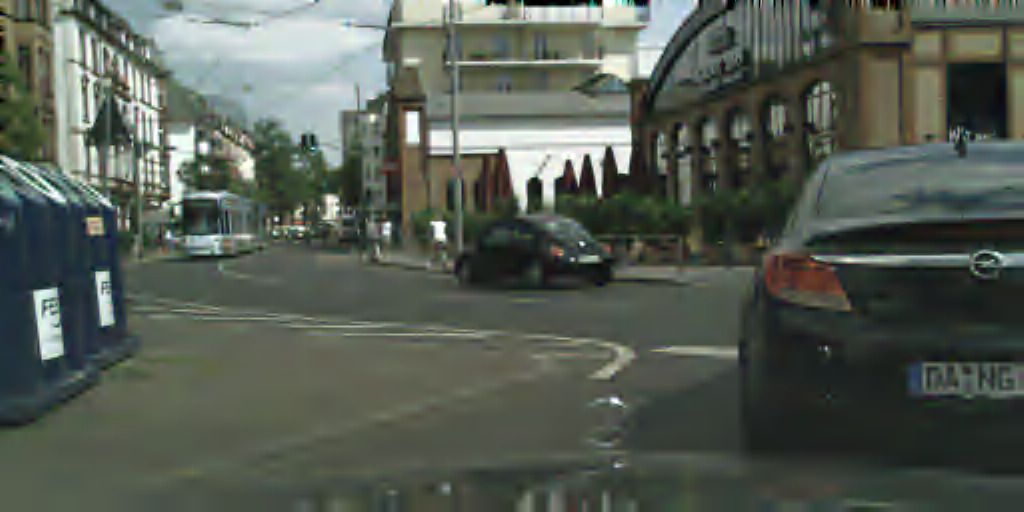}
    }
    \subfloat{
        \hspace{-.095in}
        \includegraphics[width=1.\columnwidth]{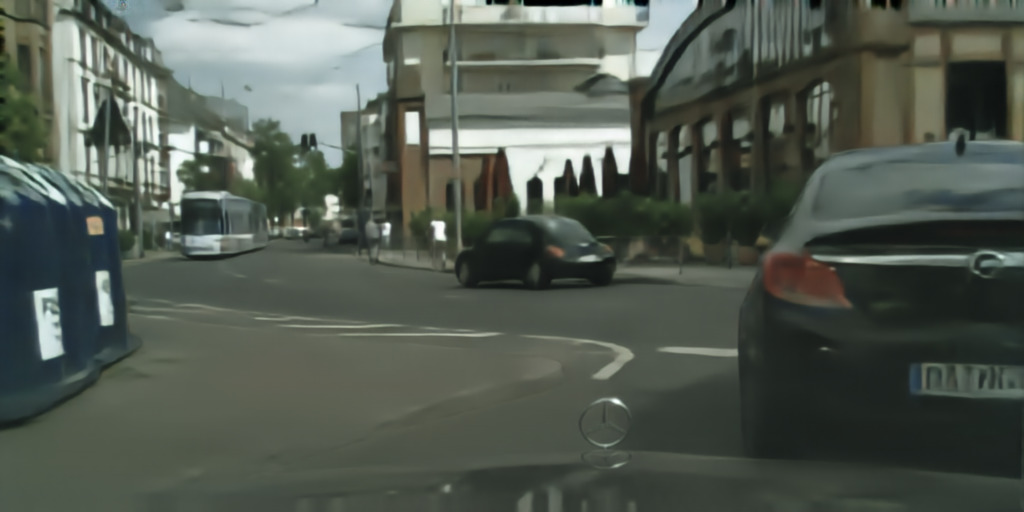}
        \rotatebox{90}{\scriptsize{(j) JP2-SE (0.16 bpp)}}
    }
    \vspace{-11pt}
    \subfloat{
        \hspace{-.17in}
        \rotatebox{90}{\scriptsize{(k) Original}}
        \includegraphics[width=1.\columnwidth]{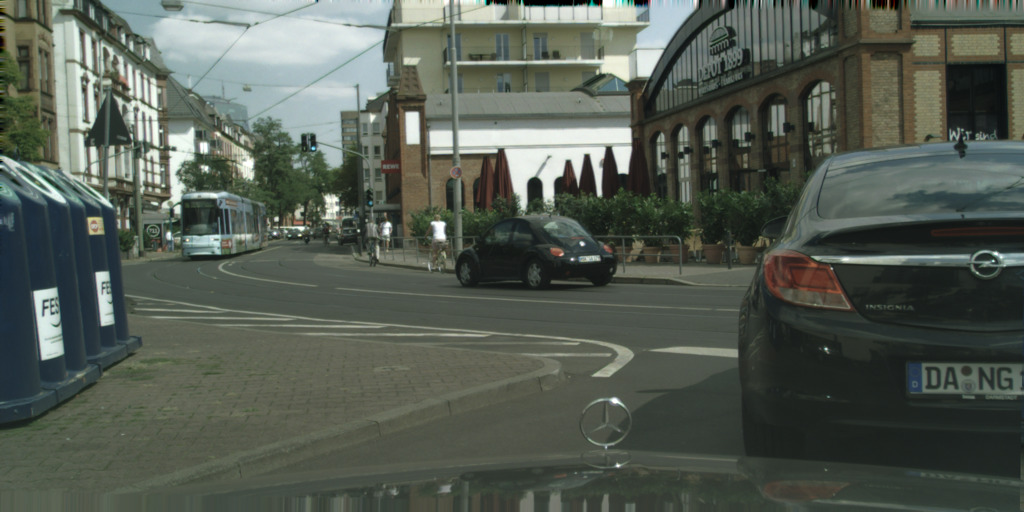}
    }
    \subfloat{
        \hspace{-.095in}
        \includegraphics[width=1.\columnwidth]{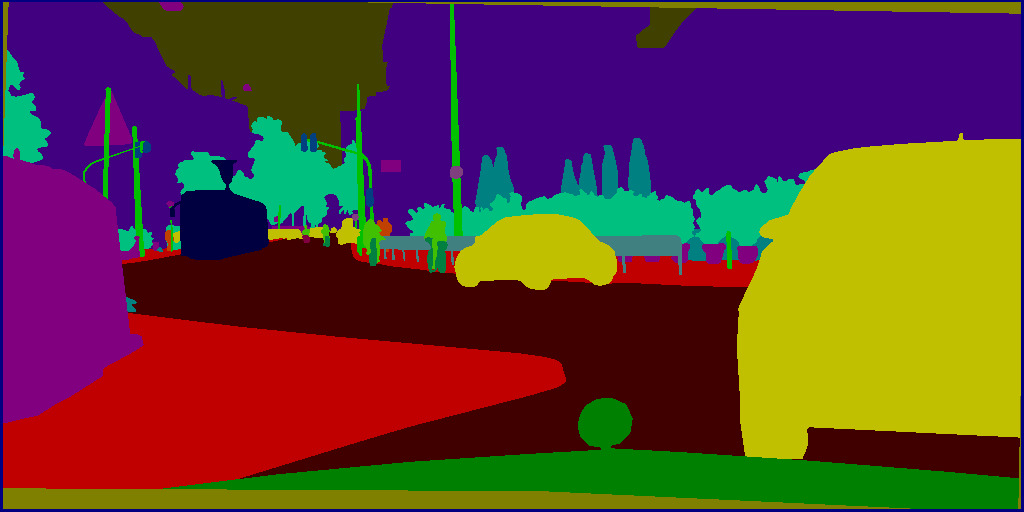}
        \rotatebox{90}{\scriptsize{(l) Semantics (class seg. map)}}
    }
    \caption{
        More visualizations for Cityscapes.
        In contrast to Fig.~\ref{afig3}, \ref{afig6}, and \ref{afig12}, here we mainly demonstrate codec performance in medium bitrate, in which the perceptual quality difference is still visible.
    }
    \label{afig9}
\end{figure*}

See Fig.~\ref{afig6}, \ref{afig12}, and~\ref{afig9} for more visualizations on Cityscapes.
In Fig.~\ref{afig6} and \ref{afig12}, the codecs operate in their respective low bitrate settings whereas for Fig.~\ref{afig9}, we demonstrate codecs operating under relatively higher bitrates.
In these images, we see the same trend we saw in images from the main text.
Specifically, the SE codecs produce results with similar distortion but better perception quality, which can be clearly seen especially in semantic boundaries: No color bleeding is observed from the SE codecs.
In contrast, this unnatural and visually unappealing phenomenon is pervasive in the originals when working in the illustrated bitrate settings.
This is perhaps because in these low to medium bitrates, the compressed hidden code cannot perfectly preserve all low-level features.
In this case, the SE codecs can use their knowledge on what low-level visual features are more likely to appear in each semantic region to make an educated guess on the features given semantics.
In contrast, the originals do not have this extra degree of freedom and can only use information in the hidden code, which is highly incomplete and therefore leads to visually unappealing reconstructions.


\section{Visualizations for ADE20k}
\begin{figure*}[h]
    \thisfloatpagestyle{empty}
    \vskip -.8in
    \centering
    \subfloat{
        \hspace{-.17in}
        \rotatebox{90}{\scriptsize{(a) JPEG (0.31 bpp)}}
        \includegraphics[width=1\columnwidth]{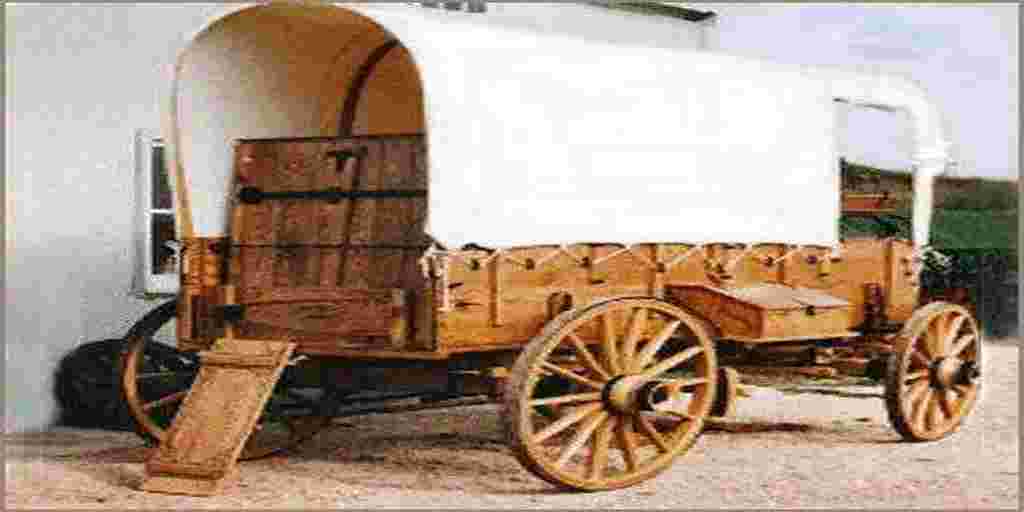}
    }
    \subfloat{
        \hspace{-.095in}
        \includegraphics[width=1\columnwidth]{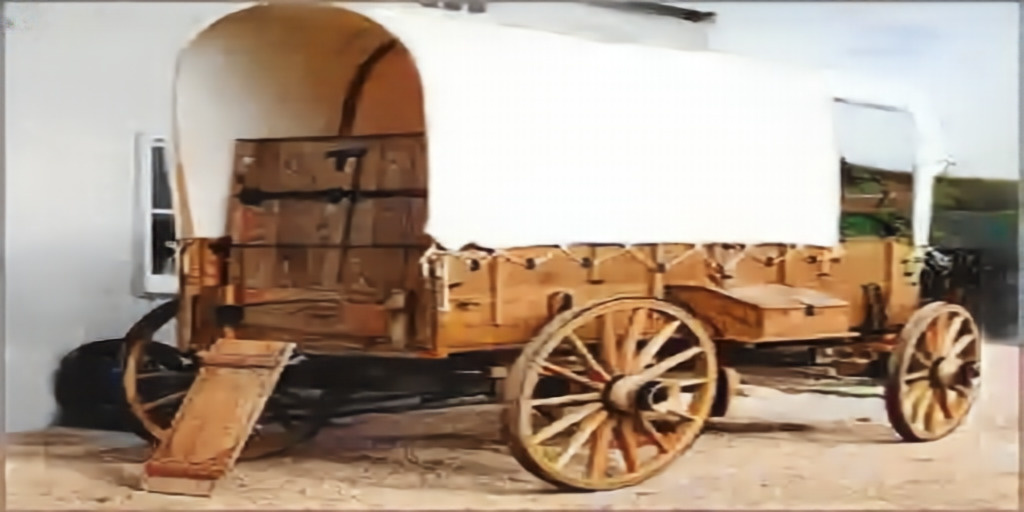}
        \rotatebox{90}{\scriptsize{(b) JPEG-SE (0.24 bpp)}}
    }
    \vspace{-11pt}
    \subfloat{
        \hspace{-.17in}
        \rotatebox{90}{\scriptsize{(c) BPG (0.09 bpp)}}
        \includegraphics[width=1\columnwidth]{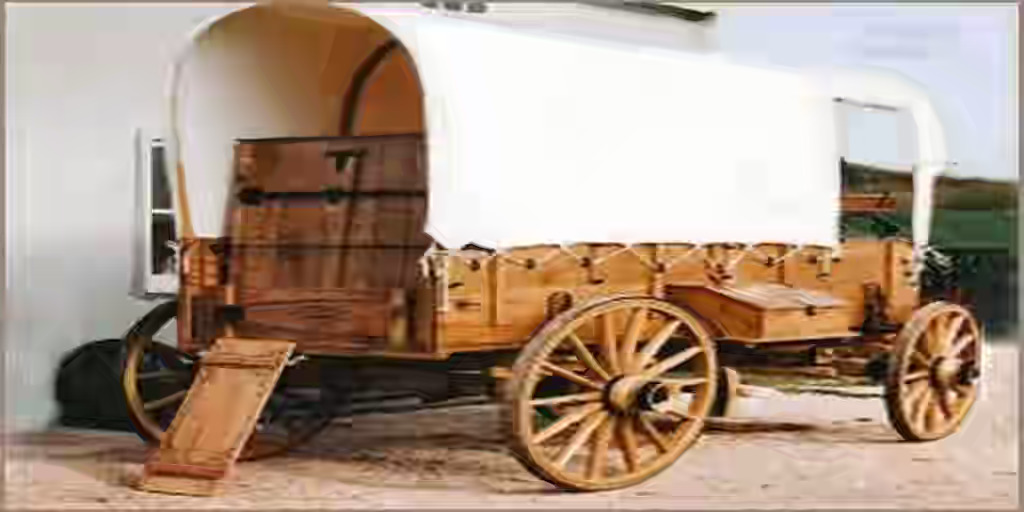}
    }
    \subfloat{
        \hspace{-.095in}
        \includegraphics[width=1\columnwidth]{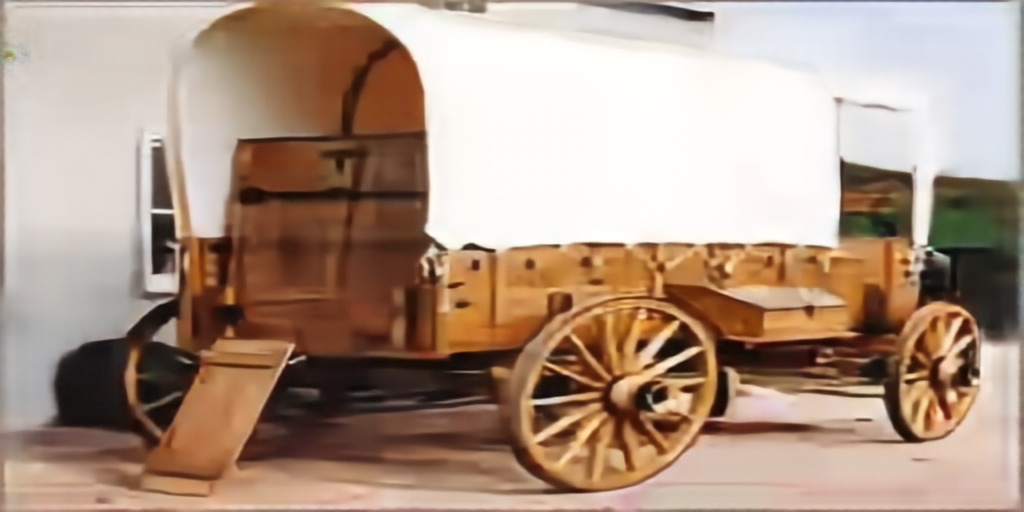}
        \rotatebox{90}{\scriptsize{(d) BPG-SE (0.06 bpp)}}
    }
    \vspace{-11pt}
    \subfloat{
        \hspace{-.17in}
        \rotatebox{90}{\scriptsize{(e) WebP (0.17 bpp)}}
        \includegraphics[width=1\columnwidth]{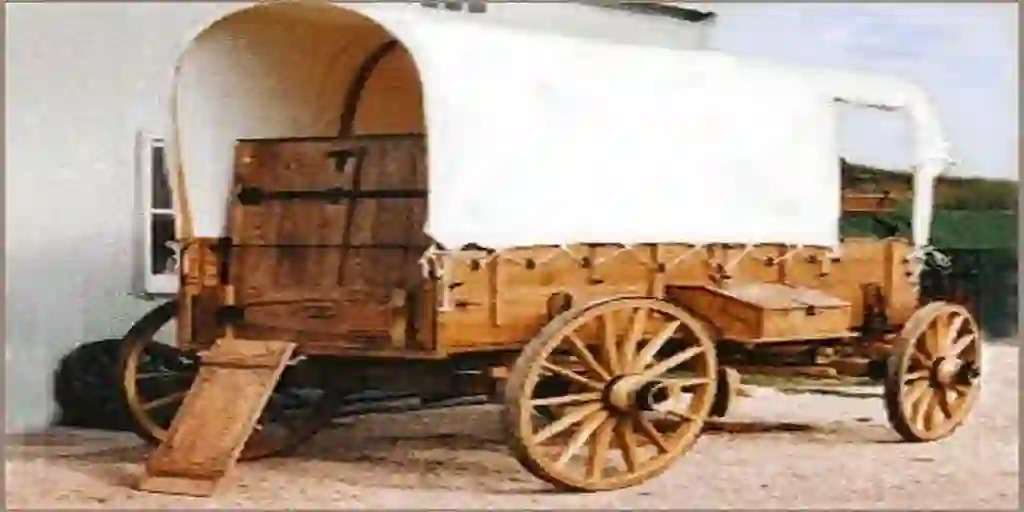}
    }
    \subfloat{
        \hspace{-.095in}
        \includegraphics[width=1\columnwidth]{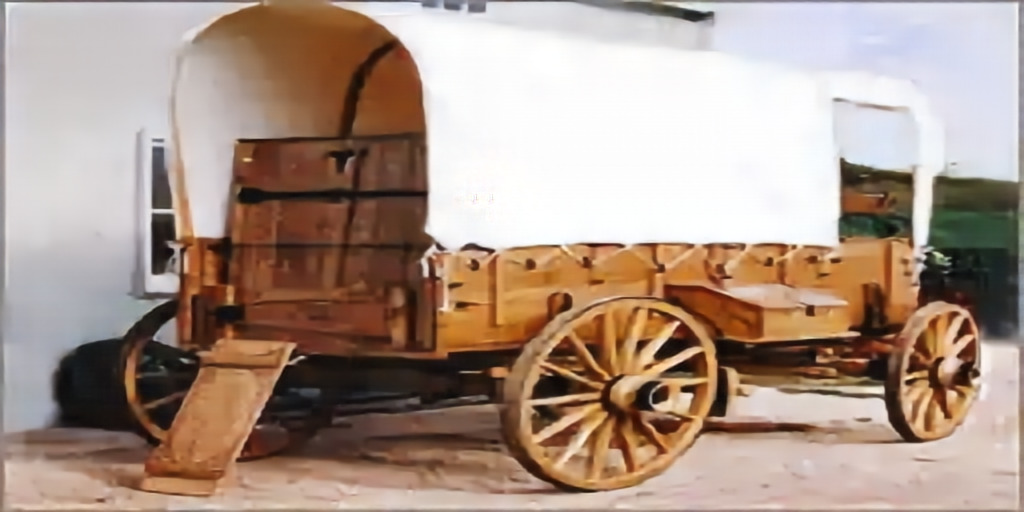}
        \rotatebox{90}{\scriptsize{(f) WebP-SE (0.13 bpp)}}
    }
    \vspace{-11pt}
    \subfloat{
        \hspace{-.17in}
        \rotatebox{90}{\scriptsize{(g) learned (0.09 bpp)}}
        \includegraphics[width=1\columnwidth]{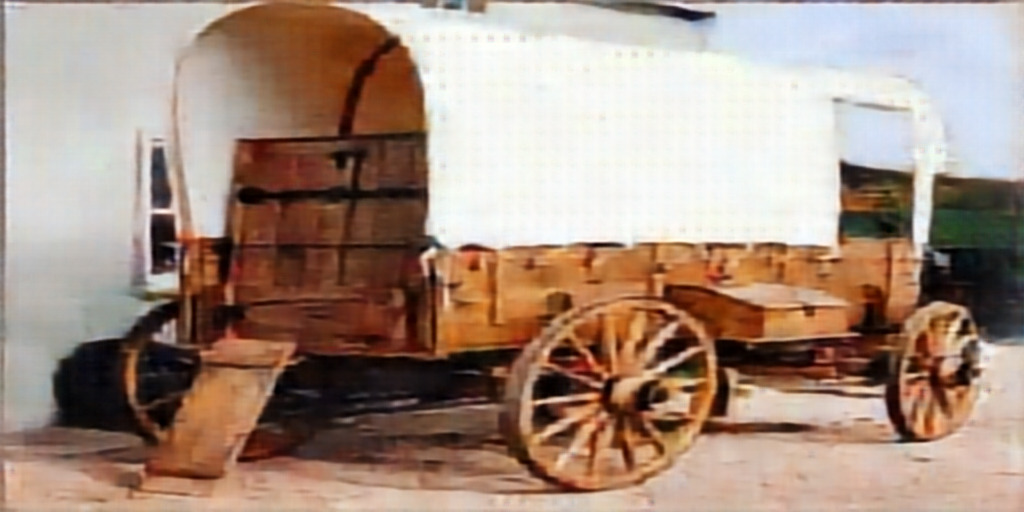}
    }
    \subfloat{
        \hspace{-.095in}
        \includegraphics[width=1\columnwidth]{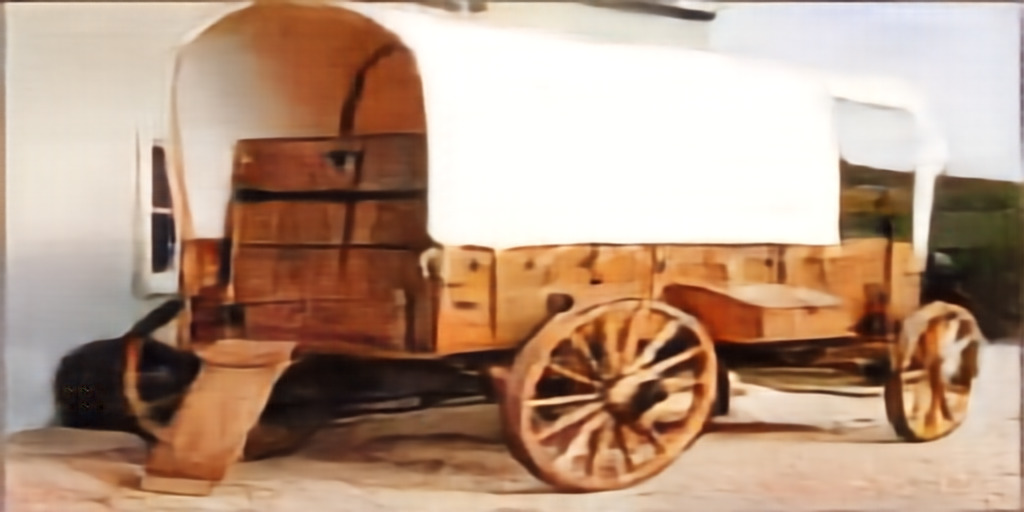}
        \rotatebox{90}{\scriptsize{(h) learned-SE (0.07 bpp)}}
    }
    \vspace{-11pt}
    \subfloat{
        \hspace{-.17in}
        \rotatebox{90}{\scriptsize{(i) JP2 (0.06 bpp)}}
        \includegraphics[width=1\columnwidth]{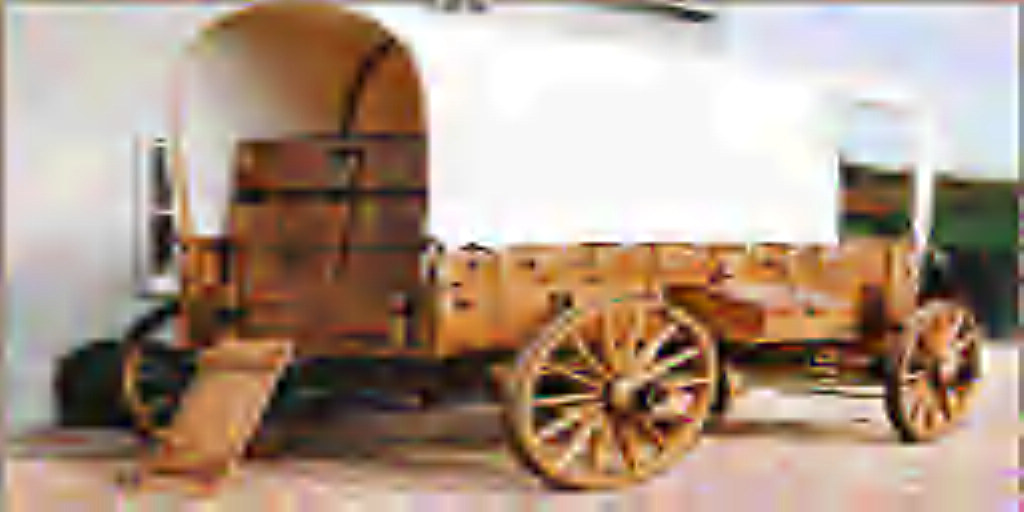}
    }
    \subfloat{
        \hspace{-.095in}
        \includegraphics[width=1\columnwidth]{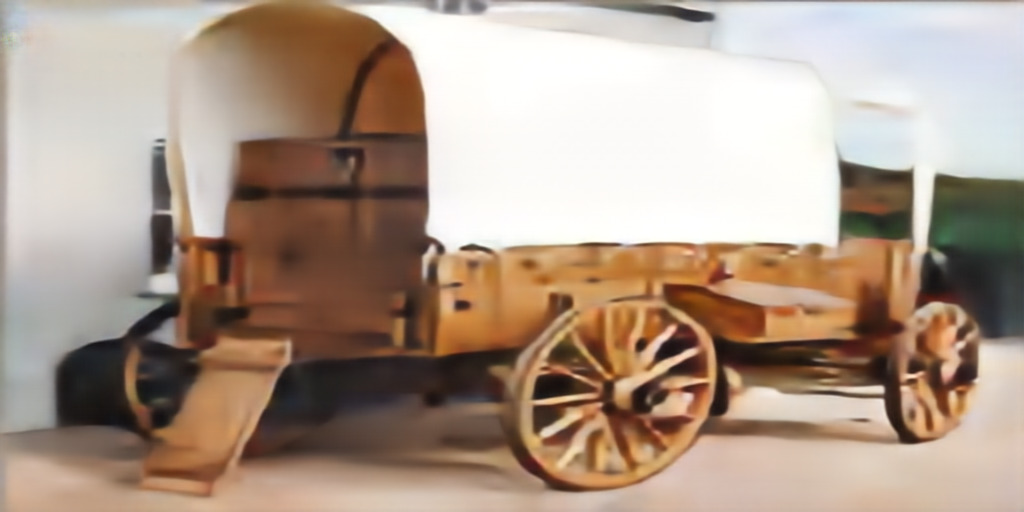}
        \rotatebox{90}{\scriptsize{(j) JP2-SE (0.07 bpp)}}
    }
    \vspace{-11pt}
    \subfloat{
        \hspace{-.17in}
        \rotatebox{90}{\scriptsize{(k) Original}}
        \includegraphics[width=1\columnwidth]{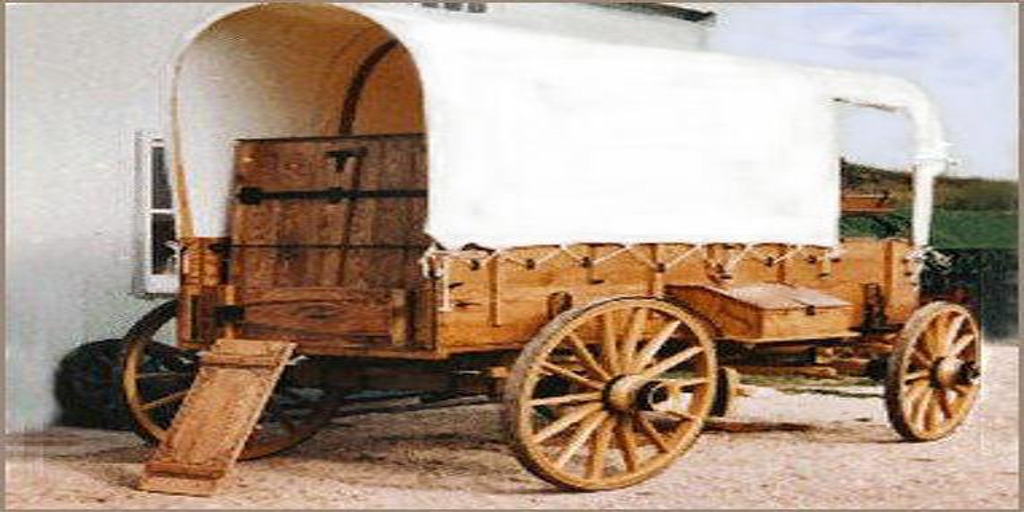}
    }
    \subfloat{
        \hspace{-.095in}
        \includegraphics[width=1\columnwidth]{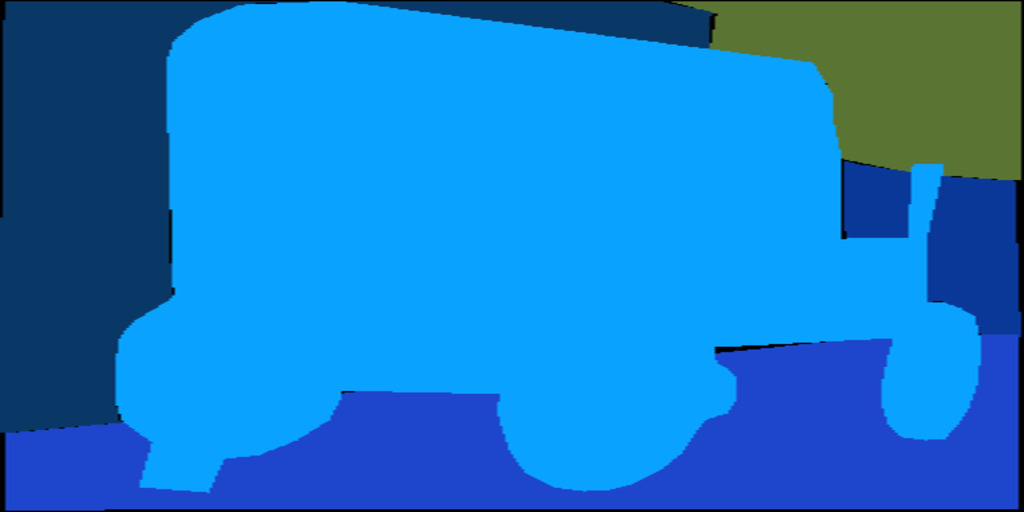}
        \rotatebox{90}{\scriptsize{(l) Semantics (class seg. map)}}
    }
    \caption{
        Visualizations for ADE20k.
        Note how the original image is clearly noisy but all the SE codecs filtered out the noise whereas some originals tried to reconstruct noise as well when given enough bpp allowance (e.g., WebP and JPEG).
    }
    \label{afig13}
\end{figure*}

\begin{figure*}[h]
    \thisfloatpagestyle{empty}
    \vskip -.7in
    \centering
    \subfloat{
        \hspace{-.17in}
        \rotatebox{90}{\scriptsize{(a) JPEG (0.30 bpp)}}
        \includegraphics[width=1\columnwidth]{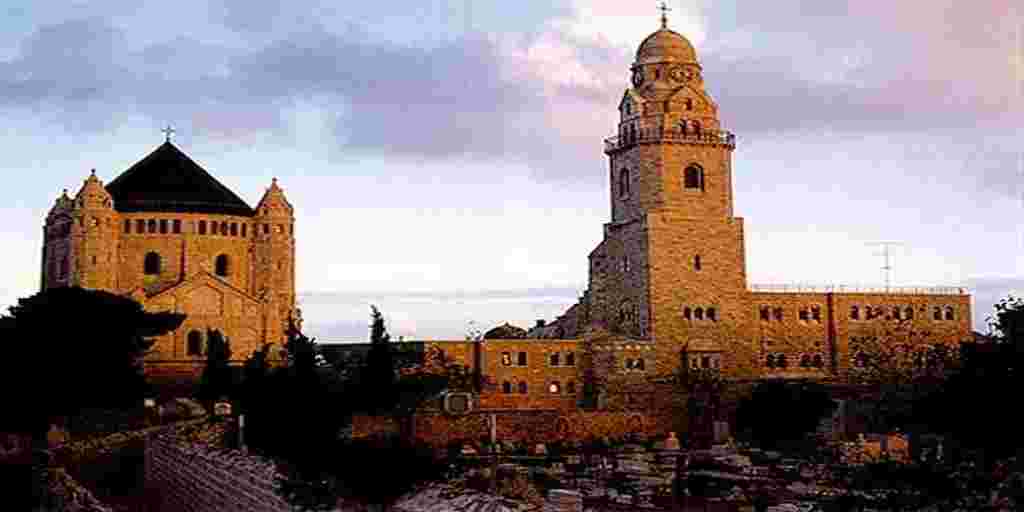}
    }
    \subfloat{
        \hspace{-.095in}
        \includegraphics[width=1\columnwidth]{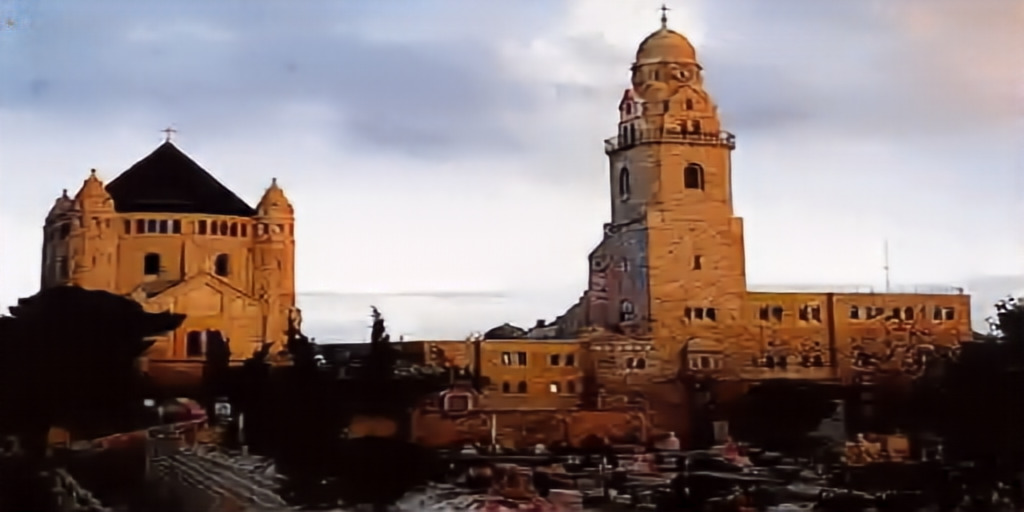}
        \rotatebox{90}{\scriptsize{(b) JPEG-SE (0.25 bpp)}}
    }
    \vspace{-11pt}
    \subfloat{
        \hspace{-.17in}
        \rotatebox{90}{\scriptsize{(c) BPG (0.09 bpp)}}
        \includegraphics[width=1\columnwidth]{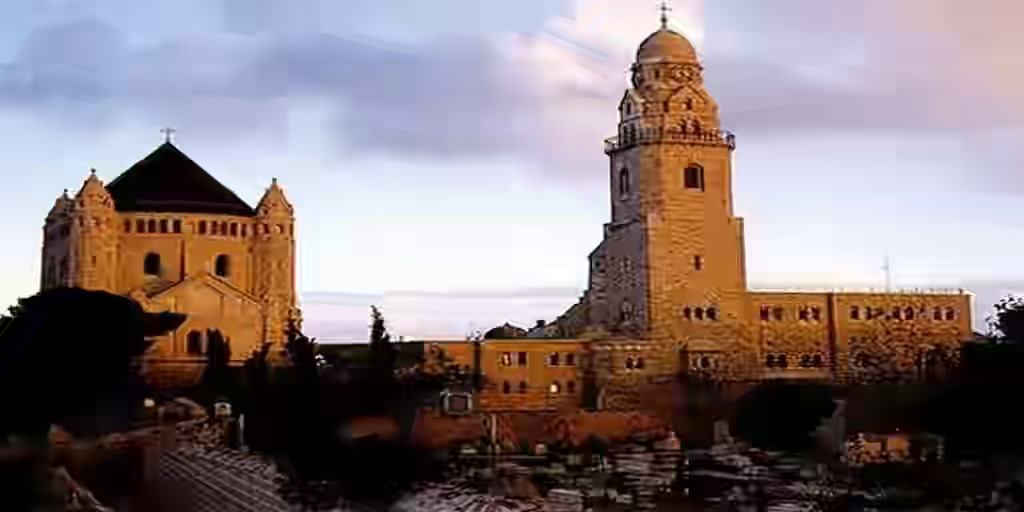}
    }
    \subfloat{
        \hspace{-.095in}
        \includegraphics[width=1\columnwidth]{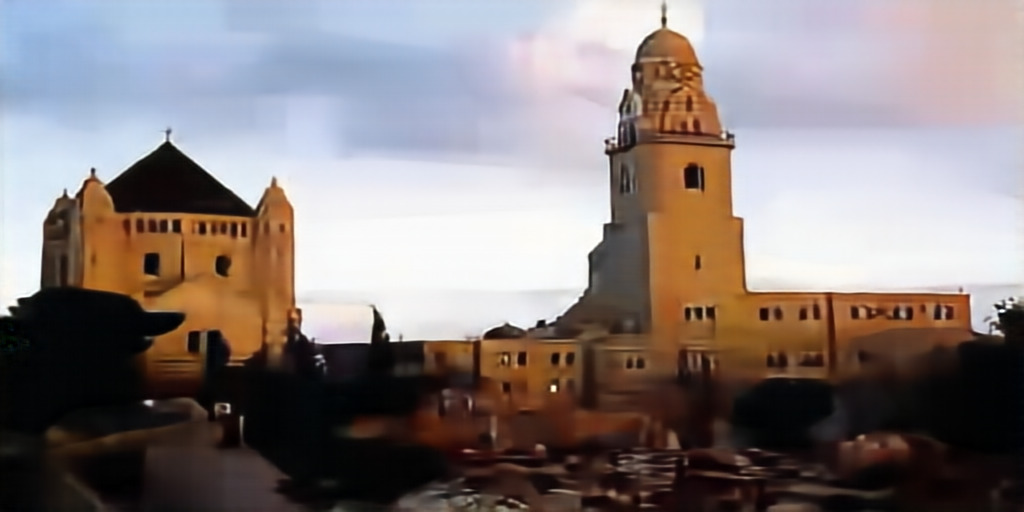}
        \rotatebox{90}{\scriptsize{(d) BPG-SE (0.07 bpp)}}
    }
    \vspace{-11pt}
    \subfloat{
        \hspace{-.17in}
        \rotatebox{90}{\scriptsize{(e) WebP (0.16 bpp)}}
        \includegraphics[width=1\columnwidth]{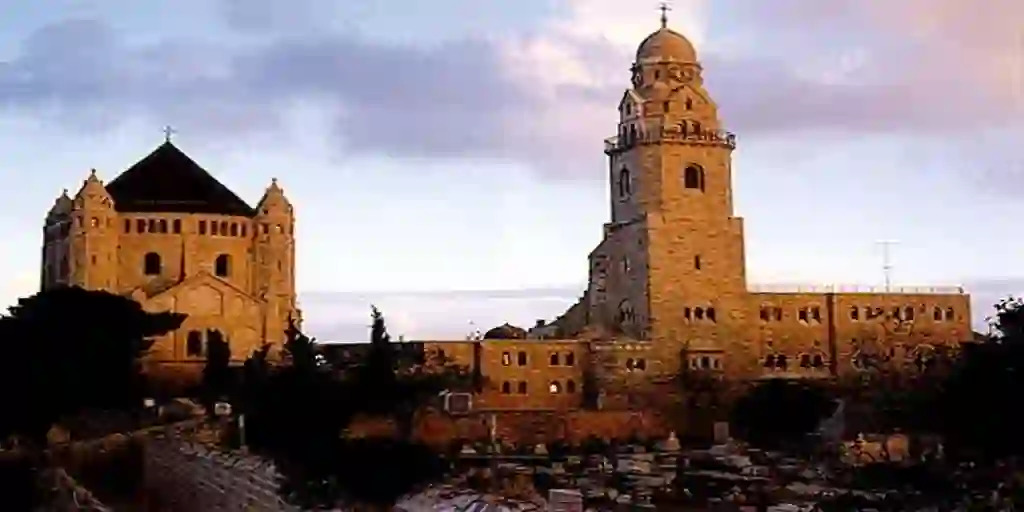}
    }
    \subfloat{
        \hspace{-.095in}
        \includegraphics[width=1\columnwidth]{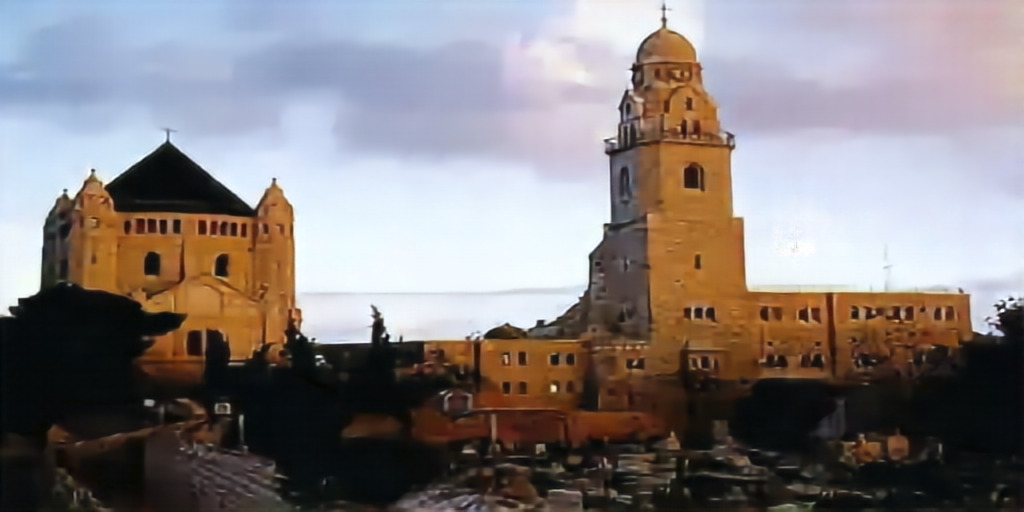}
        \rotatebox{90}{\scriptsize{(f) WebP-SE (0.14 bpp)}}
    }
    \vspace{-11pt}
    \subfloat{
        \hspace{-.17in}
        \rotatebox{90}{\scriptsize{(g) learned (0.08 bpp)}}
        \includegraphics[width=1\columnwidth]{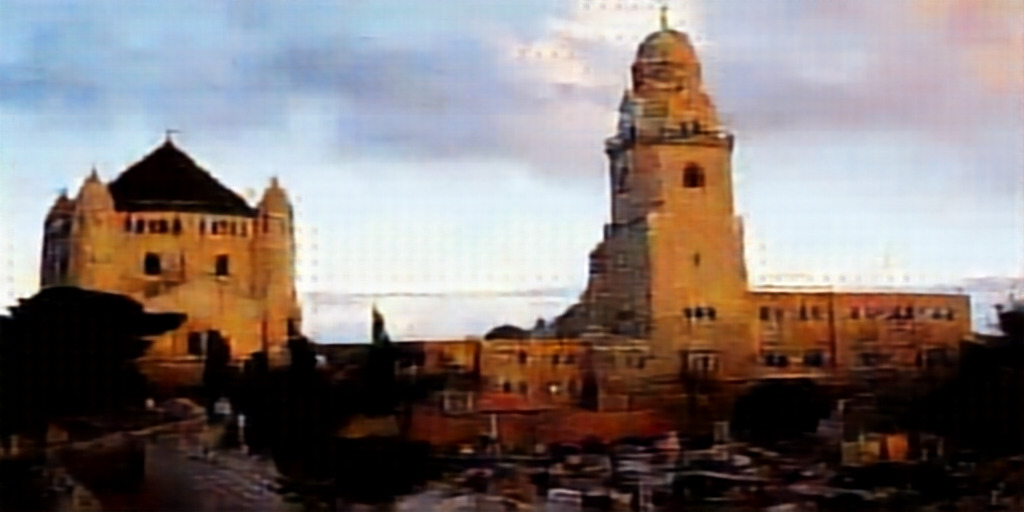}
    }
    \subfloat{
        \hspace{-.095in}
        \includegraphics[width=1\columnwidth]{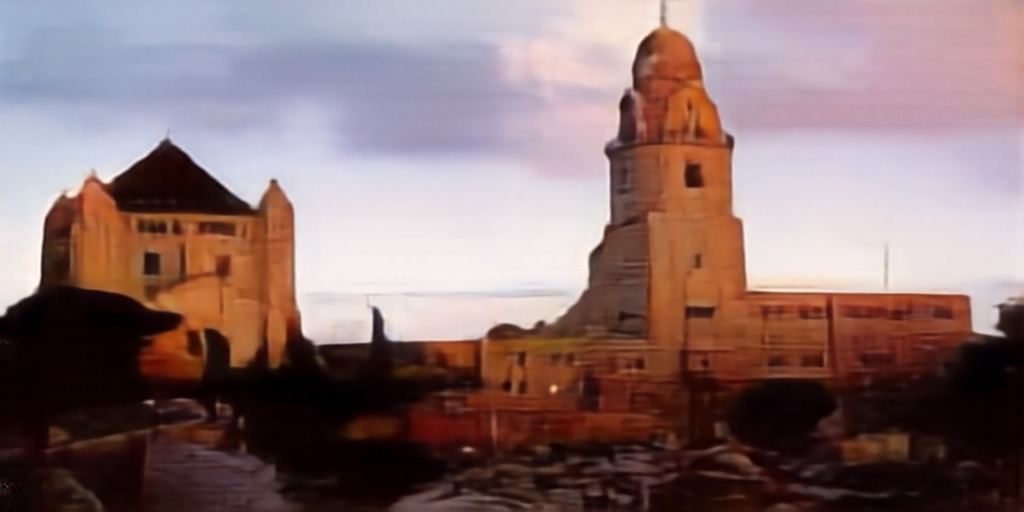}
        \rotatebox{90}{\scriptsize{(h) learned-SE (0.08 bpp)}}
    }
    \vspace{-11pt}
    \subfloat{
        \hspace{-.17in}
        \rotatebox{90}{\scriptsize{(i) JP2 (0.06 bpp)}}
        \includegraphics[width=1\columnwidth]{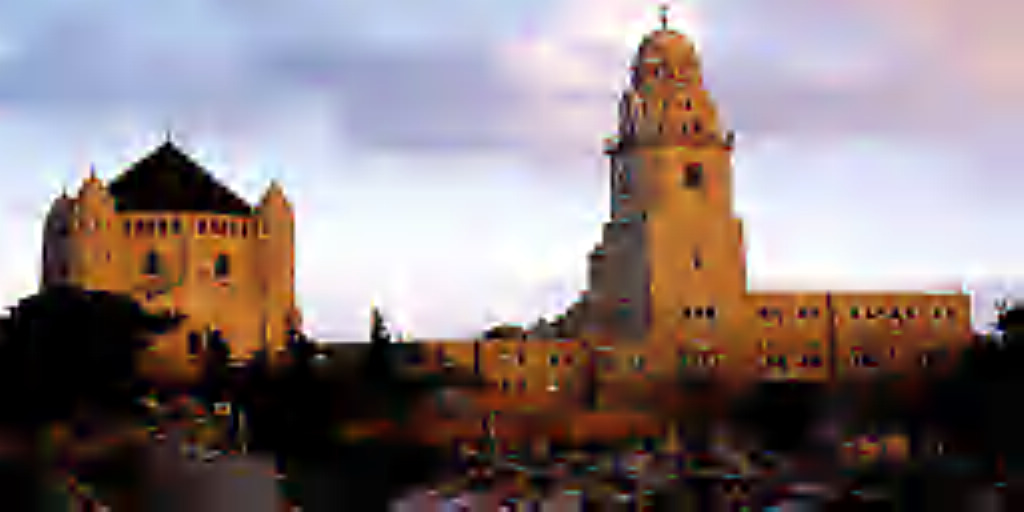}
    }
    \subfloat{
        \hspace{-.095in}
        \includegraphics[width=1\columnwidth]{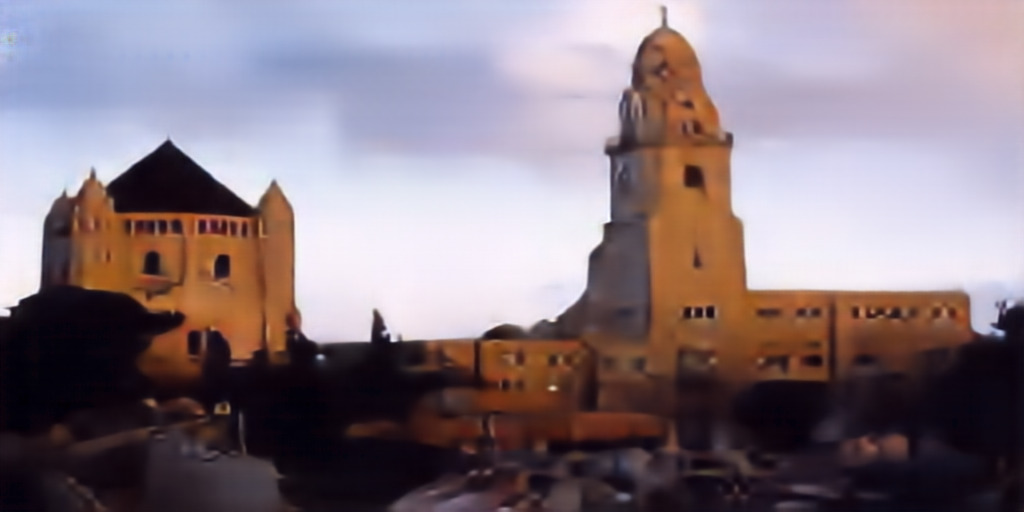}
        \rotatebox{90}{\scriptsize{(j) JP2-SE (0.09 bpp)}}
    }
    \vspace{-11pt}
    \subfloat{
        \hspace{-.17in}
        \rotatebox{90}{\scriptsize{(k) Original}}
        \includegraphics[width=1\columnwidth]{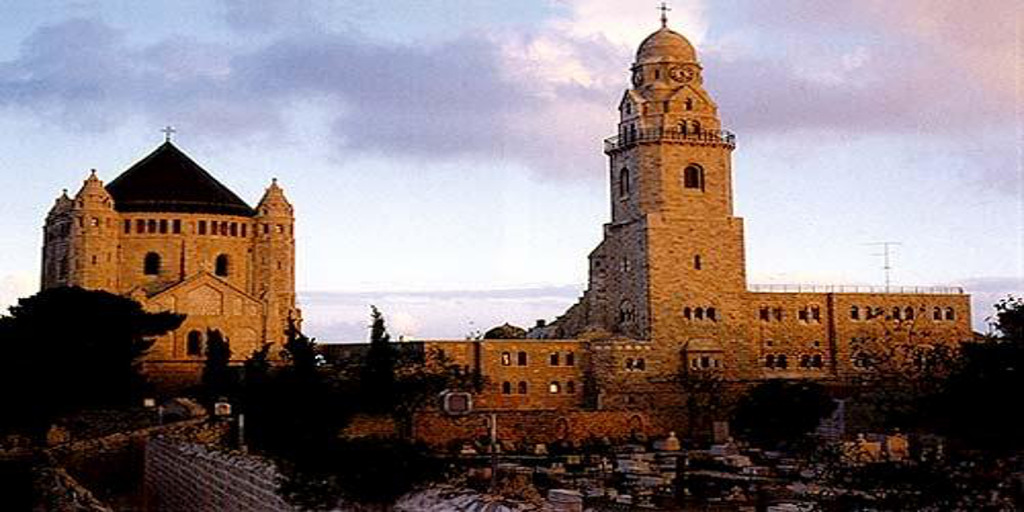}
    }
    \subfloat{
        \hspace{-.095in}
        \includegraphics[width=1\columnwidth]{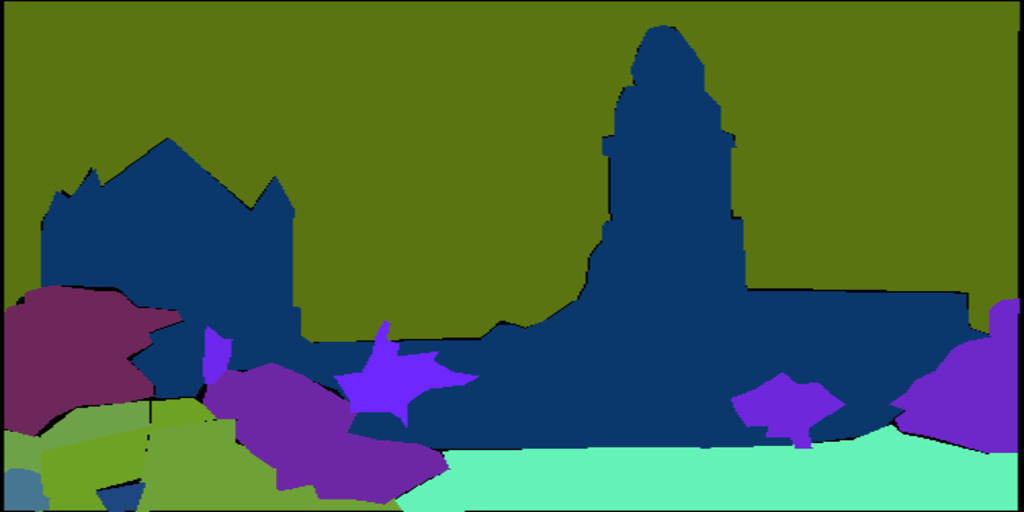}
        \rotatebox{90}{\scriptsize{(l) Semantics (class seg. map)}}
    }
    \caption{
        Visualizations for ADE20k.
    }
    \label{afig10}
\end{figure*}

\begin{figure*}[h]
    \thisfloatpagestyle{empty}
    \vskip -.7in
    \centering
    \subfloat{
        \hspace{-.17in}
        \rotatebox{90}{\scriptsize{(a) JPEG (0.33 bpp)}}
        \includegraphics[width=1\columnwidth]{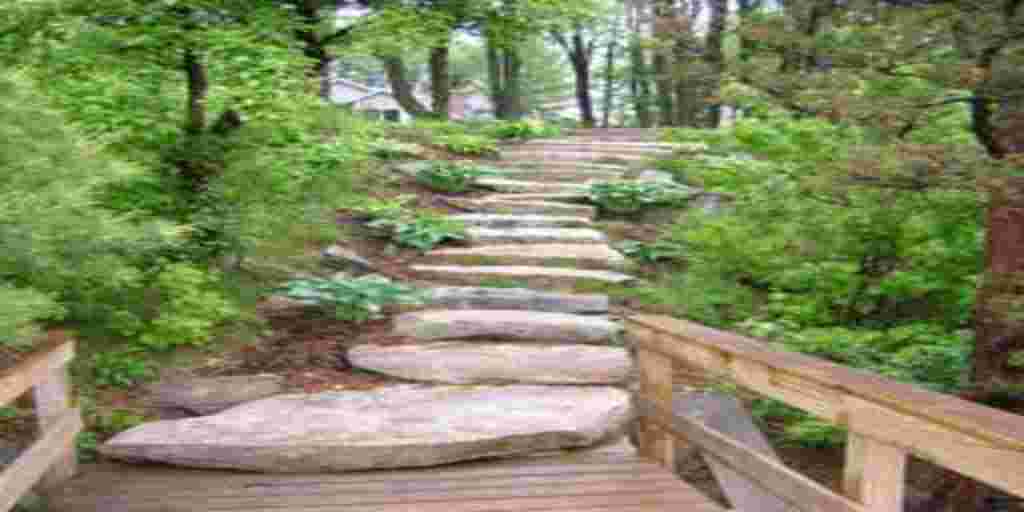}
    }
    \subfloat{
        \hspace{-.095in}
        \includegraphics[width=1\columnwidth]{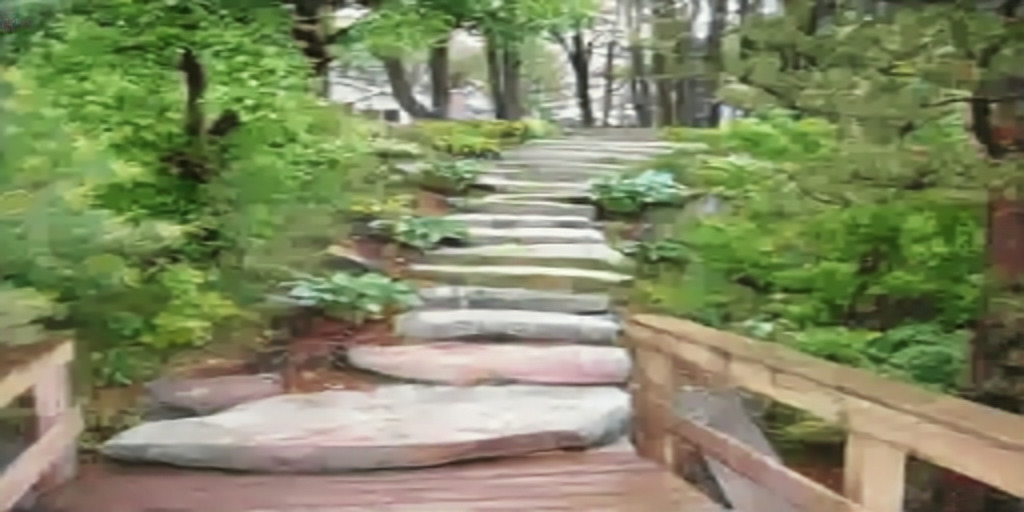}
        \rotatebox{90}{\scriptsize{(b) JPEG-SE (0.27 bpp)}}
    }
    \vspace{-11pt}
    \subfloat{
        \hspace{-.17in}
        \rotatebox{90}{\scriptsize{(c) BPG (0.11 bpp)}}
        \includegraphics[width=1\columnwidth]{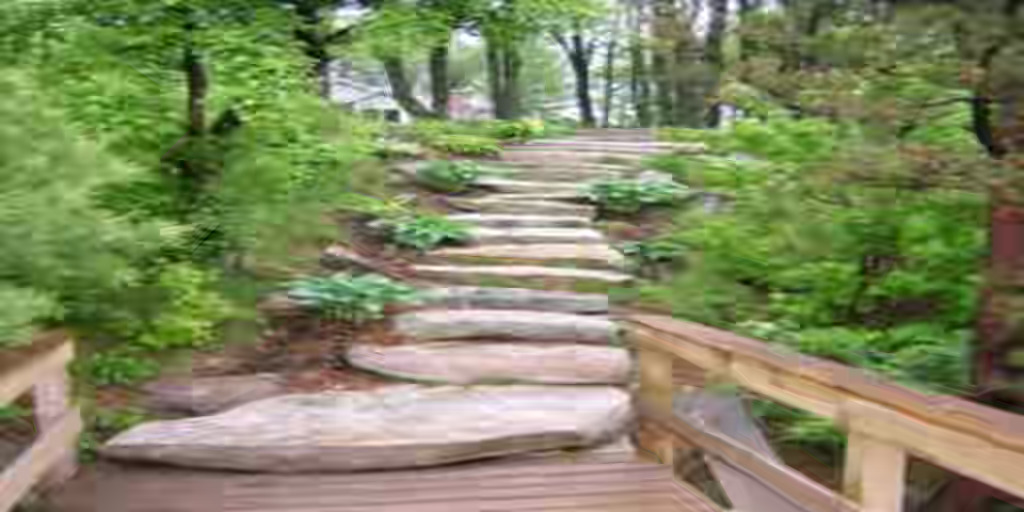}
    }
    \subfloat{
        \hspace{-.095in}
        \includegraphics[width=1\columnwidth]{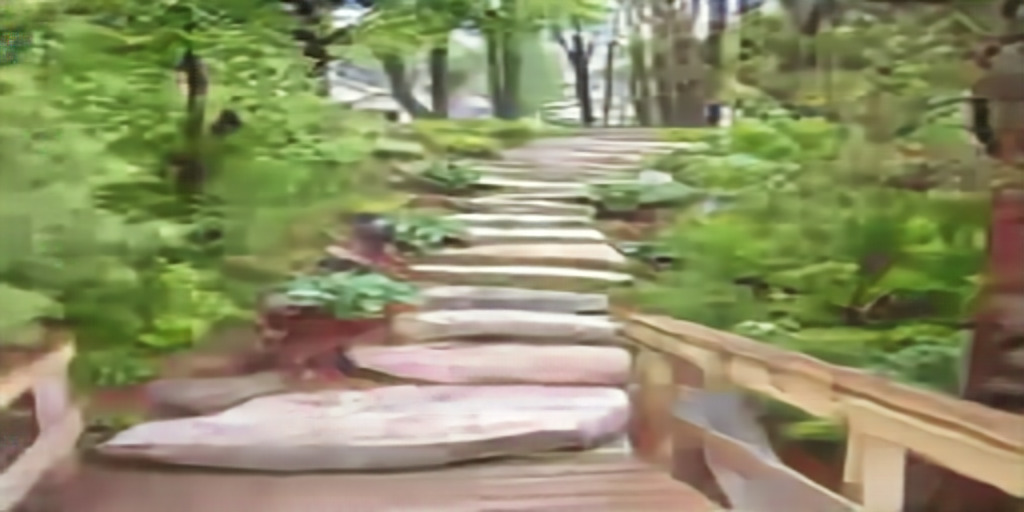}
        \rotatebox{90}{\scriptsize{(d) BPG-SE (0.08 bpp)}}
    }
    \vspace{-11pt}
    \subfloat{
        \hspace{-.17in}
        \rotatebox{90}{\scriptsize{(e) WebP (0.21 bpp)}}
        \includegraphics[width=1\columnwidth]{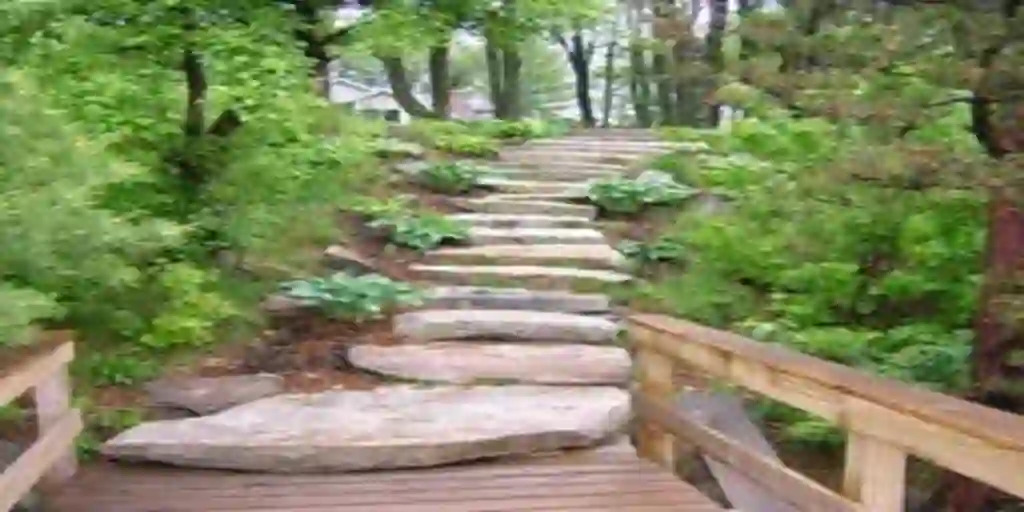}
    }
    \subfloat{
        \hspace{-.095in}
        \includegraphics[width=1\columnwidth]{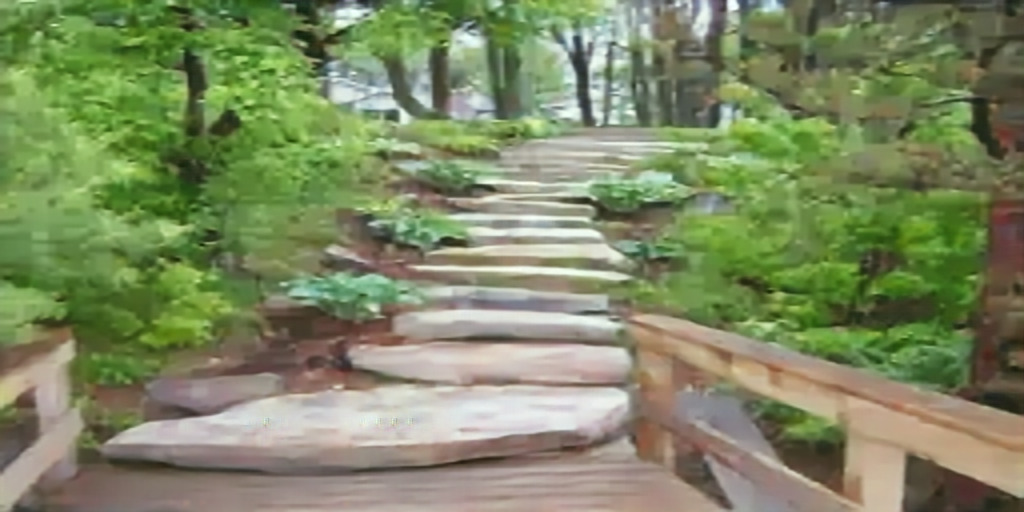}
        \rotatebox{90}{\scriptsize{(f) WebP-SE (0.17 bpp)}}
    }
    \vspace{-11pt}
    \subfloat{
        \hspace{-.17in}
        \rotatebox{90}{\scriptsize{(g) learned (0.11 bpp)}}
        \includegraphics[width=1\columnwidth]{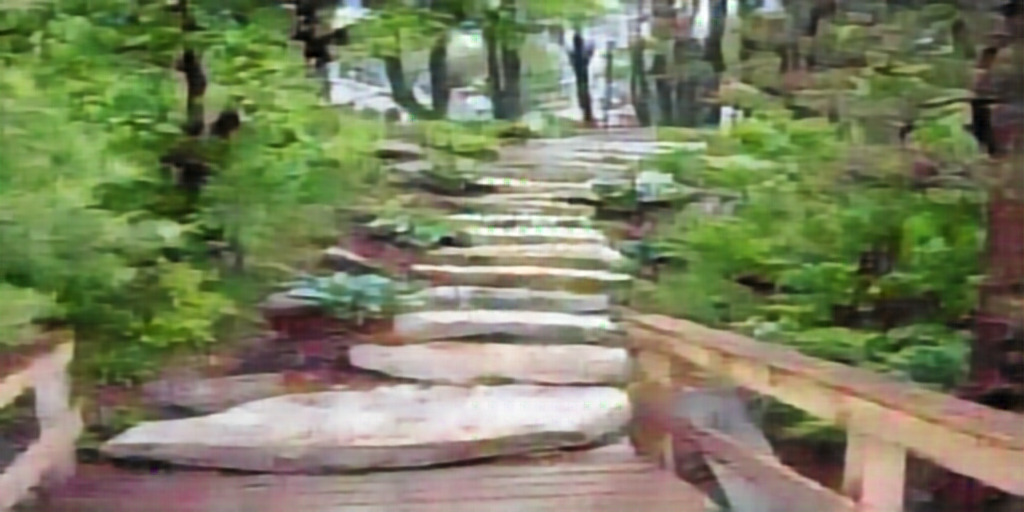}
    }
    \subfloat{
        \hspace{-.095in}
        \includegraphics[width=1\columnwidth]{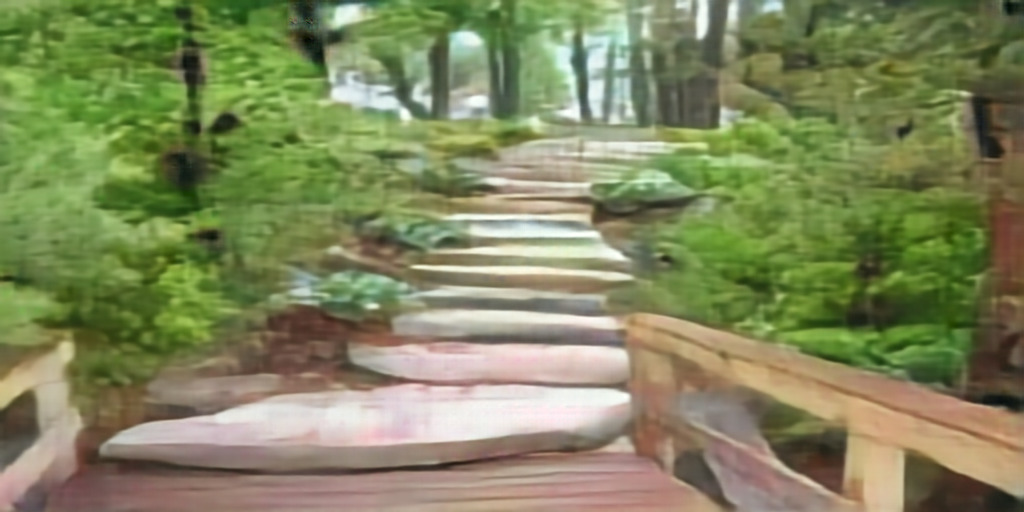}
        \rotatebox{90}{\scriptsize{(h) learned-SE (0.10 bpp)}}
    }
    \vspace{-11pt}
    \subfloat{
        \hspace{-.17in}
        \rotatebox{90}{\scriptsize{(i) JP2 (0.06 bpp)}}
        \includegraphics[width=1\columnwidth]{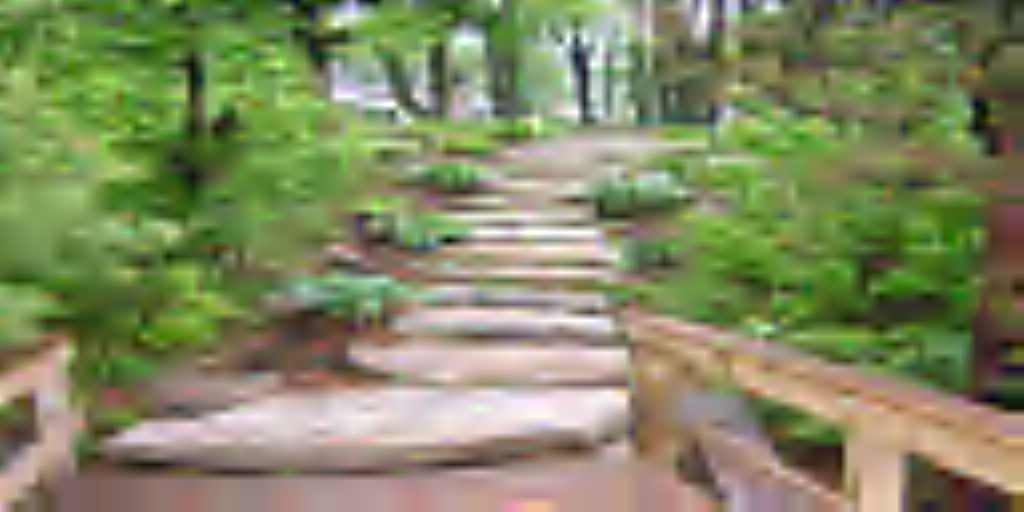}
    }
    \subfloat{
        \hspace{-.095in}
        \includegraphics[width=1\columnwidth]{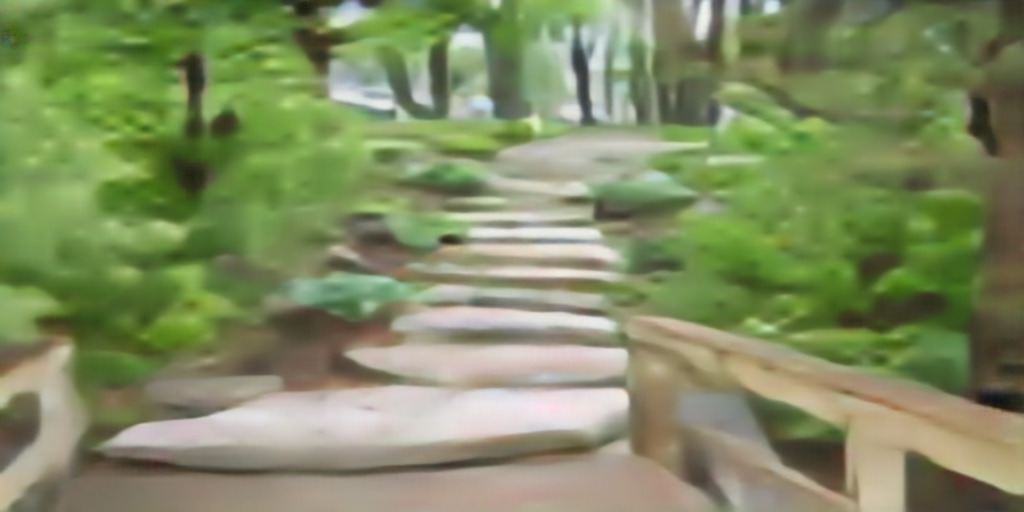}
        \rotatebox{90}{\scriptsize{(j) JP2-SE (0.09 bpp)}}
    }
    \vspace{-11pt}
    \subfloat{
        \hspace{-.17in}
        \rotatebox{90}{\scriptsize{(k) Original}}
        \includegraphics[width=1\columnwidth]{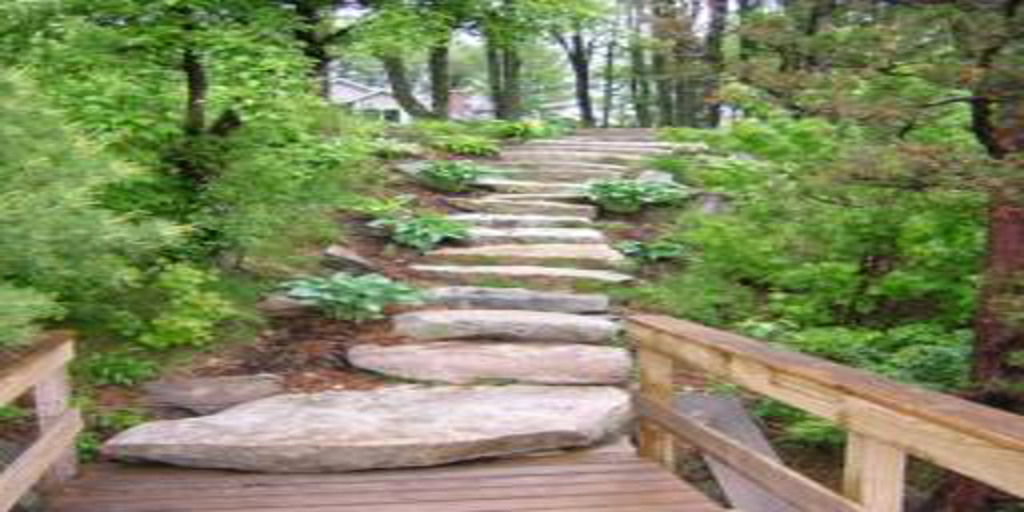}
    }
    \subfloat{
        \hspace{-.095in}
        \includegraphics[width=1\columnwidth]{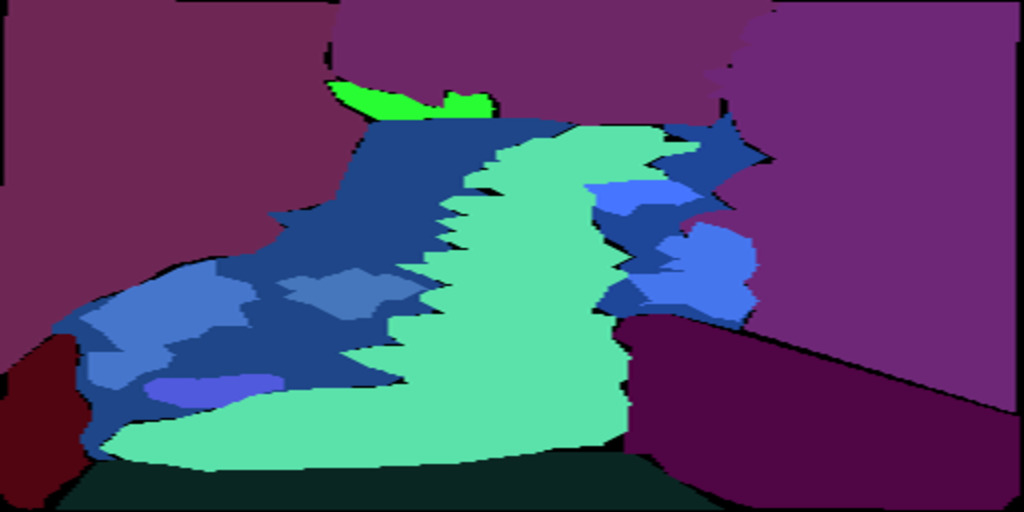}
        \rotatebox{90}{\scriptsize{(l) Semantics (class seg. map)}}
    }
    \caption{
        Visualizations for ADE20k.
    }
    \label{afig11}
\end{figure*}

See Fig.~\ref{afig13}, \ref{afig10}, and \ref{afig11} for visualizations from ADE20k.

\end{document}